%
\input mtexsis.tex
\input psfig
\def\widesinglespaced{%
   \baselineskip=\normalbaselineskip
   \multiply\baselineskip by 110
   \divide\baselineskip by 100
   \setRuledStrut
   \setTableskip}%
\preprint
\thicksize=\thinsize    
\widesinglespaced
\hfuzz=1.0pt
\superrefsfalse
\showsectIDtrue

%
\def\lsim{ \vcenter{\hbox{$\buildrel{\displaystyle <}\over\sim$}} }
\def\gsim{ \vcenter{\hbox{$\buildrel{\displaystyle >}\over\sim$}} }
%
\def\BB {{{\rm B}_B}}

\def\wc {{g^2 C_F\over 16\pi^2}}
\def\K {{\kappa}}
\def\latl {{$24^3\times 39$}}
\def\latA {{$16^3\times 39$}}
\def\latfps {{$16^3\times 25$}}

\def\bspfour {{$\beta\!=\!6.4$}}
\def\bspthr {{$\beta\!=\!6.3$}}
\def\bsptwo {{$\beta\!=\!6.2$}}
\def\bspone {{$\beta\!=\!6.1$}}
\def\bsix {{$\beta\!=\!6.0$}}
\def\bfpn {{$\beta\!=\!5.9$}}
\def\bfps {{$\beta\!=\!5.7$}}
\def\ainv {{a^{-1}}}
\def\ainvfpi {{a^{-1}_{f_\pi}}}
\def\ainvMrho {{a^{-1}_{M_\rho}}}
\def\ainvsigma {{a^{-1}_{\sigma}}}
\def\fBstat {{f_{B}^{\rm stat}}}

\def\lamqcd {{\Lambda_{QCD}}}

\def\Gax {{G_{A}}}
\def\GaxW {{G_{A}^{W}}}
\def\GaxWpm {{G_{A}^{W\pm}}}

\def\Gaxstat {{\hat{G}_{A}}}
\def\GaxstatSpm {{\hat{G}_{A}^{S\pm}}}
\def\GaxstatWpm {{\hat{G}_{A}^{W\pm}}}
\def\Gaxstatbare {{\hat{G}_{A}^{(0)}}}
\def\Gxx {{G_{B}}}
\def\GxxW {{G_{B}^{W}}}
\def\GxxWpm {{G_{B}^{W\pm}}}

\def\Gxxstat {{\hat{G}_{B}}}
\def\GxxstatSpm {{\hat{G}_{B}^{S\pm}}}
\def\GxxstatWpm {{\hat{G}_{B}^{W\pm}}}

\def\Zax {{\zeta_{A}}}
\def\Zaxp {{\zeta'_{A}}}
\def\Zaxstat {{\hat{\zeta}_{A}}}
\def\Zaxstatone {{\hat{\zeta}_{A}^{(1)}}}
\def\Zaxstattwo {{\hat{\zeta}_{A}^{(2)}}}
\def\Zxx {{\zeta_{B}}}
\def\Zxxp {{\zeta'_{B}}}
\def\Zxxstat {{\hat{\zeta}_{B}}}

\def\pproj {{\left({1+\gamma_0 \over 2}\right)}}
\def\mproj {{\left({1-\gamma_0 \over 2}\right)}}
\def\pmproj {{\left({1\pm\gamma_0 \over 2}\right)}}
\def\Udt {{U^{\dagger}_{0}}}
\def\sumx {{\sum_{\vec{x}}}}

\def\cE {{{\cal E}}}
\def\dE {{\delta\!{\cal E}}}
\def\dm {{\delta\!m}}
\def\cP {{{\cal P}}}
\def\cO {{{\cal O}}}

\def\cS {{{\cal S}}}
\def\ema {{e^{am}}}
\def\ematad {{e^{a\tilde m}}}

\def\vx {{\vec{x}}}
\def\vy {{\vec{y}}}
\def\vz {{\vec{z}}}
\def\chisq {{\chi^2}}
\def\chisqdof {{\chi^2/{\rm dof}}}
\def\chisqmin {{\chi^{2}_{\rm min}}}

\def\vs {{\rm vs.}}
\def\mone {{\tilde m}}
\def\mtwo {{\tilde m_2}}
\def\mthree {{\tilde m_3}}
\def\sigdotB {{\vec{\sigma}\!\cdot\!\vec{B}}}


\def\etal {{\it et al.}}
\def\ibid {{\it ibid.}}
\def\PRL {{Phys.~Rev.~Lett.}}
\def\PRD {{Phys.~Rev.~D}}
\def\physlett {{Phys.~Lett.}}
\def\nucphys {{Nucl.~Phys.}} 
\def\NP {{Nucl.~Phys.}} 
\def\nucphyssupp {{Nucl.~Phys.~B (Proc. Suppl.)}} 
\referencelist
\reference{eichten-lat87}
E.~Eichten, in {\sl Field Theory on the Lattice},
\journal \nucphyssupp;4,170(1988)
\endreference
%
\reference{HQsymmetry}
M.B.~Voloshin and M.A.~Shifman, \journal Sov.\ J.\ Nucl.\ Phys.;
45,292(1987);
H.D.~Politzer and M.B.~Wise, \journal \physlett;206B,681(1988);
N.~Isgur and M.B.~Wise, \journal \physlett;232B,113(1989);
\journal \physlett;237B,527(1990);
H.~Georgi, \journal \physlett;240B,447(1990)
\endreference
%
\reference{bernard-tasi}
For a review of this and related problems, see: 
C.~Bernard, {\it Weak Matrix Elements on and off the Lattice}, 
in {\sl From Action to Answers}, edited by T.~DeGrand
and D.~Toussaint, World Scientific Press, Singapore (1990);
C.~Bernard and A.~Soni, {\it Lattice Approach to Electroweak
Matrix Elements}
in {\sl Quantum Fields On the Computer}, 
edited by M.~Creutz, World Scientific Press,
Singapore (1993);
C.~Sachrajda, \journal \nucphyssupp;30,20(1993)
\endreference
%
\reference{bdhs}
C.~Bernard, T.~Draper, G.~Hockney, and A.~Soni, \journal \PRD;38,3540(1988)
\endreference
%
\reference{gavela-fb}
M.B.~Gavela, L.~Maiani, G.~Martinelli, O.~P\`ene, and S.~Petrarca, 
\journal \physlett;206B,113(1988)
\endreference
%
\reference{degrand-fb}
For similar work restricted to $D$ mesons see:
T.A.~DeGrand and R.D.~Loft, \journal \PRD;38,954(1988)
\endreference
%
\reference{alexandrou}
C.~Alexandrou, F.~Jegerlehner, S.~G\"usken, K.~Schilling, and R.~Sommer,
\journal \physlett;256B,60(1991)
\endreference
%
\reference{elc-stat}
C.~Allton, C.~Sachrajda, V.~Lubicz, L.~Maiani and G.~Martinelli,
\journal \nucphys;B349,598(1991)
\endreference
%
\reference{eichten-lat90}
E.~Eichten, G.~Hockney and H.B.~Thacker, in {Lattice 90}, 
\journal \nucphyssupp;20,500(1990)
\endreference
%
\reference{lat90}
C.~Bernard, J.~Labrenz and A.~Soni, \journal \nucphyssupp;20,488(1990)
\endreference
%
\reference{us-lat92}
C.\ Bernard, J.\ Labrenz, and A.\ Soni,
\journal \nucphyssupp;30,465(1993)
\endreference
%
\reference{hill2}
E.~Eichten and B.~Hill, \journal \physlett;240B,193(1990)
\endreference
%
\reference{wise}
M.B.~Wise, {\it New Symmetries of the Strong Interaction},
Lectures presented at the Lake Louise Winter Institute,
February 17-23, 1991 (Caltech preprint CALT-68-1721)
\endreference
%
\reference{boucaud}
Ph.~Boucaud, Lin Chyi Lung and O.~P\`ene, \journal \PRD;40,1529(1989)
and \journal \PRD;41,3541(1990)(E)
\endreference
%
\reference{hill1}
E.~Eichten and B.~Hill, \journal \physlett;234B,511(1990)
\endreference
%
\reference{boucaud2}
Ph.~Boucaud, J.P.~Leroy, J.~Micheli, O.~P\`ene and G.C.~Rossi, CERN-TH 6599/92
\endreference
%
\reference{ZA-pert}
B.~Meyer and C.~Smith, \journal \physlett;123B,62(1983);
G.~Martinelli and Y.C.~Zhang, \journal \ibid;123B,443(1983);
R.~Groot, J.~Hoek, and J.~Smit, \journal \NP;B237,111(1984)
\endreference
%
\reference{mackenzie-lat92}
P.~Mackenzie, \journal \nucphyssupp;30,35(1993);
and hep-lat/9212014, talk given
at DPF92, to appear in the proceedings
\endreference
%
\reference{kronfeld-lat92}
A.S.~Kronfeld, \journal \nucphyssupp;30,445(1993)
\endreference
%
\reference{lepage-lat91}
P.~Lepage, 
\journal \nucphyssupp;26,45(1992)
\endreference
%
\reference{lepmac-viability}
G.~Peter Lepage and Paul B. Mackenzie, FERMILAB-PUB-19/355-T (9/92)
\endreference
%
\reference{NRQCD}
B.A.~Thacker and G.~Peter Lepage, \journal \PRD;43,196(1991)
\endreference
%
\reference{smearing}
The introduction of smeared sources and early numerical studies can
be found in:
A.~Billoire, E.~Marinari and G.~Parisi, \journal \physlett;162B,160(1985);
P.~Bacilieri, \etal, \journal \NP;B317,509(1989).
Smeared sources for the static approximation
were first used in\cite{eichten-lat89}
\endreference
\reference{wall-source}
The wall source technique was first used with staggered fermions.
See R.~Gupta, G.~Guralnik, G.W.~Kilcup, and S.R.~Sharpe,
\journal \PRD;43,2003(1991)
\endreference
%
\reference{eichten-lat89}
E.~Eichten,  G.\ Hockney, and H.B.\ Thacker,
\journal \nucphyssupp;17,529(1990)
\endreference
%
\reference{newalexandrou}
C.~Alexandrou,  S.\ G\"usken, F.\ Jegerlehner, K.\ Schilling, and R.\ Sommer,
CERN-TH 6692/92; \journal \nucphyssupp;30,453(1993)
\endreference
\reference{lepage-tasi}
G.P.~Lepage, {\it The Analysis of Algorithms for Lattice Field Theory},
in {\sl From Action to Answers}, edited by T.~DeGrand
and D.~Toussaint, World Scientific Press, Singapore (1990)
\endreference
%
\reference{lat91}
C.~Bernard, C.M.~Heard, J.~Labrenz and A.~Soni, 
\journal \nucphyssupp;26,384(1992)
\endreference
%
\reference{eichten-lat91}
E.~Eichten, \journal \nucphyssupp;26,391(1992)
\endreference
%
\reference{toussaint}
D.~Toussaint, {\it Error Analysis of Simulation Results: A Sample Problem},
in {\sl From Action to Answers}, edited by T.~DeGrand
and D.~Toussaint, op. cit
\endreference
%
\reference{numrec}
W.~Press, B.~Flannery, S.~Teukolsky and W.~Vetterling,
{\sl Numerical Recipes: The Art of Scientific Computing},
Cambridge University Press, 1986, ISBN 0-521-38330-7
\endreference
%
\reference{gupta-dyn}
See, for example, R.\ Gupta, C.F.\ Baillie, R.G.\ Brickner, G.W.\ Kilcup,
A.\ Patel and S.R.\ Sharpe, \journal \PRD;44,3272(1991)
\endreference
%
\reference{seibert}
D.~Seibert, CERN-TH.6892/93, hep-lat/9305014
\endreference
%
\reference{quenched-chiral}
C.\ Bernard and M.\ Golterman, \journal \nucphyssupp;30,217(1993);
S.\ Sharpe, \journal \nucphyssupp;30,213(1993)
\endreference
%
\reference{cb_mg}
C.\ Bernard and M.\ Golterman, \journal \nucphyssupp;26,360(1992);
Phys. Rev. D46 (1992) 853
\endreference
%
\reference{grinstein-chiral}
B.~Grinstein, E.~Jenkins, A.~Manohar, M.J.~Savage, and M.~Wise,
\journal \nucphys;B380,369(1992)
\endreference
%
\reference{wise-chiral}
M.B.~Wise, \journal \PRD;45,2188(1992)
\endreference
%
\reference{jfd}
G.~Burdman and J.~Donoghue, \journal \physlett;280B,287(1992)
\endreference
%
\reference{hashimoto}
S.~Hashimoto and Y.~Saeki, Mod. Phys. Lett.A7:387(1992)
\endreference
%
\reference{eichten-etal}
A.\ Duncan, E.\ Eichten, A.X.\ El-Khadra,
J.M.\ Flynn, B.R.\ Hill and H.\ Thacker, 
\journal \nucphyssupp;30,433(1993)
\endreference
%
\reference{simone}
J.~Simone, presented at {\it Lattice '92} for the UKQCD Collaboration,
\journal \nucphyssupp;30,461(1993)
\endreference
%
\reference{bali-schilling}
G.S.\ Bali and  K.\ Schilling,
\journal \PRD;47,661(1993)
\endreference
%
\reference{GF11}
F.\ Butler, H.\ Chen, J.\ Sexton, A.\ Vaccarino and D.\ Weingarten,
IBM-RC-18617, Dec.\ 1992
\endreference
%
\reference{sharpe}
G.W.\ Kilcup, S.R.\ Sharpe, R.\ Gupta and A.\ Patel, \journal \PRL;64,25(1990)
\endreference
%
\reference{mark2}
J.~Adler {\it et al.}, \journal \PRL;60,1375(1988)
\endreference
%
\reference{expt}
WA75 Collaboration
(S. Aoki, \etal), CERN-PPE/92-157 (Sept., 1992).
Submitted to Prog.\ Theor.\  Phys
\endreference
%
\reference{elc-6.4}
A.\ Abada, {\it et al.},
\journal \nucphys;B376,172(1992)
\endreference
\endreferencelist
\pageno=1


\pubcode{UW/PT-93-06;\ \ Wash.\ U.\ HEP/93-30;\ \ BNL-49068}
\titlepage
\title 
A Lattice Computation of the Decay Constants of $B$ and $D$ Mesons.
\endtitle
\author
Claude W. Bernard
Department of Physics, Washington University, St. Louis, MO 63130
\endauthor
\author
James N. Labrenz
Department of Physics FM-15, University of Washington, Seattle, WA 98195
\endauthor
\and
\author 
Amarjit Soni
Department of Physics, Brookhaven National Laboratory, Upton, NY 11973
\endauthor
\abstract
A lattice calculation of the pseudoscalar decay constant of heavy-light
mesons is reported. Results are obtained (in the quenched approximation)
from lattices at \bspthr\ through a procedure that interpolates between
the static approximation of Eichten and the conventional ("heavy" Wilson
fermion) method. The previously observed discrepancy between these two
approaches has been resolved: we find the scaling quantity
$f\sqrt{M}$ to be significantly smaller than previous calculations had
indicated ({\it e.g.} at \bsix); in addition, we discuss a modification 
which is required in normalizing the conventional amplitude to correct
for large-$am$ lattice errors. This change guarantees that
$f\sqrt{M}$ will smoothly approach its value in the static limit.  
\null From the numerical interpolation of the static and intermediate-mass
results, we find, in units of MeV, 
$f_B=187(10)\pm34\pm15$, $f_{B_s}=207(9)\pm34\pm22$,
$f_D=208(9)\pm35\pm12$ and $f_{D_s}=230(7)\pm30\pm18$,
where the first error is statistical and the second two are
estimates of systematics due to 1) fitting and large-$am$ effects
and 2) scaling. The ratios are better determined: 
$f_D/f_{D_s}$, $f_B/f_{B_s}$, $f_B/f_D$
and $f_{B_s}/f_{D_s}$ are all 0.90 within a total error of less than 0.05.  
The purely static values are: $f_B^{\rm stat}= 235(20)\pm 21\ \MeV$,
$f_{B_s}^{\rm stat}= 259(19)\pm 19\ \MeV$, and
$f_B^{\rm stat}/f_{B_s}^{\rm stat}= 0.90(2)\pm 0.02$.
Finally, using lattices at \bspthr, \bsix\ and
\bfps\ and extrapolating to the limit of zero lattice spacing,
we have computed $f_K/f_\pi = 1.08\pm.03\pm.08$ in 
the quenched approximation, where the first error includes 
statistical and fitting
errors, and the second is an estimate of the error in extrapolation
to the continuum limit.
\endabstract
\endtitlepage

\section{Introduction}

The calculation of transition amplitudes for heavy-light mesons
has recently been a topic of great interest, from both a theoretical
and a phenomenological standpoint. Theoretically, the static
limit, in which the mass of the heavy quark is taken {\it a priori}
to infinity, has been used to define an effective theory for quantum
chromodynamics (QCD) which has additional flavor and spin
symmetries\cite{eichten-lat87}\cite{HQsymmetry}.
In continuum calculations, these symmetries
have been used to compute relations among various
heavy-meson form factors. On the lattice, the symmetry limit
is equally well defined: the heavy quark is replaced with a static 
color source, so only the light quark degree of freedom is
subject to the usual constraint imposed by the lattice cutoff,
$m_q\ll 1/a$, where $m_q$ denotes the quark mass and $a$ the grid
spacing. This approach, introduced in\cite{eichten-lat87}, provides
a method for explicit calculation of heavy-light amplitudes,
such as the static limit of the pseudoscalar decay constants $f_{Qq}$.
(We use $Q$ to denote a heavy quark, such as $b$ or $c$,
and $q$ a light quark, $u$,$d$ or $s$.)

\null From the point of view of phenomenology, a great deal of emphasis has
in particular been placed on $f_B$. Notwithstanding the physical process
through which it is defined ({\it i.e.}, the leptonic decay of the $B$), which
may in the future be accessible to experiment, $f_B$ already plays
an important role in Standard Model (SM) physics.
As a characteristic example, consider the usual parameterization of
the $B\!-\!\overline{B}$ mixing parameter, $x_{bd}\equiv \Delta M/\Gamma$.
\null From the evaluation of the top-quark-dominated box diagram, one finds 
$$
        x_{bd} \sim h(m_t)|V_{td}|^2f^{2}_B \BB,
\EQN mixing
$$
where $h$ is a rapidly increasing function of $m_t$, the top quark mass, 
$V_{td}$ is a CKM matrix element, and $\BB$ is the so-called
``B parameter'' relating the full matrix element of the weak interaction
Hamiltonian to its value in the vacuum insertion approximation%
\cite{bernard-tasi}. 
The measurement of $x_{bd}$ only determines
the fundamental SM parameter $V_{td}$ if $f_B$ and $\BB$ are known.

With respect to lattice QCD, there are essential (technical) differences
which affect the calculation of these two quantities. Because it is computed
as a dimensionless ratio, $\BB$ is more amenable than $f_B$
to the ``conventional'' method ({\it i.e.}, using one light and one
heavy Wilson quark). These calculations have indicated that vacuum
insertion ({\it i.e.}, $\BB\simeq1$) becomes a fairly
good approximation at moderately heavy quark masses; the extrapolation of
this result to the $B$ meson is then straightforward\cite{bdhs}\cite{gavela-fb}.
The large-$am$ normalization effects (see Sects. \use{sect.largeam1} and
\use{sect.largeam2}) cancel in the ratio that defines $\BB$.
For $f_B$ such leading-order lattice errors will no longer cancel.
Nonetheless, a similar strategy is possible in principle:
one may simulate moderately heavy mesons where large-$am_Q$
lattice errors are presumed to be small and look for 
asymptotic behavior in $f$.
Indeed, in the static limit one has\cite{eichten-lat87}
$$
        \phi \equiv f\sqrt{M} \subrightarrow{M\to\infty} {\rm const.}
\EQN scalelaw
$$
(up to corrections logarithmic in $m_Q$, the heavy-quark mass%
\cite{HQsymmetry}). This may then provide a guide for extrapolating
to the physical $B$ meson\cite{bdhs}\cite{gavela-fb}\cite{degrand-fb}.

In early calculations at moderate couplings
({\it i.e.}, \bfps, 5.9 and 6.0), it became evident
that there was a significant discrepancy between the two approaches:
$\fBstat$, the decay constant computed in the static limit, was much
larger than extrapolations from lighter masses indicated it should be%
\cite{alexandrou}\cite{elc-stat}\cite{eichten-lat90}\cite{lat90}. 
It was difficult to know whether one or both of the methods suffered
from large systematic errors, and because of this, interpolating
between them---as a way of computing the finite-mass corrections to
the scaling law \Ep{scalelaw}---was certainly unreliable.
In this paper we demonstrate the elimination of this discrepancy;
thus we are able to interpolate smoothly between the intermediate-mass
regime of the conventional calculation and the infinite-mass limit
of the static one in order to compute physical amplitudes.
The discrepancy has disappeared as a result of
two essential differences. First, as compared to the results mentioned
above, we find a smaller static result for $\phi$ from our computation
at \bspthr. Second, we find that ``large-$am_Q$'' errors must be
corrected for---even when only ``moderately'' heavy quarks
are used---in order for the conventional amplitude to smoothly approach
the correct static limit as the heavy-quark mass $m_Q$ is increased.

This paper is organized as follows. In Section 2 we discuss some essential
details of the calculation, emphasizing in particular how the correction
of large-$am_Q$ errors in the conventional method ensures that it
(approximately) match, in the limit $m_Q\rightarrow\infty$, with the
static result. The numerical techniques used
throughout the calculation are described in Section 3. In section 4,
we discuss the details of
the analysis and present results for $f_B$, $f_{B_s}$, $f_D$, $f_{D_s}$ and
their jackknifed ratios, including estimates for various systematic errors.
We summarize and make concluding remarks in Section 5.
Some of these results have been reported previously\cite{us-lat92}.

\section{Lattice Decay Constants\label{chpt.calc}}

\vskip -0.3cm
\subsection{The Conventional Method\label{sect.construction}}

The decay constant $f_P$ for a pseudoscalar $P$
(with arbitrary quark masses $m_Q$ and $m_q$) is defined as the vacuum
to single-particle matrix element of the axial-vector current:
$$
   \bra{0}A_{\mu}^{\rm cont}(x)\ket{P(\vec{p})} = 
                -if_Pp_{\mu}e^{-ip\cdot x}, 
\EQN fp 
$$
where our normalization is such that $f^{\rm expt}_\pi=132$ MeV.
We compute $f_P$ through the evaluation of two lattice 
correlation functions, 
$$\EQNalign{
   \Gax(t-t_0) &\equiv \sum_{\vec{x}}T
                \bra{0} A_0(\vec{x},t)\chi^\dagger(\vec{0},t_0) \ket{0},
\EQN gax;a \cr
\noalign{ \hbox{and}}
   \Gxx(t-t_0) &\equiv \sum_{\vec{x}}T
                \bra{0} \chi(\vx,t)\chi^\dagger(\vec{0},t_0) \ket{0}.
\EQN gax;b \cr}
$$
For the conventional calculation, we write $A_\mu$, the lattice axial
current, and $\chi$, the pseudoscalar interpolating operator, as 
$$\EQNalign{
        A_\mu(x) &= \overline{q}(x)\gamma_\mu\gamma_5Q(x);
\EQN currents;a \cr
        \chi(x)  &= \overline{q}(x)\gamma_5Q(x),
\EQN currents;b \cr}
$$
where the fields $q(x)$ and $Q(x)$ are the ``light'' and ``heavy'' 
fermionic fields respectively, defined via the standard Wilson 
action with quark hopping parameters $\K_q$ and  $\K_Q$.
In this section, we choose $\chi$ to be the local operator given above,
primarily to simplify the discussion of various normalization issues
which are independent of the source type. 
For the actual computations, we have used both local sources and 
extended (or ``smeared'') sources in a fixed gauge, the latter as a
means of improving the numerical precision of the results.
The specific techniques will be discussed in Section 3; their
actual implementation in the computations, in Section 4.

Evaluating the Wick contractions of the quark fields,
\Eqs{gax} are written as a configuration average of contracted
light- and heavy-quark propagators, $S_q$ and $S_Q$. For example,
$$
  \Gax(t) \phantom{a} = \phantom{a}
        -\left< \sumx \Tr[ S_{q}^\dagger(\vx,t;\vec{0},0;U)
        \gamma_0 S_Q(\vx,t;\vec{0},0;U)]
        \right>_{\{U\}}.
\EQN gaxcomp
$$
The essential formula for the calculation follows from the insertion
of a complete set of states between the interpolating operators in \Eq{gax}.
Excited states are exponentially damped, and thus at large times
(with $t_0\equiv0$) we obtain
$$\EQNdoublealign{
   \Gax(t) & \subrightarrow{|t| \to \infty}
        {\bra{0} A_0 \ket{P}\bra{P} \chi^\dagger \ket{0} \over 2M_P}
        e^{-aM_P|t|} & \equiv \Zax e^{-aM_P|t|},
\EQN zee;a \cr
\noalign{ \hbox{and}}
   \Gxx(t) & \subrightarrow{|t| \to \infty}
        {\bra{0} \chi \ket{P}\bra{P} \chi^\dagger \ket{0} \over 2M_P}
        e^{-aM_P|t|} & \equiv \Zxx e^{-aM_P|t|}.
\EQN zee;b \cr}
$$
\null From \Ep{fp} and \Ep{zee}, $f_P$ is given by
$$
\phi_P =  f_P\sqrt{M_P} = 
        C_A\phantom{;}\sqrt{2\zeta^{2}_{A} \over \Zxx}\phantom{;}a^{-3/2}.
\EQN fplatt
$$
$C_A$ is the normalization constant between the continuum
current $A_{\mu}^{\rm cont}$ and its lattice counterpart $A_\mu$.
Although we have illustrated the computation of the correlators 
using the conventional method (i.e., \Eq{gaxcomp}), \Eq{fplatt},
with suitable modifications to the constant $C_A$, holds also for
the static-quark approach. We discuss the details of the static and
conventional normalizations in the following sections.

\subsection{The Static Effective Theory \label{sect.SQET}}

In this section we discuss the implementation of the static-quark 
method, emphasizing in particular how the lattice calculation is
normalized to obtain continuum physics.
We begin by writing the discretized effective action\cite{hill2},
$$
        \cS^{(0)} = \sum_{x} \overline{h}(x) \pproj
        \left[ h(x) - U^{\dagger}_0(x-\hat{0})h(x-\hat{0})\right],
\EQN bareS 
$$
where $h(x)$ denotes the static {\it quark} (as opposed to
anti-quark) field at site $x$.
In the particle's rest frame the lower two components of $h$ vanish,
a condition enforced by the constraint equation\cite{wise}
$$
        \gamma_0h = h.
\EQN constraint
$$
The quark and anti-quark are decoupled; we omit the analogous 
formulas pertaining to the latter.

Renormalization requires the addition of a mass counterterm 
$a\dm \overline{h}h$ to the action \Ep{bareS}, yielding\cite{hill2}
$$\EQNalign{
        \cS^{(\delta m)} & = \sum_{x,y}\overline{h}(x)S^{-1}_h(x,y)h(y),
\EQN Sm \cr
\noalign{\hbox{where}}
        S^{-1}_h(x,y) & = (1+a\dm) \pproj
        \left[ \delta^4_{x,y} - {1 \over 1+a\dm}
        U^{\dagger}_0(y)\delta^4_{x-\hat{0},y} \right].
\EQN ginv \cr}
$$
The matrix $S_{h}^{-1}$ is then inverted for the
static-quark propagator in the background gauge configuration $U$:
$$\EQNalign{
   S_h(x,y)\hphantom{a} =& \hphantom{a}{1\over (1+a\dm)} 
        \hphantom{a}e^{-{\rm ln}(1+a\dm)(t_x-t_y)}
        \hphantom{a}\theta(t_x-t_y) \hphantom{a}\delta^3_{\vec{x},\vec{y}} 
        \pproj \cr
        & \times \left\{ \Udt(\vec{y},t_y)\Udt(\vec{y},t_y+1)
        \Udt(\vec{y},t_y+2)...\Udt(\vec{y},t_x-1) \right\}
\cr
        \equiv& {1\over (1+a\dm)} \hphantom{a}e^{-{\rm ln}(1+a\dm)(t_x-t_y)}
        \hphantom{a}S^{(0)}_h(x,y).
\EQN statprop \cr}
$$
We denote quantities computed in the static
approximation with a ``hatted'' version of the notation
used in Section \use{sect.construction}.
The static-light current, $\hat A_\mu$, and the 
interpolating operator, $\hat\chi$, are defined via the substitution 
$Q(x) \rightarrow h(x)$ in \Eqs{currents}, and the construction of
the correlation functions $\Gaxstat$ and $\Gxxstat$ is analogous to \Eqs{gax}.
To compute the correlators, the heavy-quark propagator is simply replaced 
by the static one, as given by \Ep{statprop}.  The calculation of $\hat\phi$ 
then follows according to \Eqs{zee} and \Ep{fplatt}. 

However, since the heavy-quark mass has been formally removed from the theory
the correlation functions $\Gaxstat$ and $\Gxxstat$ no longer
fall exponentially with the meson mass $M_P$, as in \Eqs{zee},
but instead with the bare and divergent ``binding energy,'' $\cE_0=M_P-m_Q$.
In the Wilson case, the bare mass, or its equivalent in terms
of the hopping parameter, appears in the action, so the mass
divergence can be absorbed in the usual way.
Here, {\it i.e.}, in \Eq{bareS}, the bare mass is absent---it has been absorbed 
in the definition of the field---and the counter term must be explicitly
added.  In the literature this issue has been addressed concurrently with
the calculation of $\hat C_A$, the normalization constant for the 
static-light axial current ($A^{\rm cont}_{\mu}=\hat C_A \hat A_\mu$),
because that calculation must include in some way the mass renormalization 
$a\dm$, which enters into the overall normalization 
of the heavy-quark propagator through \Eq{statprop}.

Two groups 
(Boucaud et al\cite{boucaud} and Eichten and Hill\cite{hill1}\cite{hill2})
have calculated the renormalization constant which matches
$\hat A_\mu$ to the full axial current of QCD to one loop in perturbation
theory. We write this constant as the product $\hat Z_AC(a,m)$, where
$$\EQNalign{
        C(a,m) &= 1 + \wc({3\over2}\ln(a^2m^2)),
\EQN Cam \cr
\noalign{\hbox{and}}
        \hat{Z}_A &= 1 - \wc(J+2),
\EQN za-stat \cr}
$$
where $C_F=4/3$ for the case of $SU(3)$.
For reasons which will be made clear below, we have split off into
$C(a,m)$ the dependence on the heavy quark mass.
Before accounting for the effect of mass renormalization, {\it i.e.},
the term $1/(1+a\dm)$ in \Eq{statprop}, Eichten and Hill 
find\cite{hill2} $J=30.35$.  
(After including mass renormalization, one gets J=20.38 --- see below.)
They propose two methods for the complete matching.

In order to illustrate their approach, let us consider the 
point-source correlator\Footnote\ddag{
	In practice we always use a smeared interpolating operator for
	the static computations. The result of this discussion, however,
	is unaffected by this modification; here we use a point source
	for simplicity.}
$$
  \Gaxstat(t) = \sumx
  \bra{0} \hat{A}_0(\vec{x},t),\hat{\chi}^{\dagger}(\vec{0},0) \ket{0}.
\EQN gaxstat
$$
Again specifying $t>0$, we obtain
$$\EQNalign{
  \Gaxstat(t) \phantom{a} &= \phantom{a}
        \left< \sumx \Tr[ \gamma_0 S_h(x,0) \gamma_5S_q(0,x)\gamma_5] 
        \right>_{\{U\}}
\cr
        &= \phantom{a}
        \left< \sumx \phantom{a} \delta^3_{\vec{x},\vec{0}}
        \Tr[ {\cal P}(t,0) 
        \pproj S^{\dagger}_q(x,0) ] \right>_{\{U\}}
        {1\over (1+a\dm)} \phantom{a} e^{-\ln(1+a\dm)t}
\cr
        &\equiv \phantom{a}
        \Gaxstatbare(t)
        \phantom{a}e^{-a\dE}
        \phantom{a}e^{-a\dE t},
\EQN gaxstatcomp \cr}
$$
where $\Gaxstatbare(t)$ is the ``bare'' correlator which follows from
the action \Ep{bareS},
$\cP(t,0)$ stands for the appropriate product of link matrices, as
given in \Ep{statprop}, and we have defined 
$a\dE = \ln(1+a\dm)$.
The two scenarios of Ref.\cite{hill2} are then:
\bigskip
\vfill\eject

\noindent
$\underline{\rm Method\ (1)}$ \quad Parameterize ({\it i.e.}, fit) 
$\Gaxstatbare$ as $\Zaxstat e^{-a\cE_{0}t}$.  
Then \Ep{gaxstatcomp} is written as
$$
        \Gaxstat(t)=(\Zaxstat e^{-a\dE}) e^{-a(\cE_0+\dE)t},
$$
which shows explicitly how the linear divergence in $\cE_0$ is 
removed to define the physical binding energy,
$\cE = \cE_0 + \dE \simeq \cE_0 + \dm$. 
But in this case, the residue of $\Gaxstat$ includes
a correction to the bare fitted value. In other words, we should
use $\Zaxstatone$ instead of $\Zaxstat$ in the static analog of
\Eq{fplatt}, where
$$
        \Zaxstatone = \Zaxstat e^{-a\dE}.
\EQN method1 
$$
\bigskip

\noindent
$\underline{\rm Method\ (2)}$ \quad Parameterize $\Gaxstatbare$ as 
$\Zaxstattwo e^{-a\cE_{0}(t+1)}$
so that \Ep{gaxstatcomp} is instead written as
$$
        \Gaxstat(t)=\Zaxstattwo e^{-a(\cE_0+\dE)(t+1)}
$$
The physical binding energy emerges as before, and it is 
trivial to see that the residue now satisfies the equation
$$
        \Zaxstattwo = \Zaxstat e^{a\cE_0}=
        \Zaxstat e^{a(\cE-\dE)} = \Zaxstatone e^{a\cE}.
\EQN method2
$$
(Note that
$\Zaxstat$ is still {\it defined} by the fit of method (1).)
\bigskip

\Eq{method2} demonstrates explicitly that the two methods differ
at finite $a$.
Method (2), chosen in Ref.\cite{hill2}, appears to us
to be an unnatural 
choice which introduces $\cO(a)$ terms in order to accommodate a suggestive
form of the propagator.
Method (1), however, simply parallels the standard procedure for 
Wilson quarks. This can be seen as follows.
A rescaling of the static field in the action \Ep{Sm}, given by
$$
        h \rightarrow (1+a\dm)^{-\half}h,
\EQN stat-rescale
$$
normalizes the diagonal term of \Ep{ginv} to unity, 
an analogous procedure to the quark field rescaling
$$
        \psi \rightarrow ({1 \over 2\kappa})^{-\half} \psi
\EQN wilson-rescale
$$
which leads to the standard form of the Wilson action.
In the latter case, quark-mass renormalization is achieved through
tuning the hopping parameter $\kappa$ such that the correct physical 
spectrum is reproduced, and a factor $\sqrt{2\kappa}$ for each quark
field is ultimately replaced in physical amplitudes. 
If the physical binding energy $\cE$ of the heavy-light state
were experimentally measurable,
a similar tuning procedure, instead of a perturbative calculation,
could be used to determine $a\dm$ in the static case.
Either with tuning or with perturbation theory, the rescaling  
of the static field(s) in physical amplitudes 
following  \Ep{stat-rescale} is equivalent to method (1). 

Therefore, $\Zaxstat$ and $\Zxxstat$ will denote the residues 
computed from a fit to the bare correlators; we use them in the 
static analog of \Eq{fplatt}. 
The remaining factor of $e^{-a\dE/2} \simeq 1-a\dm/2$
is absorbed into the perturbative renormalization of the static-light 
axial current. From\cite{hill2} we find
$$
        a\dm = -\wc\phantom{;} 19.95
\EQN adeltam
$$
and thus obtain
$$
        \hat{Z}_A = 1 - \wc\phantom{;} 22.38.
\EQN Zastat
$$ 
(A rigorous treatment of the ambiguity discussed above
has been given by Boucaud et al\cite{boucaud2}, and
they show that method (2) is in fact inconsistent with
an $\cO(a)$-improvement of the light-quark action.
\Eq{Zastat} is in agreement with their result.)
Including the factor $\sqrt{2\K}$ for the light quark, we thus obtain
$$
        \hat C_A = \sqrt{2\K_q}\hat{Z}_A C(a,m_Q).
\EQN CAhatold
$$

\subsection{The Naive Large-Mass Limit\label{sect.hopping-limit}}

The conventional normalization for the light meson decay constants
is given by
$$
        C_A = Z_A \sqrt{2\K_Q}\sqrt{2\K_q},
\EQN CAold
$$
where $Z_A$ is the perturbative renormalization of the lattice
axial current\cite{ZA-pert},
$$
        Z_A = 1 - .133g^2,
\EQN za-conv
$$
and the factors of $\sqrt{2\K}$ undo the initial re-scaling of the
Wilson quark fields, indicated by \Eq{wilson-rescale}.
Since we wish to interpolate between the static and conventional
methods, it is instructive to consider the latter in the naive 
limit $m_Q\rightarrow \infty$ ({\it i.e.}, $\K_Q\rightarrow 0$). 
In this case the propagator can be approximated 
by the leading term in the hopping-parameter expansion,
$$\EQNalign{
        S_Q(x,0) &\subrightarrow{\K_Q \to 0} (2\K_Q)^{t}
        \hphantom{;}\delta^3_{\vec{x},\vec{0}} 
        \hphantom{;}\theta(t)
        \pproj \cP(t,0)
\EQN hpe \cr
        &=\ e^{-\ln(1/2\K_Q)t}\ \times\ S^{(0)}_h(x,0).
\cr}
$$
Using \Ep{hpe} to compute the bare (unnormalized) heavy-light
correlation function, we obtain identically the bare
static-light result,
but with a heavy-quark rest mass contributing to the inverse 
correlation length:
$$\EQNalign{
        \Gax &\rightarrow \Zaxstat e^{-(aM_P)t},
\EQN gaxlim;a \cr
        \Gxx &\rightarrow \Zxxstat e^{-(aM_P)t},
\EQN gaxlim;b \cr
\noalign{\hbox{and}}
        aM_P &\subrightarrow{\K_Q\to0} \ln({1\over 2\K_Q}).
\EQN bigmass \cr}
$$
Therefore, 
using \Ep{CAold} and  \Ep{CAhatold}, 
one finds the limiting expression
$$
        \phi_P \subrightarrow{\K_Q\to0} \hat{\phi}_P
        \phantom{;}e^{-aM_P/2}\phantom{;}(1+\cO(\alpha_s)),
\EQN noema
$$
where the factor of $e^{-aM_P/2}$ just comes from the $\sqrt{2\K_Q}$
in \Ep{CAold}.
Thus, when $\K_Q$ is pushed to absurdly small values, a pure lattice
artifact (the factor $e^{-aM_P/2}$ in \Ep{noema}) will dominate in the
heavy-light amplitude. 

\subsection{The Leading-Order Large-$am$ Correction\label{sect.largeam1}}

\Eq{noema} demonstrates that the systematic lattice error
when $am_Q \gg 1$ can be removed in the static limit by
a trivial redefinition of the normalization constant $C_A$.
This modification can be formulated in terms of a
large-$am$ corrected normalization for the Wilson quark.
In order to derive it, let us first consider the free theory.
Computing the spatially-summed propagator both from the Wilson action 
(with $U=1$) and in the continuum, one finds the normalization 
equation
$$
  2\kappa e^{am} \sum_{\vec{x}} \bra{0} \psi(x)\overline{\psi}(0) 
    \ket{0}^{\rm latt}
   = \int d^3\vec{x} \bra{0} \psi(x)\overline{\psi}(0) \ket{0}^{\rm cont},
\EQN propnorm
$$
where
$$
  am = \ln(1 + {1\over 2\kappa} - {1\over 2\kappa_{c}^{(0)}});\quad\quad
        \kappa_{c}^{(0)}={1\over 8}.
\EQN ambare
$$
Based on this relation, or on equivalent arguments, a change in the 
tree-level Wilson quark normalization factor of 
$$
        \sqrt{2\K}\rightarrow\sqrt{2\K\ema} 
\EQN leading-correction
$$
has been
suggested\cite{mackenzie-lat92}\cite{kronfeld-lat92}\cite{lepage-lat91}.
Note that \Eq{propnorm} is valid for arbitrary $am$, but that it
is a zero-momentum (or static) relation. 
In the regime where $am\ll1$, the zero-momentum projection is irrelevant,
because the usual factor of $\sqrt{2\K}$ is recovered.
Conversely, since a typical simulation of light hadrons requires that
$\ainv\gg\lamqcd$, the condition $am\gsim1$ implies that $m_Q\gg\lamqcd$, 
and therefore that the heavy-quark dynamics in this regime are
non-relativistic.  
Within the context of a non-relativistic expansion of the heavy-quark
propagator, we thus refer to the factor $\ema$ in \Ep{leading-correction}
as the large-$am$ correction for the leading-order term. 
It guarantees a smooth transition between the light- and static-quark
regimes. In the remaining part of this section, we discuss
the inclusion of interactions, and in the following 
section we examine the large-$am$ lattice errors in the next-to-leading 
terms of the quark propagator in this limit. 

A full one-loop correction to \Ep{propnorm} for arbitrary $am$ has not
been computed. However, in lattice perturbation theory, the primary 
contribution to a variety of short-distance dominated quantities 
is produced by tadpole graphs. Lepage and Mackenzie have demonstrated
that a mean-field approach can be used to improve the matching between
lattice operators and their continuum counterparts; in a rough sense,
this technique sums tadpole contributions to all orders in 
$\alpha_s$\cite{lepmac-viability}. (The sense is rough because
``summing the tadpoles'' is not a gauge-invariant procedure
beyond leading order, but the mean field approach of
\cite{lepmac-viability} is gauge invariant.)
The ``tadpole correction'' is easily made
to the free-theory equation \Ep{propnorm} because it is independent
of the quark momentum. In the mean-field language, one substitutes
the link matrices in the Wilson action with the mean value $u_0$. 
Computationally, the theory is then equivalent to the free one
with a hopping parameter shifted by 
$$
        \kappa \rightarrow u_0\kappa.
\EQN K-shift
$$
Thus tadpole improvement of the leading-order correction ``$\ema$'' is
given by $am \rightarrow a\tilde{m}$, where, from \Ep{ambare},
$$
  a\tilde{m} = \ln({1\over 2u_0\kappa} - 3).
\EQN am
$$
Several defining relations for the quantity $u_0$ are possible; all
give similar results, either when computed non-perturbatively or to one 
loop in perturbation theory using a boosted coupling\cite{lepmac-viability}.
We use the definition 
$$
        u_0 = {\K^{(0)}_c\over \K_c} = {1\over 8\K_c}, 
\EQN unot
$$
where $\K_c$ is the critical hopping parameter, as determined by the 
extrapolation to zero pion mass. 

Now let us return to $C_A$, the normalization constant for the axial current.
The tadpole-improved large-$am$ correction requires the
substitution for the quark-line normalization factor, 
$\sqrt{2\K} \rightarrow \sqrt{2u_0\K e^{a\tilde m}}$. However, if used
in \Eq{CAold} as it stands, this will double count the 
leading tadpole perturbative correction, which already contributes
to $Z_A$. Let us write the perturbative renormalization in the form
$$
        Z_A = Z_1 Z_{2a}^{-1} Z_{2b}^{-1},
$$
where $Z_1$ is the vertex correction, $Z_{2a}$ is the
wave-function renormalization from the continuum-like self-energy
graph, and $Z_{2b}$ is the wave-function renormalization from the
tadpole. It is easily verified that to one loop in perturbation theory,
$u_0$, the tadpole contribution to $1/8\K_c$, is the inverse of 
$Z_{2b}$, 
when computed in Feynman gauge.
Thus, in $C_A$, one may either define a new constant $Z_{A}'$ by removing the
factor of $Z_{2b}$, or else eliminate the $u_0$ which accompanies the
leading factors of $\sqrt{2\K}$. We use the latter definition so that
in the light-quark limit, where $\ematad\rightarrow1$,
the traditional normalization constant is recovered. Since we evaluate
$Z_A$ using a boosted coupling, there is numerically little difference 
between the two approaches.

The lattice-continuum normalizations are then given explicitly by
$$\EQNdoublealign{
   C_A(\K_Q,\K_q) &= Z_A\sqrt{2\K_Qe^{a\tilde m_Q}} 
                \sqrt{2\K_qe^{a\tilde m_q}} \qquad &{\rm (conventional)};
\EQN CAnew \cr
   \hat C_A(\K_q) &= \hat{Z}_A C(a,m_Q) \sqrt{2\K_q e^{a\tilde m_q}}
                \qquad &{\rm (static)},
\EQN CAhatnew}
$$
where $a\tilde m_{Q,q}$ are given by \Ep{am} with $\K=\K_{Q,q}$.
With these definitions, the limiting expression \Ep{noema} becomes
$$
        \phi_P \subrightarrow{\K_Q\to0} \hat{\phi}_P (1+\cO(\alpha_s)).
\EQN ema
$$

Finally, let us consider this limit in detail in order to evaluate
the extent of the mismatch at $\cO(\alpha_s)$.
In $C_A$ we can associate with each quark line a factor of
$$
        Z_{2}^{-\half}\sqrt{2\K e^{a\tilde m}},
$$
where $Z_2\equiv Z_{2a}Z_{2b}$.
As a simplification, we use the perturbative relation 
$u_0=1/Z_{2b}$, despite the fact that in practice we evaluate $u_0$
non-perturbatively.  We then find
$$\EQNalign{
        Z_{2}^{-\half}\sqrt{2\K_Qe^{a\tilde m_Q}} &=
         Z_{2a}^{-\half}\sqrt{u_02\K_Q} 
         [{1\over2u_0\K_Q} - 3 ]^\half
\cr
        & \subrightarrow{\K_Q \to 0} Z_{2a}^{-\half}.
\cr}
$$
In the limit $\K_Q\rightarrow0$, there is a full cancellation of tadpole 
corrections, leaving only the continuum-like wave function renormalization.

In the static-light renormalization $\hat Z_A$, 
the same ``tadpole cancellation'' already exists
between the mass renormalization term discussed in Section \use{sect.SQET}
and the wave function renormalization; {\it i.e.},
$$
        (1 - {a\dm^{(b)}\over 2})\hat{Z}_{2b}^{-\half} = 1
$$
(the label ``$b$'' again denotes the tadpole contribution).
This relation can be verified explicitly from the calculations
in\cite{hill2}. The remaining mismatch thus results purely from 
non-tadpole perturbative corrections---both the wave function of 
the heavy (static) quark and the vertex graph for both theories.
Although the non-tadpole coefficients are small 
in the conventional calculation, they are not particularly so
in the static one.
At \bspthr\ for example, using $\ainv\sim3$~GeV, $m_Q\sim5$~GeV and 
$g^2\sim 1.6$, we find
$$
        \hat C_A/C_A(\K_Q\rightarrow0) \sim 1-\wc 11.2 \sim 0.85.
$$
Since we do not approach the limit $\K_Q\rightarrow0$ in 
practice, we expect that the $15\%$ difference seen here is
an overestimate of the actual error incurred when interpolating between
moderately heavy Wilson quarks ({\it i.e.}, $am_Q\sim1$) and the static limit. 
The systematic error from this effect is estimated concurrently with
large-$am$ errors in Sect. \use{sect.analysis2}.

\subsection{Large-$am$ Errors at Next-to-Leading Order\label{sect.largeam2}}

By interpolating between moderately heavy mesons and the static limit,
we would like to compute the deviation from the scaling law \Ep{scalelaw}
in the mass range of the $D$ and $B$ mesons. 
To do this, we adopt the phenomenological parameterization
$$
        \phi_P = c_0 (1 + {c_1 \over M_P} + {c_2 \over M_P^2}  +
        \cO({1\over M_P^3}))
\EQN analysis
$$
and analyze the lattice results (both conventional heavy-light 
and static) in the form $\phi\ \vs\ 1/M_P$.
If the agreement between the two methods is good,
we should find $c_0\approx\hat\phi$; $c_1$ will characterize the
finite-mass correction. (We discuss the logarithmic correction to
\Ep{analysis} in the following section.)

However, the removal of large-$am$ lattice errors described in the 
previous section has only been made to leading order for a non-relativistic
Wilson quark ({\it i.e.}, for the static component). It is possible to demonstrate
explicitly---again by taking the infinite-mass limit---that the conventional
calculation will not behave according to \Ep{analysis} because
lattice effects distort the dynamics of the heavy quark.
In the hopping parameter expansion, corrections to the leading
behavior \Ep{hpe} are of order $2\K_Q$; thus, using \Ep{bigmass}, 
one finds that the limiting expression on the lattice will be
$$
        \phi_P \subrightarrow{\K_Q\to0} \hat{\phi}_P(1 + Ce^{-aM_P}).
$$
At very large quark mass, the $1/M_P$ terms which should be
present in the conventional amplitude
will thus be exponentially suppressed by lattice artifacts.

In the non-relativistic limit, the Wilson action can be rewritten
by separating the quark and anti-quark components through a
Foldy-Wouthuysen transformation.
After including the correct leading-order normalization for the
field $\psi$, the action for quarks is expanded, in terms of a
two-component spinor $\phi$, in a discretized version of the following
form\cite{lepage-lat91}\cite{kronfeld-lat92}\cite{mackenzie-lat92}:
$$
        S_W = \phi^{\dagger}[m + D_0 + {1\over2m_2}\vec{D}^2
                + {g\over4m_3}\sigdotB + \dots ]\phi.
\EQN NRaction
$$
In the limit that $am\ll1$, the mass parameters in this expansion
will all be approximately equal, and continuum physics will be well
approximated at sufficiently weak couplings. However, 
\Ep{NRaction} is obtained without placing any restriction on $am$,
and when $am\gg1$, ratios of the masses $am$, $am_2$ and $am_3$ diverge
exponentially. We refer to this effect as the large-$am$ error at
next-to-leading order.

In the free theory, or in the mean-field approximation, the masses
$am_2$ and $am_3$ can be easily calculated. The kinetic mass, $am_2$, 
results from the dispersion relation,
$$
        E(\vec{k}) = m + {\vec{k}^2 \over 2m_2} + \dots,
$$
which is obtained from the calculation of the free Wilson quark 
propagator in the non-relativistic limit. We find
$$
        am_2 = {e^{am}\sinh{am} \over \sinh{am} +1}.
\EQN mtwo
$$
The mean-field substitution $am\rightarrow a\mone$ in \Ep{mtwo}
defines $a\mtwo$, the tadpole-improved kinetic mass.
The coefficient of the $\sigdotB$ term can be obtained by
splitting $1/2m_2$, the coefficient of the $\vec{D}^2$ term, into
two pieces, one which is contributed by the naive lattice action and 
one which is contributed by the Wilson term. The former is equivalent
to $1/2m_3$, because the Wilson term adds no $\sigdotB$ piece to the action. 
The result is\cite{kronfeld-lat92}
$$
        am_3 = am_2(\sinh am + 1),
\EQN mthree
$$
and tadpole corrections are again included via the mean-field
method to define $a\mthree$.

The non-relativistic limit of the Wilson action \Ep{NRaction} is
similar in form to the non-relativistic QCD (NRQCD) action proposed
recently by Lepage and Thacker for the study of heavy-quark bound states%
\cite{NRQCD}. In NRQCD, the action is discretized {\it a priori} in this form;
the coefficients of the higher-dimensional operators
are free to be tuned in order to match the effective theory to full QCD.
Although the standard Wilson action does not afford this freedom, it is,
under certain conditions, suitable for the simulation of non-relativistic
quarks even when $am\gg1$\cite{lepage-lat91}. For example,
in the case of quarkonium ({\it i.e.}, a heavy-heavy meson), the $\sigdotB$
interactions are suppressed by a factor of the velocity
squared and thus can be neglected in the lowest-order calculation.
This fact follows from the non-relativistic nature of the bound state.
Furthermore, the rest mass $m$ only contributes an overall constant to
the action; it is irrelevant in the calculation of transition amplitudes.
(In NRQCD, $m$ is typically removed from the theory from the beginning.)
Thus, if one computes an arbitrary heavy-heavy amplitude 
as a function of the meson pole mass, say $A(M)$, then to lowest order
$M$ sees the the rest mass $\mone$, whereas $A$ sees the kinetic mass $\mtwo$.
This error can be approximately corrected by adjusting the meson mass:
$$
        aM\ \rightarrow\ aM'\ =\ aM + (a\mtwo-a\mone).
\EQN mshift
$$
Then $A(M')$ gives the correct functional dependence, within additional
$\cO(\alpha_s)$ corrections which are not included by the tadpole
approximation.

In the current heavy-light analysis, we make the
shift \Ep{mshift} in the pole mass $M_P$ to remove large-$am$ errors
at next-to-leading order to the extent that we can.
But because in this bound state the light quark is relativistic,
the $\sigdotB$ interactions of the heavy quark are no longer suppressed.
At large $am$, we are still faced with the problem that 
$a\mthree\not=a\mtwo$.
Thus, although the mass shift removes the exponential suppression of
the $1/M$ term in the amplitude, the subsequent analysis of the 
lattice results according to \Eq{analysis} will not necessarily produce the 
correct coefficient $c_1$. The systematic error induced by this effect
in the calculation of $f_B$ and $f_D$ will depend on a number 
of factors---{\it e.g.}, the overall size of the deviations from the
asymptotic limit \Ep{scalelaw} and
the extent to which $a\mthree$ differs from $a\mtwo$ in the mass
region where the conventional method is used. We estimate this error
from the numerical results in Section \use{sect.analysis2}.

\subsection{Logarithmic Corrections\label{sect.logM}}

Finite logarithms in the matrix element of the full-QCD axial 
current become the source of an additional ultraviolet divergence in the
static limit\cite{HQsymmetry}. This fact results in an explicit
logarithmic dependence on the heavy quark mass in the static-light
axial renormalization constant, which we have written as the
factor $C(a,m)$ given by \Eq{Cam}.
If sufficiently fine lattices could be constructed---so that $am_Q\ll1$
even as the scaling region \Ep{scalelaw} were approached---then the 
logarithmic corrections would be entirely contained in the
lattice matrix element of the conventional calculation.
In the opposite extreme, where $\K_Q\rightarrow0$ and
$am_Q\gg1$, the logarithmic dependence is lost entirely,
as part of the ``non-tadpole'' $\cO(\alpha_s)$ corrections 
which have not been included.

Anticipating improved computations, where the former of these
two limits may be approached, we analyze the numerical data assuming
the presence of the full leading logarithmic correction. To do this,
we eliminate the factor $C(a,m_Q)$ from the renormalization constant
of the static-light axial current given by \Eq{CAhatnew}
($\hat\phi$ is thus well-defined at $1/M_P=0$)
and divide it, in the form $C(a,M_P)$, from the conventional 
amplitude. In this form, $\phi$ is fit to a quadratic in $1/M_P$,
as implied by \Eq{analysis}. To compute a physical result, {\it e.g.}
$\phi_B$, we interpolate the fit to the appropriate mass and then
multiply by the logarithmic correction, {\it e.g.} $C(a,M_B)$.
In so far as the logarithmic dependence on the quark mass may be
absent from the lattice results in the region where $am_Q$ is large,
this procedure too will involve some systematic error. 
This error is akin to the mismatch in the constant terms of the
leading perturbative correction discussed at the end of
Section \use{sect.largeam1} but is considerably smaller.  
At \bspthr, for example, $C(a,M_B)$ itself is a correction of only
a few percent. The procedure described in Sect. \use{sect.analysis2}
is an estimate of the cumulative effect of all such 
systematic effects which alter $\phi_P$ in the region of large-$am_Q$.

\section{Numerical Techniques\label{chpt.numerical}}

\vskip -0.3cm
\subsection{Smeared Sources\label{sect.sources}}

The basic techniques and advantages of using non-local, or ``smeared''
sources in a fixed gauge have been well established in the recent
literature\cite{smearing}.
In this section we discuss briefly our use of Coulomb-gauge, ``wall-source''
(WS) propagators\cite{wall-source}, which we use to compute our best
results from the conventional method at \bspthr.
We also use, for the static limit, a method of smearing which
utilizes standard point-source propagators, which we will refer to
as ``cube'' smearing for reasons which will become obvious in the 
following section.  For the conventional method, at all couplings other
than \bspthr, we use a standard point-source construction of the
correlators, which was given explicitly in Section \use{sect.construction}.

The wall-source propagator $\tilde S$ is defined by
$$
        \tilde S(\vec{x},t;t_0;U) = 
        {1\over V} \sum_{\vec{y}} \langle \psi(\vec{x},t) 
        \overline{\psi}(\vec{y},t_0) \rangle_{U},
\EQN wallprop
$$
and is computed from the Wilson action
$\cS_{W} = \overline{\psi}S^{-1}\psi$ through the solution
of the equation
$$
        \sum_z S^{-1}(x;z;U)\tilde S(z;t_0;U) = {1\over V} \delta_{t,t_0} 
                \sum_{\vy} (\delta^{3}_{\vec{x},\vec{y}}),
\EQN wallinv
$$
where $V$ represents the spatial volume of the lattice, and
we have suppressed color and spin indices.
The technique is trivially implemented by the replacement
of the usual single-site delta function with the sum of delta 
functions in the matrix inversion program.

One must then construct a smeared interpolating operator $\chi$
so that the correlation functions analogous to those defined by \Ep{gax}
reduce to contractions involving only wall-source propagators.
Since the gauge-dependent overlap function
$\bra{0}\chi^\dagger\ket{P}$ is removed from the final physical
amplitude, the extraction of the decay constant and the normalization
issues discussed in the previous section remain unaffected.
For reasons which we explain shortly, we construct a pair of operators, 
denoted by the two signs in the following definition:
$$
        \chi^{\dagger}_{W\pm}(t) = {1\over V^{2}} \sum_{\vx,\vy}
        \overline{Q}(\vx,t)\pmproj\gamma_5 q(\vy,t).
\EQN chiwall
$$

They are used as follows.  We compute WS propagators 
from a ``left'' source $t_{0L}$ ({\it i.e.}, near the edge defined as $t=0$)
for the upper two source spin components only.
Similarly, we compute propagators from a ``right'' source $t_{0R}$
(near the opposite edge) for the lower two source spins.
Note that we have used Dirichlet, not periodic, boundary conditions 
in the time direction.
We then obtain the wall-source correlators by replacing $\chi$,
as defined by \Ep{currents;b}, with the smeared version \Ep{chiwall} 
in \Eqs{gax}. Thus we compute,
$$\EQNalign{
        \GaxWpm(t,t_0) &= 
        \left< \sum_{\vx} \Tr\left[
        \tilde S_{q}^{(\pm)\dagger}(\vx,t;t_0;U)
        \gamma_0 \tilde S_Q^{(\pm)}(\vx,t;t_0;U) 
        \right]\right>_{\{U\}}\phantom{aaa}
\EQN gaxWScomp;a \cr
\noalign{\hbox{and}}
        \GxxWpm(t,t_0) &= 
        {1\over V} \left< \Tr\left[
        \sum_{\vx}\tilde S_q^{(\pm\pm)\dagger}(\vx,t;t_0;U)
        \sum_{\vy}\tilde S_Q^{(\pm\pm)}(\vy,t;t_0;U)
        \right]\right>_{\{U\}}\phantom{aaa}
\EQN gaxWScomp;b \cr}
$$
where we use the abbreviated notation
$$
        S^{(\pm)} \equiv S \pmproj, \quad{\rm and}\quad
        S^{(\pm\pm)} \equiv \pmproj S \pmproj. 
\EQN projop
$$
Note that $\GxxWpm(t,t_0)$ has just two spins at both source and sink ---
as usual, $\GxxWpm(t,t_0)$ must be ``diagonal'' (same source and sink)
so that the overlap of the interpolating field
with the lowest state can be determined.
For $t_0=t_{0L}$ we use the
source $\chi_{+}$ so that both of the correlation functions 
need only the upper two source spins of the propagators; they
are analyzed for forward propagation $t>t_0$.
Similarly, for $t_0=t_{0R}$ we use the source $\chi_{-}$.
These correlators need only the lower two spins and
are analyzed for backward propagation $t<t_0$.
So the amount of computing required in terms of propagator
generation is the same as for the standard single-source approach:
we have doubled the number of source times and halved the number
of spins.

This scheme has been chosen in anticipation of the B parameter
calculation, where the matrix element is computed from a
lattice ``figure-eight'' graph containing a local four-fermion 
operator at its intersection point and meson interpolating operators
at either edge.  We have altered our usual convention for propagator
generation---{\it i.e.}, placing the source at the center of the
lattice---because in this case the local operator of interest must
be constructed 
using the {\it sink} point of the wall source propagators. They
must therefore originate from either side of the lattice center.
For the decay constant calculation, however, this scheme has
provided two sets of propagators for each configuration,
$\tilde S^{(+)}(\vx,t;t_{0L};U)$
and $\tilde S^{(-)}(\vx,t;t_{0R};U)$.

After averaging over configurations,
the forward- and backward-moving correlators
are simply related through time reversal symmetry\cite{bernard-tasi}.
Numerically, we may either (1) invoke this symmetry 
and define the per-configuration two-point function as the average
of the two sources, or (2) treat the sources as two
independent sets of configurations and ``double'' the
statistical sample. 
The choice will clearly affect the estimate of the statistical errors
on the correlation functions
and therefore also the central values and errors of the fitted
parameters. The doubled method has a distinct advantage when using
covariant fits as we have done, since in this case the total number 
of time slices
included in the fit is limited by the number of configurations.
Since our wall sources at $t_{0L}$ and $t_{0R}$ are widely separated,
we expect that the forward-backward pairs for a given configuration 
will not be strongly correlated. 
To verify this, we have analyzed the 
data using the standard (time-reversal averaging) method
when the type of fitting allows for it. 
We have not seen any significant difference in 
the statistical errors of final results between these two methods. 

\subsection{Smearing Techniques for Static Quarks\label{sect.smeared}}

It has been observed in previous static computations that point-source
correlation functions, such as that defined by \Ep{gaxstat}, fail to show
ground-state dominance at sufficiently early times and at large times
are saturated with noise.
This problem has prompted the use of smeared sources which, it is hoped,
may be tuned to have the best possible overlap with the ground state,
so that it dominates sufficiently in the early timeslices.
Qualitatively, at least, this program has been successful%
\cite{eichten-lat89}\cite{alexandrou}\cite{elc-stat}%
\cite{eichten-lat90}\cite{lat90}\cite{newalexandrou}.
Nevertheless, it is interesting to note the source of the problem
in terms of a simple analysis of the Euclidean correlators. 

Lepage has
outlined a method for estimating
signal-to-noise in discussing the proton on the lattice\cite{lepage-tasi}.
First, consider $G_\pi(t)$, a single-configuration contraction of two
light-quark propagators. The computed estimate of the ``pion'' correlation
function is defined by the average over $N$ configurations, which at large
times falls exponentially: $\langle G_\pi(t) \rangle \sim Ze^{-aM_\pi t}$.
The expectation for the noise is given by the formula 
$\sigma_{G}^2 = {1\over N}(\langle G_{\pi}^2\rangle-\langle G_\pi\rangle^2)$.
Since both $\langle G_{\pi}^2\rangle$ and $\langle G_\pi\rangle^2$
overlap with a two-pion state, the signal-to-noise ratio (SNR) is
a constant at large times:
$$
        \lim_{t\to\infty}\; {\langle G \rangle \over \sigma_G} \;\sim\;
        \sqrt{N}\qquad{\rm (pion).}
$$
In contrast, consider a similar analysis of the static-light correlator%
\cite{lat91}\cite{lepage-lat91}\cite{eichten-lat91}
In this case, the signal falls exponentially with the (divergent) 
mass parameter $a\cE_0$: $\langle G \rangle \sim Ze^{-a\cE_0 t}$.
The noise, however, is dominated in the large-time region by 
$\langle G^{2}\rangle$, which has a non-zero overlap with a single
pion, as depicted in \Fig{cheeseburger}. The static lines effectively
cancel, and $\sigma_{G}^{2}\sim e^{-aM_\pi t}$. The SNR is thus given by
$$
        \lim_{t\to\infty}\; {\langle G \rangle \over \sigma_G} \;\sim\;
        \sqrt{N}e^{-a(\cE_0-{M_\pi \over 2})t}\qquad{\rm(static-light)}.
\EQN sigtonoise
$$
Typically, the binding energy $a\cE_0$ is much larger than half the
pion mass. For example, at \bsix\ with $\K_q=.155$, we find
$a\cE_0\simeq0.6$, whereas the corresponding pion mass is
$aM_\pi\simeq0.3$. An essential point to note is that the use of
smeared sources may optimize the ground-state signal in the earlier
timeslices, but it will not change the behavior described
by \Eq{sigtonoise}.

We have used two types of sources for the static computations. The first
is a wall source (used only at \bspthr) and is a straightforward
adaptation of the technique outlined in the previous section. The heavy
Wilson quark field $Q(x)$ in the interpolating operator \Ep{chiwall} 
is replaced with the static field $h(x)$ (or, for backward-movers,
the analogous anti-quark field); the sums necessary to
construct the correlation functions $\GaxstatWpm$ and $\GxxstatWpm$
are easily computed due to the
trivial nature of the static-quark propagator. 

The second type of smeared operator is introduced in order to 
accommodate 
point-source, rather than wall-source, light-quark propagators. 
We refer to it as a ``cube'' and define it by
$$
        \hat{\chi}^{\dagger}_{n+}(\vx,t) = {1\over n^3} \sum_{\vz}^{[V_n]}
        \overline{h}(\vx+\vz,t)\gamma_5 q(\vx,t),
\EQN chicube
$$
where $[V_n]$ denotes the sites of the cube centered around $\vx$.
We have written only the static {\it quark} form of the operator;
the anti-quark construction is analogous.  As before,
we compute the forward-moving correlators from $\chi_{n+}$ and 
the backward-moving correlators from $\chi_{n-}$.
They are given by
$$\EQNalign{
        \GaxstatSpm(t,t_0) &= 
        {1\over n^3} \left< \sum_{\vz} \Tr\left[
        S_{q}^{(\pm)\dagger}(\vz,t;\vec{0},t_0;U)
        \cP^{\pm}(\vz,t,t_0;U)
        \right]\right>_{\{U\}}\phantom{aaaa}
\EQN gaxPWcomp;a \cr
\noalign{\hbox{and}}
        \GxxstatSpm(t,t_0) &= 
        {1\over n^6} \left< \sum_{\vz,\vz'} \Tr\left[
        S_{q}^{(\pm)\dagger}(\vz+\vz',t;\vec{0},t_0;U)
        \cP^{\pm}(\vz',t,t_0;U)
        \right]\right>_{\{U\}},\phantom{aaaa}
\EQN gaxPWcomp;b \cr}
$$
where we have included the spin structure ${1+\gamma_0\over2}$
$\mproj$  of the static quark (anti-quark) propagator through
the notation \Ep{projop} applied to the light-quark propagator.

\subsection{Fitting and Jackknife Analysis\label{sect.fitting}}

We compute statistical errors using the jackknife technique.
Our discussion here will be limited to a brief description of
1) the inclusion of correlations in the numerical data---{\it i.e.},
the use of ``covariant'' fits---and
2) the specific parameterization ({\it i.e.} fitting function) for
the correlation functions.
To facilitate this, let us define any jackknifed quantity $Q$
with the notation  $Q\equiv\{\overline{Q},Q_i\}$, where 
$\overline Q$ results from an analysis of the full average of
$N$ configurations, and $Q_i$ from an analysis of the $i$'th 
``subset average'' of configurations, where one or more
has been removed.

The (estimated) covariance matrix used in the $\chisq$-minimization
fitting routine is then defined through a straightforward
generalization of the standard formula\cite{toussaint}.
For an $m$-elimination jackknifed quantity $Q(k)$,
where $k$ denotes some external parameter dependence,
we write it as
$$
        C_Q(k_1,k_2) = ({N\over m}-1)
        \left< (Q_i(k1)-\overline Q(k1))(Q_i(k2)-\overline Q(k2))\right>.
\EQN covar
$$
By definition, correlation functions are computed by averaging
over individual configurations. In this case, and for $m=1$, \Eq{covar}
reduces identically to the standard covariance formula of 
Ref.\cite{toussaint}. This, of course, is not the case for a quantity
which is computed {\it from} an average, such as a fitted mass.

In fitting a quantity $Q(k)$ to some theoretically expected
functional form, $F(k,\vec{A})$ ($\vec{A}$ is the set of parameters
to be fit), the fit is jackknifed, but the fits
to $\overline Q$ and the $Q_i$ use the same inverse covariance
matrix $C^{-1}_Q$.
Each level of fits ({\it e.g.}, correlation functions,
$f(\K_q)\ \vs\ 1/\K_q$, {\it etc}.) takes into account correlations
in the numerical data and produces a meaningful $\chisqdof$
($\chisq$ per degree of freedom).
This method thus gives an indication of how different systematic
errors affect the calculation at each stage, and, as a practical
matter, clearly requires many fewer configurations than would the inclusion
of all correlations simultaneously.

Our fits to the correlators $\Gax$ and $\Gxx$ are ``coupled,''
{\it i.e.}, made simultaneously and with identical mass parameters.
Thus at the stage of a single lattice meson we compute all
correlations, including those between the two channels $\Gax$ and $\Gxx$.
To denote this, we assign a dummy index $a$ to the time coordinate.
It takes on the values $a=(A,B)$---{\it i.e.}, 
we define $G(t^A)\equiv\Gax(t^A)$ and $G(t^B)\equiv\Gxx(t^B)$.
The meaning of the covariance matrix $C_G(t^{a}_1,t^{b}_2)$ is
then immediately apparent. 
Depending on the type of correlator, we compute one of two types of
fits (the choice is specified in Section \use{chpt.results}).
We denote them loosely as the ``single-state'' fit,
$$\EQNalign{
        \chisqmin [ G(t^a) \!-\! \zeta_a e^{-aMt^a} ] 
                &\longrightarrow M,\Zax,\Zxx
\EQN ex3 \cr
\noalign{ \hbox{and the ``two-state'' fit,}}
        \chisqmin [ G(t^a) \!-\! (\zeta_a e^{-aMt^a}+\zeta'_a e^{-aM't^a}) ]
        &\longrightarrow M,M',\Zax,\Zaxp,\Zxx,\Zxxp,
\EQN ex6
\cr}
$$
where the quantities on the right denote jackknifed fit parameters.
One should keep in mind that we double the configuration set using the
backward movers. Thus our typical sample size, $N_{\rm config}\simeq 20$,
allows, in principle, a maximum of $2N_{\rm config}\simeq 40$ time slices
in $G(t^a)$
to be fit. In fact, to produce stable fits we have found it
necessary to stay well below this limit. For example, for the two-state
fit we typically chose $N_{t^A} \simeq N_{t^B} \simeq 10\!-\!15$.

\subsection{Fitting Errors\label{sect.fit-errors}}

There are several sources of systematic error associated
with the fitting and extrapolation procedures whose approximate size
we compute numerically. Our method for doing so is {\it ad hoc}, but it gives
an estimate of the variance one might expect as a result of a different,
although equally valid, choice of analysis parameters and, to some extent,
techniques.  We proceed as follows. We
perform additional analyses where
one or more of the input parameters or analysis procedures is altered 
from its ``best-fit'' setting. For each computed quantity,
this produces a set of ``alternate'' results. The ``fitting error''
is computed as the (symmetric) standard deviation of the alternates
from the best-fit, for those alternates which pass a cut on the last
relevant $\chisqdof$  ($\chisqdof \le 2$).
The alternate analyses are made with various
combinations of the following changes:
{
\EnvLeftskip=.6cm
\EnvRightskip=0cm
\itemize
\itm
Extrapolations to the chiral and strange limits:
In each analysis, the central values are obtained using
the three lightest quarks (except at \bsix\ where we omit
$\K=.156$). Alternates are computed using only the lightest two quarks
for the extrapolation to the pion and/or the kaon.
In several cases (at \bsix), the heavier two were also used.
\medskip
\itm
Type of fits:
For the wall-source, heavy-light correlators at \bspthr, 
we have used the two-state fit for central values 
(Sects. \use{sect.heavy} and \use{sect.static}).
In these cases we sample a single-state fit over the latest-possible time
interval to include ``finite-time'' effects ({\it e.g.}, higher-state contamination).
\medskip
\itm
Jackknifing: 
We use the single-elimination jackknife to compute central values.
Higher-elimination jackknifes ({\it i.e.}, 2- and sometimes 4-elimination)
are sampled for the effect of a poorly estimated covariance matrix,
resulting from our use of moderate numbers of configurations.
In addition, we compare the time-reversed ``folded'' and ``unfolded'' 
analyses in cases where the number of configurations permits.
On all covariant analyses, the central values were computed as unfolded.
\medskip
\itm
Fit intervals on the correlation functions:
The time intervals for the central values were chosen based on
an examination of the effective mass plots and, in the case
of covariant fits, the $\chisqdof$.
All analyses ({\it i.e.}, the best and those modified per the changes above)
are made using several other fit intervals, which are shifted 
from the optimal choice. 
In principle, these shifts are a test for
the presence of finite-time effects.
However we find that in practice the variance is largely due to
the statistical errors in the evaluation of the covariance matrix.
When non-covariant fits are used, the shifts typically have a negligible
effect on final results in comparison to the jackknife error.
\enditemize
}

Since our configuration samples are generally not that large,
some effort has been required to achieve a reasonable 
level of stability in the covariant fits. To invert the covariance
matrix, we use LU, followed by a correction for accumulated roundoff
error\cite{numrec}. This error is potentially a serious problem because
the covariance matrix is often close to singular, a situation which is
further exacerbated by the fact that the elements themselves may span
several orders of magnitude, since the correlators fall exponentially.
The latter problem we remedy by taking the log of the data and 
fitting to the log of the function.
Prior to this, we rescale the correlators so that the fitted parameters
(the $M$'s and $\zeta$'s) are as nearly as possible of the same order of
magnitude. This helps prevent additional roundoff error in the inversion
of the Hessian matrix during the fit updates.

\null From this we generally obtain satisfactory results, although we feel
that, to a certain extent, numerical errors remain which are not
reflected by the jackknife. Unfortunately, we do not have the large
statistical sample which would be necessary to better quantify
this statement. However, the techniques listed above provide a rough
estimate of the size of this effect.
Other groups have performed fairly in-depth studies on covariant 
fitting techniques, using various types of smeared sources to compute
the correlation functions (for example, see\cite{gupta-dyn}).
Our findings are qualitatively  similar.

\section{Analysis and Results\label{chpt.results}}

We now discuss the details of the computations, present qualitative
and quantitative results, and estimate bounds for various systematic
errors. We begin by providing the essential parameters pertaining
to the gauge configuration and propagator generation, 
and then present the analysis in various stages: light mesons
({\it i.e.} the pion and kaon), conventional heavy-light states,
static-light, the combined analysis for $1/M_P$ corrections to
the asymptotic scaling law \Ep{scalelaw} and the calculation of
physical amplitudes, and finally bounds for large-$am$ and scale
errors.  In practice, the calculation was
split up in this fashion, but in such a way so that results
passed from one stage to the next ({\it e.g.}, $\ainv$, $\K_c$, {\it etc}.)
could be included in the jackknife analysis for statistical errors.

In the sections pertaining to the fits to the correlation
functions, we refer extensively to the effective mass as a way of 
determining the point of ground-state saturation. For a correlator
$G(t)$, the basic definition of the effective mass is
$$
       aM(t) = \ln({G(t)\over G(t+1)}).
\EQN effmass
$$
Where a single state dominates $G(t)$, $aM(t)$ has a plateau.
(We actually use single-exponential, uncorrelated fits over three
timeslices---{\it i.e.}, centered at $t$---to which we associate the fitted
mass parameter $M(t)$. In practice, there is very little difference
between the results of this method and the discrete logarithmic derivative
given above.)

\subsection{Simulation Parameters\label{simulation}}

\noindent
$\underline{\beta\!=\!6.3}$
\bigskip

We have generated twenty $24^4$ Coulomb-gauge lattices at \bspthr,
one set of five with the pseudo-heatbath algorithm
(6K passes to thermalize, 2K passes separation), and three sets of five 
with a combination ``3-6 over-relaxed + 1 pseudo-heatbath'' algorithm
(4K total passes to thermalize, 2-2.2K total passes separation).  
Two sets had hot starts, and two had cold starts.
\null From these lattices we have generated both wall-source
and point-source propagators after extending the gauge configurations
in the time direction using the periodicity. 
In all propagator generation,
we have employed spatially periodic and fixed time 
boundary conditions.
In three of the sets, relative temporal shifts were introduced to 
further randomize the sample.
Wall-source propagators were made at $24^3\times55$ for two source
locations, $\tilde S^{(+)}$ from $t_{0L}=13$ and $\tilde S^{(-)}$ 
from $t_{0R}=41$ (sites labeled $t=(0,54)$ and using the notation of
Section 3).
Point-source propagators were made at $24^3\times61$
with a central source ($t_0=30$).
In both cases, the time boundary condition was Dirichlet.

In order to measure the overlap functions of the wall-source correlators,
the source must be placed sufficiently far
from the lattice edge in order to avoid any distortion from the boundary.
Checks to determine a sufficient displacement were 
made as follows.
Two propagators, sizes  $24^3\times55$ and $24^3\times45$,
were computed at $\K=.150$ from the same configuration, 
with the wall source in each case at the same trial
position $t_{0L}$ as measured from the ``left'' edge of the lattice.
\null From these propagators, degenerate correlators $\GxxWpm$
were computed; call them $G_{55}$ and $G_{45}$.
\null From each we computed the local masses $M(t)$, using \Ep{effmass},
and the local amplitude, $\zeta(t) = G(t)\times\exp(tM(t))$, 
where $t$ labels the distance from the source.
The corresponding residues $\zeta_{55}(t)$ and $\zeta_{45}(t)$
were compared as $t$ was increased until
the point $t=\tilde{t}$ where the difference exceeded $1\%$.
The number of slices between and including that labeled by 
$\tilde{t}$ and the ``right'' edge of the shorter lattice
was taken as the necessary source-displacement $T$.
We found $T=13$ and thus used $t_{0L}=13$ and $t_{0R}=41$ for the 
$24^3\times55$ propagators.
Since mass fluctuations are amplified exponentially in the correlator
residues, this test is far more stringent than the analogous comparison
of the local masses themselves.
\bigskip

\noindent
$\underline{\beta\!=\!6.0}$
\medskip

At \bsix\ we have constructed and analyzed two independent data sets. 
The first consists of nineteen configurations of gauge size
$16^3\!\times\!40$. Eight of the nineteen were generated using
the pseudo heat bath (PHB) algorithm, 2K to thermalize, 1K separation.
The remaining eleven configurations were generated using
an Overrelaxed/Metropolis (OM) algorithm with a 600-sweep separation.
Point-source propagators for light-quark masses only
were generated at $16^3\!\times\!39$ with a central source.

The second set consists of eight configurations of gauge
size $24^3\!\times\!40$. These have 2K OM sweeps for thermalization
and 1K sweeps for subsequent separations. 
Point-source propagators were generated for heavy and light quark masses
at $24^3\!\times\!39$ with a central source.

\bigskip
\noindent
$\underline{\beta\!=\!5.7}$
\medskip

At \bfps\ a single analysis was made on the combined data set
of 32 configurations described in the following table:
\bigskip
\vbox{
\halign{\hfil\qquad\qquad #\quad & #\quad & #\hfil\cr
Configs. & Gauge                & Quark \cr
16 PHB   & $16^3\!\times\!24$   & $16^3\!\times\!25$ \cr
 4 PHB   & $16^3\!\times\!24$   & $16^3\!\times\!33$ \cr
12 PHB   & $16^3\!\times\!32$   & $16^3\!\times\!33$ \cr}
}
\bigskip
\noindent
Propagators were generated with a central point source for
light quark masses only.
The analysis is limited by the smaller time dimension,
so we refer to them as $16^3\!\times\!25$. 
In each case the configurations were separated by 1K passes.

\subsection{Light Mesons\label{sect.light}}

An analysis of the light pseudoscalars was done on each of
the data sets listed above in order to compute 
$f_K/f_\pi$ and to obtain the (jackknifed)
parameters needed for the heavy-light computations.
In this section, we discuss several aspects of this
analysis. First, we focus on the quality of the data at \bspthr,
discuss some details of the fits to the correlation functions,
and compare wall sources with point sources. 
(We have used the former only at \bspthr.)
Second, we discuss the extrapolation of the raw numbers
obtained from the correlator fits to physical values of the quark mass.
Lastly, we present results for $f_K/f_\pi$
and estimate its scaling error at \bspthr.
The numerical results are provided in \Tbl{tab-chir63W} -- \Tbl{tab-chir57J}.

The covariant single state fit \Ep{ex3} was used to compute 
the ground state masses and decay amplitudes in the chiral regime.
Statistical errors were obtained from the single elimination
jackknife, treating the forward- and backward-moving
correlators as separate configurations.
We show the effective masses and best fit 
for the $\K\!=\!.150$, wall-source pion at \bspthr\ in \Fig{fig-emWpion63}.
The mass of the wall-point correlator ($\GaxW$) 
appears to be asymptotic after a displacement of approximately $t\!=\!12$,
however it contains fluctuations which make an assessment of
systematic finite-time effects difficult.
We found that the most satisfactory fits in terms of 
minimum $\chisqdof$ were obtained from intervals displaced further
from the source; for final results we chose $t_{\rm min}=15$.
Note also that the wall-wall correlator ($\GxxW$) is much
noisier than 
the wall-point.
This is due to the extra summation in \Eq{gaxWScomp;b}. It produces
a relative increase in the number of contractions between quark propagators
whose spatial separation at the source and sink is much
larger than the size of the state, and the correlation function is thus subject 
to large unphysical fluctuations in the gauge links on a given time-slice.
This feature has previously been found to be true also of Wuppertal 
sources\cite{gupta-dyn}.

Due to the differences in the construction of our wall- and point-source
propagators it is difficult to make a clear comparison between the
two methods. Qualitatively, as seen in \Fig{fig-emPpion63},
the point-source effective masses approach ground-state saturation
more slowly. However, because the central placement of the point source
results in a smaller temporal length, we have used a similar fit
interval as for the wall-source analysis in order to remain sufficiently
far from the boundary. 
Clearly, however, the plateaus in the point source case are not very convincing,
and a large systematic error would result if we were forced to 
rely on this data alone.
Results extracted using $t=(15,21)$ are
given in \Tbl{tab-chir63P} and may be compared to the 
wall-source results (from fits over $t=(15,24)$)
in \Tbl{tab-chir63W}. The two methods produce consistent results within
errors.  Due to the better quality of the effective-mass plateaus,
we use the wall sources for our final results.
In addition, the errors in $af_\pi$ from the wall-source analysis
are smaller, which results in a better scale determination.

The analysis of the light mesons proceeds according to 
a standard extrapolation scheme, motivated by the leading order
predictions of chiral perturbation theory.
Define the lattice quark mass as 
$am_0(\K) = {1\over 2\K} - {1\over 2\K_c}$, and denote
the fitted masses and decay constants of the degenerate 
mesons as $aM(\K)$ and $af(\K)$.
We fit $(aM)^2$ linearly in the quark mass---{\it i.e.}, linearly in
$1/\K$---and extrapolate the fit to $aM=0$ where we extract the
critical hopping parameter, $\K_c$.
Similarly, we fit $af$ linearly in $1/\K$ and extrapolate to 
$1/\K=1/\K_c$ to compute $af_\pi$.
The scale is determined from $a^{-1}_{f_\pi}=132\ \MeV/(af_\pi)$.
The above procedure can be iterated, extrapolating instead to 
the physical pion mass in order to determine $\K_c$.
Although we in fact do this, the resulting shift in $\K_c$ and 
$\ainv$ is in all cases smaller than the statistical error.

An additional level of fits and extrapolations is required
in order to determine $\K_s$, the strange quark hopping parameter,
and to compute $f_K$. 
For each combination of light quarks ($\K_1,\K_2$)
we compute $(aM)^{2}_{\K_1,\K_2}$ and $af_{\K_1,\K_2}$.
We fit both quantities linearly in $1/\K_2$ and extrapolate 
to $1/\K_2=1/\K_c$.
We then fit $(aM)^2_{\K_1,\K_c}$ linearly in $1/\K_1$ and interpolate
to the physical kaon mass in order to determine $\K_s$.
Finally, we compute $af_K$ by fitting
$af_{\K_1,\K_c}$ linearly in $1/\K_1$ and interpolating to $1/\K_1=1/\K_s$.
The four quantities $aM_\pi$, $aM_K$, $af_\pi$ and
$af_K$ are thus used to determine three parameters---$\K_c$, $\K_s$, 
and $\ainv$---and make one prediction---$f_K$.

Our fitting technique (Sect. \use{sect.fitting}) 
would in principle allow us to include the inter-kappa correlations
and thus quote a meaningful $\chisqdof$ in the chiral fits described above.
In fact we have not done so, because we find 
that the correlated fits of $(aM)^2$ and $af$ 
in many cases do not converge to a sensible result and
often give a large $\chisqdof$.
The most likely cause of this is that the jackknife procedure
underestimates the errors since it does not include systematic
effects ({\it e.g.}, finite volume and
finite time errors) which may affect the results
at different quark masses in a non-uniform way.
Assuming this to be the case, we turn off the correlations
to obtain the best approximation to the leading chiral behavior
which the linear fit functions are intended to model.  
Typical chiral extrapolations
(for the heavy-light case) are shown in \Fig{fig-HLmxt63W}
and \Fig{fig-HLfxt63W}.
Note that the fits to straight lines look excellent to the eye;
yet correlated fits (when they converge at all)
would still give a large $\chisqdof$.
For example,
for the degenerate light-light case, correlated fits at \bspthr\ give
$\chisqdof = 7$ for $M_\pi\ {\rm vs.}\ 1/\K$ and
$\chisqdof = 20$ for $f_\pi\ {\rm vs.}\ 1/\K$,  yet go through the error
bars of all the points.
Recent work by Seibert\cite{seibert} discusses potential problems
with covariant fits of which this may be an example.

We continue to jackknife the entire fitting procedure and typically
obtain statistical errors in the extrapolated results which are similar
in size to the errors on the individual masses or amplitudes used in the fit.
The systematic
fitting error which we compute numerically (Sect. \use{sect.fitting})
roughly compensates for this underestimate---its largest contribution 
comes from analyses where we omit the heaviest meson and 
extrapolate from only the lightest two.
Of course we cannot rule out the possibility that the problem plaguing 
this stage of the analysis stems from real physics---{\it i.e.},
that the numerical
data exhibit small violations of the leading-order chiral behavior, either
because of the large values of the quark mass being used or because of
problems with the quenched approximation itself\cite{quenched-chiral}.
A careful study of this issue will require larger lattices (allowing lighter 
quarks) and better statistics.

The results for $f_K/f_\pi$ are
plotted versus the lattice spacing in \Fig{fig-fKofpi}.
Here the statistical and fitting errors have been added in quadrature.
Note that we study the dimensionless ratio $f_K/f_\pi$ 
(rather than $f_K$ itself) as a matter of convenience. There is no 
reduction in the statistical error since in dimensionful quantities 
we multiply the scale, which we determine from $f_\pi$, 
within the jackknife procedure.
(That is to say, {\it all} of the decay constant computations are 
effectively made using a  jackknifed ``$f/f_\pi$'' method.)
\null From the linear fit shown in \Fig{fig-fKofpi} we find the slope
$.19\pm.07\ \GeV^{-1}$ and the intercept
$f_K/f_\pi(a\!=\!0) = 1.08\pm.03\pm.08$.
The first error is a measure of statistical and fitting errors;
the second, systematic errors in the $a\to0$ extrapolation.
We estimate the latter by
repeating the fit using a quadratic, rather than linear, parameterization,
and then symmetrizing the difference.
This quenched result is $1.6 \sigma$ less than
the experimental value (1.22).  It is interesting to note that
the sign of the disagreement is in accord with sign of the difference
between the chiral logs in the full and quenched theories\cite{cb_mg}.

\subsection{Conventional Heavy-Light Mesons\label{sect.heavy}}

We turn next to the conventional heavy-light mesons and focus
entirely on the results at \bspthr. We pair
the three light quarks $\K_L$ = .149, .150, and .1507 with the
following ``heavy'' quarks: for wall sources 
$\K_H$ = .148, .145, .140, .135, .130, .125, and .117;
for point sources $\K_H$ = .148, .145, .125, .110, and .100.
As compared to the light mesons,
the correlation functions, both wall and point, approach
ground-state saturation more slowly.
However, we found that the wall-source data was
well approximated by the two-state function \Ep{ex6}.
These fits were made over a larger number of timeslices which began
much closer to the source.
For quoted results we chose the interval $|t-t_0|$=(3,18)
on each correlator, providing a total of 32 points in the fit.
A sample of this method is shown in \Fig{fig-emWhevy130} for $\K_H=.130$.
The effective mass curve, computed from the logarithmic derivative
of the fit function, provides the usual comparison of the fit to the data.
In the alternate analyses from which we estimate fitting errors,
we shift the intervals in either direction and also use
single state fits at much larger time displacements.

The analysis of the point-source heavy-light correlators was again
worrisome due to the finite-time restriction, as can be seen from 
the effective mass plot, \Fig{fig-emPhevy125}.
Here the two-state function did not work well,
and we have computed the single-state fits to these data
as a check on the wall-source results.
We cannot expect {\it a priori} that this analysis will produce
a reasonable outcome, because we are forced to fit 
very close to the region where there are obvious, large
boundary effects.
In this respect, the agreement between the two methods
(which is within the combined statistical and fitting errors
estimated from the wall-source analysis) may be somewhat fortuitous.
Note also that the statistical errors on the bare
fitted amplitudes of the point-source analysis
are actually smaller than the corresponding results from
the wall sources for the {\it heavy-light} states
(c.f. $\K_H=.125$ for the two cases).
In the final analysis of physical quantities, however,
we still obtain smaller statistical errors from the wall sources, 
primarily because of the smaller statistical error in the scale determination.

For each heavy quark $\K_H$, we extrapolate the correlator 
masses and decay constants linearly in $1/\K_L$ to the chiral and 
strange limits of the light quark. From the extrapolated values
we recompute the amplitude $\phi_P$ which we analyze in conjunction
with the static results for phenomenological predictions.
The linear fit of $f$ rather than $\phi$ is motivated by the 
results obtained\cite{grinstein-chiral} from the application of chiral
perturbation theory to mesons containing a single heavy quark%
\cite{wise-chiral}\cite{jfd}. 
In practice, however, the curvature is slight and it makes little difference
which is used. The extrapolation points $1/\K_c$ and
$1/\K_s$, computed in the light-meson analysis, are included in the jackknife.
Since we neglect the inter-kappa correlations in these fits,
we again estimate a fitting error which includes analyses where extrapolations
are made from the lightest two quarks.
The fits and extrapolations of the heavy-light masses and decay constants 
are shown in \Fig{fig-HLmxt63W} and \Fig{fig-HLfxt63W} respectively.

\subsection{Static Results\label{sect.static}}

For the static calculation, the use of smeared sources
is necessary in order to obtain ground state dominance at
sufficiently early times, and one would thus like to 
optimize the overlap of the source with the ground-state
wave function of the heavy-light meson.
In our simulations, however, the cube sources (Sect. \use{sect.sources})
allow only a crude volume adjustment, and it is not clear {\it a priori}
that any particular cube will sufficiently eliminate the contributions 
of higher states in both correlators.  
A search for an early plateau in the effective masses,
$M_A(t)$ and $M_B(t)$, or in the ratio of correlation functions, $G_A/G_B$,
can lead to some judgement of the best smearing volume to use. 
However, for several reasons, this technique may not be entirely 
trustworthy and in practice difficult to implement.
First, the effective mass of the smeared-point correlator, $M_A(t)$,
need not be monotonic; this presents the possibility of short-range
false plateaus, which are difficult to detect because the 
SNR is
falling exponentially.
Second, the location of the true plateaus in $M_A(t)$ and $M_B(t)$ need not 
coincide; thus the ratio $G_A/G_B$ is difficult to interpret unambiguously.
Finally, as based on some ``best plateau'' criterion, there is
a noticeable variation in the optimum smearing volume as the
light-quark mass is adjusted. Unless the source tuning procedure
is sufficiently precise, different amounts of higher-state
contamination at various light quark masses will complicate the
extrapolations to $\K_c$.

In this section we discuss the static-light analysis and
present the results, first at \bspthr\ and then at \bsix. 
At \bspthr\ we have better statistics, and the smearing techniques
are more effective because the lattice is finer.
We have used both wall and cube sources, the latter
computed using a range of smearing volumes $V_s$. We choose our best results
by demanding consistency among the different sources (and analysis methods)
in order to minimize systematic errors from possible higher-state
contamination as best we can.
Our results at \bsix, from cube sources only, are presented in
comparison to other static results which have been published
in the literature. At this lattice spacing, and with our
statistics, we find that
the finite time restriction (imposed by the exponentially falling
SNR) and the limited effectiveness of the cube smearing
together prevent us from making a convincing determination 
of the amplitude $\hat\phi$\cite{lat91}.
Our results are smaller than previously reported\cite{lat90},
but there is still a strong dependence on how one chooses to analyze the data.

There is some similarity of our results at  \bsix\ here with the
results of ref.\cite{hashimoto} for difference size sources
at fixed fitting interval, but a crucial difference is that we
do not believe there are any intrinsic problems with the
cube-smearing technique.  In principle, all sources must agree
at large enough times.  The issue, however, is that with limited
statistics, a rapidly falling SNR, and only a few source
sizes to choose from on this fairly coarse lattice, it can
be difficult to go to large enough times.  The data is then
ambiguous, and guesswork will be involved in deciding the
``optimal'' available source and time interval to
analyze the data.
\bigskip\medskip

\vfill\eject
\noindent
$\underline{\beta\!=\!6.3}$
\bigskip

We consider first a given smearing volume, {\it e.g.}, $V_s=13^3$.
The basic analysis procedure is analogous to that for the conventional
method: we make a single-state, covariant fit to the two correlation 
functions over some range of time slices. The fitted parameters are the
unrenormalized binding energy, $a\cE$, and the amplitudes
$\Zaxstat$ and $\Zxxstat$ which enter into \Eq{fplatt}.
This is repeated for several values of the light quark hopping
parameter; we then make a linear fit of $\hat{\phi}\ {\rm vs.}\ 1/\K$
and extrapolate to $1/\K_c$ and $1/\K_s$, using the (jackknifed) 
parameters computed in the light-quark analysis.
As in the previous cases, we neglect the inter-kappa correlations.
The effective masses for $V_s=13^3$, 
plotted against the mass obtained from a fit over the range $t=(9,16)$,
are shown in \Fig{fig-emS13P63}. 

Whereas the size of the cube source can be set to roughly match the
size of the ground-state wave function, the size of the
wall-source is certainly  too large.
Thus higher state contamination is expected to increase\cite{lat91}.
However, the wall has an advantage in terms of statistics
(sources for both light and heavy quarks at every point on a timeslice),
and ultimately its performance must be evaluated empirically.
Although there is a large admixture of excited states in the two-point
functions (judging from the effective mass plots), we again obtain
good fits for wall-sources
by using the two-state function over an early range of time slices.
We use this method to extract the ground-state parameters needed to
compute the static amplitude.
The effective masses of the wall-source, static-light state, with
$\K_q=.150$, are shown in \Fig{fig-emSW63}. They are plotted against the 
curve computed from the logarithmic derivative of the fit function using 
the best-fit parameters as input.

A qualitative examination of the effective mass plots from the cube 
sources indicates that the optimum smearing volume is $V_s\sim 13^3-15^3$. 
Nevertheless, due to the caveats discussed above, our preference
is to obtain consistent results over a range of $V_s$.
In \Fig{fig-fvsV63} we show the results of the raw decay amplitude
({\it i.e.}, without perturbative renormalization constants) for several
cube sizes.  For each size, we show the dependence of the amplitude
on the fitted time interval of the correlator $\Gax$, while that
for $\Gxx$ has been held fixed at $t=(10,15)$.
We mark the fits which satisfy $\chisqdof<1$ with ``$\times$.''
The early-time fits clearly show systematic differences,
but as the fit interval is moved further from the source,
the results in the range $V_s\sim 13^3-17^3$ are in good agreement
(both central values and size of errors) and are consistent
with the wall-source results extracted from the two-state analysis\cite{lat91}.
If we instead use an early fit interval for $\Gxx$---fitting where
the effective mass first appears to plateau---we find that these
results are altered by at most $3\%$. (At \bsix, a similar change in
the fit interval of the smeared-smeared correlator
reveals a much larger discrepancy; see the discussion below.)
We use the single-state fit over the later-time interval
with the source $V_s=15^3$ (presented in \Tbl{tab-stat63W})
for our combined analysis from which we compute the $B$- and $D$-meson
decay constants (Sect. \use{sect.analysis1}).
The remaining systematic uncertainty associated with the source-type
and fit interval is accounted for in the fitting error which we
estimate numerically.

\bigskip\medskip

\vfill\eject
\noindent
$\underline{\beta\!=\!6.0}$
\bigskip

At \bsix\ we have poorer statistics and our results are less
conclusive. Nevertheless, it is informative to compare them
with those that have recently appeared in the
literature\cite{alexandrou}\cite{elc-stat}\cite{lat90}%
\cite{newalexandrou}.\Footnote{\dag}{
        Eichten {\it et al}.\cite{eichten-etal}\cite{eichten-lat90} have also reported 
        results using smeared interpolating operators in the static limit. 
        We do not include them here because the bare coupling was not the
        same (they used \bspone, \bfpn\ and \bfps), and this makes an
        unambiguous comparison difficult.  In addition, preliminary results
        from UKQCD were discussed at {\it Lattice '92}\cite{simone}.}
Our aim here is to examine only the numerical aspects of the
static calculation, independent of any questions of perturbative 
renormalization or choice of scale. To this end, we compile in 
\Tbl{tab-statcomp} results for the bare lattice quantity 
$\tilde\phi a^{3/2}\equiv\sqrt{2\K}\hat{\phi}^{(0)}a^{3/2}$, where
$$
        \hat{\phi}^{(0)}a^{3/2} \equiv \sqrt{2\hat\zeta^{2}_{A}\over\hat\Zxx}.
$$
In general we choose the light-quark hopping parameters 
$\K=.154$ and $\K=.155$; the ELC group used slightly different values,
however in each case we extrapolate the results
linearly in $1/\K$ to the chiral limit (which we choose to be $\K_c=.157$)
so that a reliable comparison can be made.
(We emphasize that all extrapolations to $\K_c$ are ours, using our
values for $\K_c$, $a^{-1}$, and the perturbative corrections to get
$\fBstat$.)
As a rough guide,
the statistical error assigned to the extrapolated result
has been taken from the lightest quark used in each case.
To compute a physical value of $f_B$ from the static calculation,
the ELC and Wuppertal groups have used the
perturbative renormalization $\hat{Z}_A=0.8$ and
scale estimates of $\ainv=2.0\GeV$ and $2.3\GeV$ respectively.
In our Lat '90 results, we left open the choice of $\hat Z_A$ and
used $\ainv=1.75\GeV$. These differences have been removed in computing
$\tilde\phi$. From this ``bare'' result, we compute
$$
        \fBstat \equiv \hat Z_A C(a,M_B) {\tilde\phi \over \sqrt{M_B}}
\equiv C(a,M_B) {\hat\phi \over \sqrt{M_B}},
$$
with $\ainv=2.3$ GeV, $M_B=5.28$ GeV and $\hat Z_AC(a,M_B)=0.70$;
the latter obtained using $g^2=1.77$ (Sect. \use{sect.analysis3}).

The Wuppertal collaboration has used
Coulomb-gauge exponential wave functions as the smeared interpolating
operators. In \Tbl{tab-statcomp} we denote their two-point functions
which are analogous to our point-smeared ($\Gax$) and smeared-smeared
($\Gxx$) by using our notation in quotes. 
In each of rows 1--3, the fitting
technique is somewhat different, although in all cases 
an early interval has been used and the statistical errors
are relatively small.  Comparing the chiral extrapolations,
the ELC and Wuppertal results agree at roughly the $10\%$ level.
The authors of Ref.\cite{alexandrou} attribute the slightly lower
ELC result to finite volume effects.
There is a larger discrepancy ($\sim 30\%$) between our 
preliminary results from Lat'90 (row 3) and those of the ELC group.
ELC used the same smearing technique; however they used a cube size
$V_s=7^3$, whereas we chose $V_s=5^3$.

A sample of our updated results
is presented in the lower two sections of \Tbl{tab-statcomp}. 
First, we recompute the amplitude at $V_s=5^3$ (row 4) to check the
effect of changes in the analysis procedure---the difference 
is roughly within the statistical error (c.f. rows 3 and 4).
Next, at $V_s=7^3$, our central values are in very good agreement with
those of the ELC calculation (c.f. rows 1 and 5).
Since the lattice sizes are different, this indicates that finite-volume
errors are probably negligible at this level of precision.
Continuing to $V_s=9^3$, however, we find that the amplitude again 
decreases---by approximately $20\%$ when a similar (early) fitting
interval is used. From our analysis, it is difficult to determine
which of these is the ``best'' result based on some objective criterion
such as the lowest $\chisqdof$ of the fit.

To illustrate the source of these differences, we show a sequence of
effective mass plots in \Fig{figs-emassV}, using the light quark $\K=.155$.
In each plot the local masses of the smeared-smeared and smeared-point
correlators, as well as the single mass from the coupled fit, are shown. 
(Because the number of lattices is small, we were not able to
include the effects of correlations, and the values of $\chisqdof$
which the fits produce are deceptively small.) As the smearing volume is 
increased from $V_s=5^3$ to $9^3$, 
the effective masses appear to plateau earlier in the smeared-smeared
correlator and later in the smeared-point. This illustrates a problem
with the cube source which was pointed out above; because of this
we avoid using the ratio $G_A/G_B$ in the analysis.
In addition, there is some evidence that long-range ({\it i.e.}, highly
correlated in $t$) fluctuations are causing a misleading signal at $V_s=9^3$:
the effective mass of $\Gxx$, a monotonically decreasing quantity,
appears to rise slightly at the earliest time slices.
The final analysis of these data thus requires a significant amount
of judgement. We consider two approaches, both which seem fairly reasonable.

First, we repeat the procedure that was
used at \bspthr. We use a later fit interval on the smeared-smeared
correlator $\Gxx$ and study the dependence of the amplitude
on $V_s$ and on the fit interval for $\Gax$. These results,
computed for the four values of the light quark hopping parameter,
are shown in \Fig{figs-stat60}.
At values of $V_s$ where the amplitude is independent of $t_{\rm min}$,
a plateau exists in the effective mass of the correlator $\Gax$.
By comparing the four plots, notice that using this particular
``plateau criterion'' will result in a different optimal smearing
size for different
light-quark masses---ranging from $5^3$ at $\K_q=.152$ to $9^3$
at $\K_q=.156$. Nevertheless, we again find (albeit within large
statistical errors) that at large time displacements the results
computed using different smearing volumes are in good agreement.
As at \bspthr, we choose our best result from a later fit
interval with a large smearing function ($V_s=9^3$).
An example of the outcome is provided in the final section of
\Tbl{tab-statcomp} ({\it i.e.}, row 7); we provide complete results in 
\Tbl{tab-stat60l}.
Notice that at $\K=.154$ this result is in good agreement with the 
corresponding one from $V_s=7^3$ (row 5, an earlier fitting interval):
there is a plateau for the particular combination 
$V_s=7^3$ and $\K_q=.154$. However, for $\K_q=.155$ (a lighter quark),
$V_s=7^3$ appears to be less optimal, and the difference between rows
5 and 7 becomes more pronounced. Because of this, there is an increase
in the slope of the chiral extrapolation, which causes systematic
differences to be magnified in $f_{B}^{\rm stat}$. 
This effect produces a significant amount of the difference between 
the result $\fBstat=237$ MeV in row 7 and, say, the ELC result
$\fBstat=323$ MeV in row 1.

Alternatively, let us consider a different approach to the analysis,
based on the effective-mass plot, \Fig{figs-emassV}(d). 
We assume that the early-time plateau in the smeared-smeared correlator
(which is optimal for $V_s=9^3$) reflects the least contaminated projection
onto the ground state and that the smeared-point correlator contains
higher-state contributions which become small only at large times.
Thus we fit $t=(2,9)$ for $\Gxx$ and $t=(10,14)$ for $\Gax$.
Now, the fitted mass is primarily constrained by the smeared-smeared
correlator and is significantly larger than that found in the first
analysis described above (c.f. \Figs{figs-emassV}(c) and (d)).
It does not appear to agree very well with the local masses
of the smeared-point correlator, but our premise was that this channel
contains substantial higher-state contributions and that, due to fluctuations
and the overall signal loss, its effective-mass plateau is misleading.
Thus, we compute the result given in the last row of \Tbl{tab-stat60l},
$\fBstat=364$ MeV.

\null From all of this, the first conclusion is that the ``raw'' lattice
results coming from different simulations are in rather good agreement.
As a final example of this point, note that the choice of fitting
used to produce the latter (and larger) of the two results above
is a similar approach as that used by the Wuppertal collaboration, and the
results agree well. The largest source of disagreement comes
from the choice of analysis procedure; the fact that the disagreement
exists simply implies, in our opinion, that the data are not very good.
For the reasons
discussed above, we have chosen the first of the two analyses as our
best result. However, we cannot rule out a systematic error which is roughly
given by the difference between the two. For the remainder of this
paper, we will focus primarily on our results at \bspthr, which we find
to be much less ambiguous.

\subsection{Combined Heavy-Light Analysis\label{sect.analysis1}}

In the final stage of the analysis, we combine the static and
(large-$am$ corrected) conventional results and analyze them using
the parameterization \Ep{analysis}. The following is a summary of the 
essential steps leading to and including this stage of the analysis:
\enumerate
\itm Compute the static result $\hat\phi$ for $\K_q=\K_c$ and $\K_s$.
The normalization $\hat C_A$ is given by \Ep{CAhatnew} with the factor
$C(a,m)$ removed. The light-quark extrapolations are described in 
section \use{sect.static}.
\medskip

\itm Compute $\phi_P(\K_Q)$ and $M'_{P}(\K_Q)$, again for chiral and 
strange light quarks. $C_A$ is given by \Ep{CAnew} and $M'_{P}$ is 
computed from the pole mass $M_P$ according to \Ep{mshift}. 
The light-quark extrapolations are described in section \use{sect.heavy}.
\medskip

\itm Divide out the logarithmic factor $C(a,M'_{P})$ from the
amplitude $\phi_P$ computed in step (2).
\medskip

\itm Fit the amplitudes to a quadratic in $1/M'_{P}$. 
These fits are made using the jackknife estimation of the covariance matrix 
so that correlations among the heavy (and static) quarks are included and
a meaningful $\chisqdof$ is obtained (section \use{sect.fitting}).
\medskip

\itm Interpolate the fit to physical masses ({\it e.g.} $1/M'_P = 1/M_B$)
and multiply the logarithmic correction ({\it e.g.}, $C(a,M_B)$) to obtain
the physical amplitude.
\medskip
\endenumerate

In \Fig{fig-FvsMchE63} we show the result from step (4) at \bspthr,
where the light quark has been extrapolated to the chiral limit.
The static point in all cases is taken from the source $V_s=15^3$,
and the conventional points are from the wall-source results.  From this
analysis we find a smooth interpolation between the heavy lattice mesons
and the static limit, as judged by a goodness-of-fit criterion:
$\chisqdof=1.9/3$ (heavy-chiral) and $\chisqdof=3.4/3$ (heavy-strange).
As the fit and extrapolation parameters are varied to estimate the
fitting errors (section \use{sect.fitting}), we generally find that
the quality of the final fit is not severely altered.

We obtain the deviations from the large-mass scaling behavior by
extracting the first two coefficients of the fit function,
$$
        F = c_0(1 + c_1/M + c_2/M^2).
\EQN fitform
$$
We find
$$\EQNalign{
        c_0 &= .52(4)\ \GeV^{3/2},
\cr
        c_1 &= -1.14(15)\ \GeV,
\cr
        \chisqdof &= 1.9/3\qquad {\rm (chiral\ light\ quark),}
\cr
\noalign{\hbox{and}}
        c_0 &= .58(3)\ \GeV^{3/2};
\cr
        c_1 &= -1.08(13)\ \GeV,
\cr
        \chisqdof &= 3.4/3\qquad {\rm (strange\ light\ quark).}
\cr}
$$
Interpolating these fits, using the experimentally-measured masses and
taking $M_{B_s} = M_B + (M_{D_s}-M_D)$, we obtain the decay constants
$$\EQNalign{
        f_B\            &=\     187(10)\pm12\ \MeV \cr
        f_D\            &=\     208(9)\pm11\ \MeV \cr
        f_{B_s}\        &=\     207(9)\pm10\ \MeV \cr
        f_{D_s}\        &=\     230(7)\pm10\ \MeV.}
$$
The first error is statistical, obtained by a full jackknife of the
procedure outlined above (Sect. \use{sect.fitting}); the second is the 
systematic fitting error computed as the variance of the ``alternate''
results (Sect. \use{sect.fit-errors}). The latter covers the light-quark 
extrapolations and various uncertainties in fitting the correlation functions,
including finite-time effects. The analyses which contributed to this 
estimate were required to satisfy the criterion $\chisqdof<2$ for the
final fit.

\subsection{Large-$am$ Errors\label{sect.analysis2}}

We consider now the systematic error associated with the large-$am$
corrections introduced in sections \use{sect.largeam1} and \use{sect.largeam2}.
In \Fig{fig-FvsMstrEN63} we plot the heavy-strange results, both
corrected and uncorrected (the latter are computed by removing the factors
of $\ematad$ in \Eqs{CAnew} and \Ep{CAhatnew}, and eliminating the mass 
shift \Ep{mshift}).
Using the uncorrected results, we make the following two analyses.
First, we fit the set of points corresponding to the same values
of $\K_Q$ that were used in the analysis above, including the static.
Noting that in this case $\phi$ flattens out at a value well below
the static result, we would not expect a good fit. Indeed, we find 
$\chisqdof=28/3$ (heavy-chiral) and $\chisqdof=32/3$ (heavy-strange).
However, given that $aM_P$ is $\cO(1)$ for the heaviest mesons used,
it is more legitimate to argue that when excluding the corrections one
should throw out the largest masses and attempt to interpolate between
the static limit and an intermediate-mass regime where lattice errors
might, {\it a priori}, be small enough.
This has been the traditional approach; 
an example of it is shown in \Fig{fig-FvsMstrEN63},
where we eliminate from the fit uncorrected points with $a\mone > 0.3$.
Note that we still find a discrepancy between the static 
and conventional results which is characterized by poor fits: we find
$\chisqdof=16/2$ (heavy-chiral) and $\chisqdof=16/2$ (heavy-strange).
Either $\cO(am)$ errors are still too large, or else the conventional
points are too far from the heavy-quark scaling regime to model
the data using \Ep{analysis}. 
As discussed further in the following section, we are unable to obtain
good fits, even after allowing changes in the analysis parameters
to accommodate fitting and scale systematics.
Furthermore, we emphasize here the importance of using the covariant fit
in arriving at the above conclusions:
since the conventional points are highly correlated, the trend in
the data is significant, and only the covariant fits are able to take this into
account.
By contrast, we show in
\Fig{fig-FvsMuncorr} the corresponding uncorrelated fit, where the
discrepancy between the two methods appears to be much less pronounced.
Thus, without the large-$am$ corrections applied to the 
Wilson quark, there remains a significant discrepancy between 
the static and conventional methods---even at \bspthr, using results
where lattice errors should be the smallest. 

We now attempt to characterize the systematic errors associated
with large-$am$.
Since the factor $\ema$ acts at leading order, its full effect is,
in principle, an over-estimate of the systematic error which remains
in the corrected amplitude when the heavy quark is non-relativistic.
A better approach is to consider the correction at $\cO(1/M_P)$
directly. We discuss three methods below. 

The first approach is to re-analyze the data
with the leading-order correction in place, but without making the 
shift in $M_P$, given by \Ep{mshift}. The difference between a quantity
computed from this analysis and that from the fully-corrected one 
gives some idea of the size of the sub-leading large-$am$ corrections.
Although the mass shift is significant at the largest masses,
in both analyses the fit is constrained by the static point and
by the lighter mesons, upon which the shift has a small effect.
Consequently, the error estimates turn out to be quite small---for
the amplitudes $f_B$, $f_{B_s}$, $f_D$ and $f_{D_s}$ they range between
2--3$\%$.  (With or without the mass shift we obtain good fits, 
{\it i.e.} $\chisqdof\simeq1$.)  

A second approach is to try to examine explicitly the error caused
by the suppression of the $\sigdotB$ term.
The shift in $M_P$ adjusts the heavy-quark mass to the kinetic
value $\mtwo(\K_Q)$. Using this as the definition of the quark
mass, let us write the true amplitude as
$$
        \phi_{\rm true} \simeq \hat\phi(1 + {X_{\rm kin} \over 2\mtwo}
                        + {X_{\sigdotB} \over 2\mtwo}) + \cO(1\over \mtwo^2).
\EQN phi-true
$$
Our lattice result is given instead by
$$
        \phi_{\rm latt} \simeq \hat\phi(1 + {X_{\rm kin} \over 2\mtwo}
                        + {X_{\sigdotB} \over 2\mthree}) + \cO(1\over M^2),
\EQN phi-latt
$$
where the difference between $\mtwo$ and $\mthree$ is discussed in
section \use{sect.largeam2}. In principle, we could fit the
lattice data to the form \Ep{phi-latt}, determine the
parameters $\hat \phi$, $X_{\rm kin}$, and $X_{\sigdotB}$,
and reconstruct the true amplitude from \Ep{phi-true}.
A more stable approach, numerically, would be to choose three
heavy mass points (including, perhaps, the static point) and
{\sl solve} for the three unknown parameters. The inherent systematic
errors could then be estimated by varying the
choice of the three points.  Unfortunately, when we apply this analysis here,
we find that the current data
is just not precise enough: Estimates of the difference between
$ \phi_{\rm latt}$ and $ \phi_{\rm true}$ at the $B$
vary between approximately
5 and 40\%, depending on the three points  chosen and on whether the
masses $\mtwo$ and $\mthree$ are adjusted to take into account
the difference between the heavy quark mass and the heavy-light meson mass
(which should be a higher order effect).  

It is easy to understand why the above method fails with the current data.
\null From   \Fig{fig-FvsMchE63},  one sees that the dependence $\phi_P$ on
$1/M_P$ is quite linear for the heavier masses which lead up to
the static point.  
It will therefore be hard to pull out the coefficient $X_{\sigdotB}$, which
is responsible for non-linearities in \Ep{phi-latt} and 
for the difference between
\Ep{phi-latt} and \Ep{phi-true}.

Since the first method is rather {\it ad hoc}, and since the
second may be giving indications of considerably larger errors,
we have implemented a third approach.  This approach, while
indirect, has the advantage that it estimates the combined
systematic errors due to all identified problems that affect
the large-$am$ points:  the perturbative mismatch with the
static computation (section  \use{sect.largeam1}),
the procedure for treating the logarithmic corrections 
(section \use{sect.logM}), as well as the difference between
$\mtwo$ and $\mthree$.  The idea is to eliminate all heavy
quark points with $a\mone > 0.17$ ({\it i.e.}, all but
our lightest three heavy quarks---this corresponds to eliminating
all mesons with mass $aM > .36$ for a chiral light quark)
and to fit the remaining points
in conjunction with the static value.  We take
the difference between the results of these fits and
the previously discussed central values as our estimate of
the large-$am$ systematic error.  From \Fig{fig-FvsMstrEN63}
one sees that the fit to the large-$am$ corrected points
comes quite close even to the light-mass points that were not
included in the fit.  Thus, it should be no surprise
that we get quite small error estimates if we use
the same fitting form \Ep{fitform}
on the light-mass points alone.  However, one may argue
that, without the heavy-mass points, we have no direct
evidence for the basically linear behavior of $\phi_P$ with
$1/M_P$ in the region of the $B$.  We therefore chose,
as a possible smooth
interpolation between the light-mass points and the
static limit, a quartic fit in $1/M_P$  with the coefficient
of the linear term set equal to 0. (Since there are four
remaining parameters and only four points to fit, 
we actually just solve for the parameters.)
The result is shown in
\Fig{fig-new}.  This gives a considerably larger
deviation from the central values,
and we take this deviation as a conservative estimate of the
combined large-$am$ error.
The results of this analysis are given in \Tbl{tab-decay}. 

\subsection{Scale Errors and Other Systematics\label{sect.analysis3}}

To study scale errors, one can in principle consider two
approaches. The first and preferable way is to repeat
the calculation at a variety of different lattice spacings,
thereby computing ``physical'' results as a function of
$a$ ({\it e.g.}, $f(a)$). Let us refer to any deviation from
scaling behavior ($f(a)={\rm const.}$) as a ``pure'' scale 
violation; we computed this error to be $\simeq6\%$ for $f_K$
at \bspthr\ (section \use{sect.light}).
Alternatively, at a given lattice spacing, one can simply use
several different methods to determine $\ainv$ and compare the results.
The scale error computed this way is less informative
because it includes other systematic effects (such as quenching)
in an unknown way.
In either case, there are two issues which we wish to consider:
(1) the connection between scale errors and large-$am$ effects,
and (2) the estimation of a systematic error in physical amplitudes.

Combining results at \bspthr\ and \bsix,\Footnote{\dag}{
        The analysis procedure at \bsix, \latl\ (8 configurations),
        was analogous to that at \bspthr, with the following exceptions: 
        No fits were covariant, and individual configurations were
        time-reversal averaged before jackknifing.
        In addition, for the scaling comparisons
        depicted in \Fig{fig-FvsMchE6063} and \Fig{fig-FvsMchN6063},
        we have not (at \bsix) included the scale $\ainvfpi$ in the
        jackknife analysis.
        This results in a smaller statistical error---the one that is
        typically reported in lattice simulations.
        }
we first examine the scaling behavior of the heavy-light amplitude
both with the large-$am$ corrections (\Fig{fig-FvsMchE6063}) and 
without them (\Fig{fig-FvsMchN6063}).
Assuming that the corrections remove most of the large-$am$ systematics,
there may be some evidence for scaling violations
in \Fig{fig-FvsMchE6063}, but the differences between \bsix\ and
\bspthr\ are not much larger than the statistical errors.
In the uncorrected case the difference between the results at
the two couplings is smaller in general.  
Note however that there is a significant
difference between the trends of the two simulations
at the highest masses.
At \bsix, the onset of the downward slope in the 
uncorrected amplitude is evident, presumably due to  the
``$e^{-aM_P/2}$'' behavior discussed in section \use{sect.hopping-limit}.
At \bspthr, the heavy-mass data is still rising, but with decreasing
slope---presumably a smaller-$am$ manifestation of the same
effect.  Our conclusion is that the high-mass \bsix\ data suffer  from
additional suppression relative to the \bspthr\ data, and 
that the apparent agreement between the two couplings is
accidental, indicating a cancellation between the
large-$am$ errors and true scaling violations.  However,
other interpretations are clearly possible.
Due to the limitations of our
method and the quality of the data, it is difficult to disentangle specific
effects such as this
in an unambiguous way.  Thus, the scaling behavior between
\bspthr\ and \bsix\ cannot be used to choose between the corrected and
uncorrected conventional method in the regime
$am\lsim1$; nor can a quantitative estimate of pure
scaling violations be made for this data.

Let us use the results at \bspthr\ and consider the scale
error in terms of an uncertainty in $\ainv$. If we choose $f_K$
to set the scale, the outcome can be more or less predicted from
\Fig{fig-fKofpi}. The difference between the ratio 
$f_K/f_\pi(\beta\!=\!6.3)$ and $f_K/f_\pi({\rm phys.})$ is about $7\%$;
this essentially translates directly to other quantities, although 
there are additional effects in this case since we fit 
$\phi_P\ \vs\ 1/M_P$ in physical dimensions.
Re-computing the full analysis using $\ainv(f_K)$, we find a systematic
shift of $\sim 8\%$ in the decay constants of the $D$ and $B$ mesons.

Of course, we may consider a variety of other ways to choose the
scale---the string tension, the nucleon mass, the rho mass, {\it etc}.
In principle, the most conservative bound will be obtained using
$\ainvMrho$ since here the discrepancy with $\ainvfpi$ is
typically quite large.
In this simulation we have not computed the vector mesons; however,
high-statistics studies of the string tension\cite{bali-schilling}
and the quenched hadron spectrum\cite{GF11} allow us to make an
estimate of the difference in these scales at \bspthr.
To do this, we first interpolate the string tension results
of\cite{bali-schilling} to $\beta=6.17$. We choose this coupling
because it is the weakest one used in the spectrum calculation
of\cite{GF11}, from which we obtain $aM_\rho$. Using
$\sqrt{\sigma}=440$ MeV and the physical value of $M_\rho$, we compute
the relative scale error $(\ainvMrho - \ainvsigma)/\ainvsigma \simeq .08$
This we expect to bound the corresponding error at \bspthr.
Therefore, we interpolate the string tension results to
\bspthr\ and make the scale assignment $\ainvMrho = 1.08 \ainvsigma$.
(At \bspthr\ we find $\ainvsigma=3.18$ GeV, similar to the scale we
obtain from $f_\pi$, and thus $\ainvMrho=3.44$ GeV.) 
Note that the scale error obtained this way
is slightly larger than the one determined from
$f_K$; we choose the larger value in order to make
a conservative error estimate.

We show the results of the analysis using this scale in \Tbl{tab-scalesys}.
\null From the top half of the table (large-$am$ corrected) we compute the
systematic scale error quoted with our final results. All other
aspects of this analysis were unchanged from that which produced our
central values.
The lower half of the table contains the results computed without
the large-$am$ corrections and restricted to $a\mone\le 0.3$, as described
in the previous section. The point here is that despite the
change to the scale $\ainvMrho$, there still remains a discrepancy
between the conventional and static methods, which is indicated by
the large value of $\chisqdof$ for the fit.
\medskip

Since we use $f_\pi$ to set the scale for our central values,
any uncertainty in the choice of coupling used in the perturbative
renormalization constants $Z_A$ and $\hat Z_A$ will have only a
small effect in the final results for the heavy-light decay
constants. That is to say, in the conventional amplitude $f\sqrt{M}$,
a change in $g^2$ will have no effect since a jackknifed $f/f_\pi$
method is used---the factors of $Z_A$ cancel.
There will be residual effects due to the change in the ratio
$\hat Z_A/Z_A$ and the scale which normalizes the masses $M_P$
to be used in the final fit of $\phi\ {\rm vs.}\ 1/M_P$.
Our central values are obtain using the ``boosted'' coupling
$g_{V}^2(\tilde q)$ derived from the heavy-quark potential; the momentum
$\tilde q$ is defined as the average $\ln(q^2)$ of the tadpole graph,
$\tilde q=2.58/a$\cite{lepmac-viability}. This produces the following
values:
\vfill\eject
$$\EQNdoublealign{
\beta=&5.7,\qquad  g_{V}^2 \simeq &1.95; \cr
&6.0, &1.77; \cr
&6.3, &1.62. \cr}
$$
By repeating the analysis using the bare coupling ($g^2=6/\beta$),
we find changes in the heavy-light decay constants of $\lsim 3\%$.
However, based on the studies presented in\cite{lepmac-viability},
the bare coupling is, in any case, a poor choice of expansion parameter;
we expect this to be an overestimate of the uncertainty in the final results. 
Since there exists reasonable motivation for using the boosted coupling,
and since the residual effect of changes in the coupling is so small,
we consider this uncertainty to be subsumed in our general estimate of
scale errors and do not include it as a separate source of systematic error.

We expect finite-volume errors to be negligible compared to the other
errors, even though our $24^3$ lattice has a volume
of only $\sim (1.5 {\rm \ fm})^3$.  This is because we work
only with pseudoscalar meson states at zero 3-momentum, 
which are quite insensitive to volumes of this size.  For example,
in the calculation\cite{sharpe}  of $B_K$ with staggered fermions,
no significant difference was found between the results for
$16^3$ and $24^3$ lattices at \bsix; the $16^3$ lattice has almost
the identical physical volume as our \bspthr\ lattice.
Similarly, \Tab{tab-chir60l} and \Tab{tab-chir60A} show
no difference, at the level of our statistics,
for $f_K/f_\pi$ or $\ainv$ between 
our $16^3$ and $24^3$ lattices at \bsix.
Note that finite size effects on the
heavy-light pseudoscalar states should be even
less than for light-light states, since the former
are somewhat smaller in size.
Of course, the size of the finite-volume
effects should be checked explicitly for the current computation;
that work  is in progress.

Our results for the decay constants 
and bounds for various systematic errors are summarized in
\Tbl{tab-decay}. In \Tbl{tab-ratios} we present the
jackknifed ratios and similar estimates of systematics.
For final results, we add the fitting/extrapolation
and large-$am$ errors in quadrature. They are estimated through
similar numerical procedures which are essentially independent
of each other. We list the scale error separately.
Thus, we find
$$\EQNalign{
        f_B\            &=\     187(10)\pm34\pm15\ \MeV \cr
        f_D\            &=\     208(9)\pm35\pm12\ \MeV \cr
        f_{B_s}\        &=\     207(9)\pm34\pm22\ \MeV \cr
        f_{D_s}\        &=\     230(7)\pm30\pm18\ \MeV\ ,}
$$
where the first error (in parentheses) is statistical, and
the second two are systematic, representing respectively
the large-$am$ plus fitting errors, and the scale errors.

We now briefly recall the experimental status 
regarding the measurement of these 
quantities.
For $f_D$ there only exists an upper
bound at present, $f_D < 290\ \MeV\ (90 \% {\rm CL})$ \cite{mark2}.
Recently an important first step in the experimental determination
of $f_{D_s}$ was made, yielding
$f_{D_s} = 232 \pm 45 \pm 20 \pm 48$ MeV\cite{expt}.
For now the errors in this measurement are too large
for it to have an impact on lattice calculations
of $f_{D_s}$; improved measurements are eagerly awaited.
Once a precise experimental determination of $f_D$
or $f_{D_s}$---or for that matter any one of the four
heavy-light decay constants---becomes available, then
lattice results on the ratios of the $f$'s (see \Tbl{tab-ratios}) 
could prove useful in pinning down the values of the others,
since the errors on the ratios are rather small ($<5\%$). 

\section{Conclusion\label{chpt.con}}

We have computed the heavy-light decay constants $f_B$, $f_{B_s}$,
$f_D$ and $f_{D_s}$ in the quenched approximation at \bspthr.
The results have been summarized above; in addition we present
the results for the jackknifed ratios in \Tbl{tab-ratios}.
Included in these calculations, we have computed the ratio
$f_K/f_\pi$, extrapolated to the limit of zero lattice spacing.
Here we find, for the quenched theory, $f_K/f_\pi=1.08\pm.03\pm.08$.
There are several important aspects of this work
which warrant some additional concluding remarks.

First, the reliability of the static method must be considered
independently of any comparison to the conventional calculation
(and, preferably, independently of issues of scale and renormalization).
Early results for $\fBstat$ were significantly larger than those
reported here, as emphasized in Sect. \use{sect.static}.
The most likely source of such discrepancy is that various
computational techniques ({\it i.e.}, sources, fitting, {\it etc}.) are sensitive
to different amounts of higher-state contamination in the correlation
functions.  
Unlike the pion correlator, the static-light channel suffers
from an exponentially falling SNR. To boost the ground-state signal
in early timeslices, various smearing techniques have been used,
both in this work and elsewhere. However, higher-state errors
remain a difficult issue, because the fundamental problem of noise 
at large times is not eliminated by the smearing. To minimize the error,
we have studied the dependence of the amplitude on the size
of the cube source and on the range of fitted timeslices.
Physical results must of course be independent of the source,
which is only used as a tool for the lattice calculation. And, 
while it is not guaranteed that more than one cube size will be
satisfactory, it is equally possible that none will be: a cube is
a poor approximation to the ground state wave function to begin with,
and excited-state contamination may never be sufficiently reduced.
Consistency of the results over some range of sources provides a 
reliable check that this is in fact not the case.

At \bspthr, we see agreement between
the static results from different smearing volumes---around some 
size which is presumably characteristic of the ground
state wave function---and from the wall source,
where a different (two-state) fit was used over an early range of
timeslices. The ``best'' cube size, based on the earliest plateau
in the effective masses, depends on the light-quark
mass. For our final results, we have used larger smearing volumes
(which in this sense are optimal for the lightest quarks) and later
fitting intervals, where the overall dependence on cube size is minimized.
However, the results from earlier fitting intervals are not very
different, and we included the variations in our estimate of
the systematic error from fitting.
We report our final
results from this simulation.

At \bsix, our sample is smaller and the cube sources
cannot be adjusted as precisely. 
Apparently for these reasons,
the \bsix\ static results are considerably more ambiguous.
The results from a range of different smearing volumes only
agree when we use a later fitting interval.
(The ``best'' cube size depends strongly on the light
quark mass.)  We therefore tend to prefer such later fitting
intervals, and
this preference results in a significantly smaller value of $\fBstat$
(237 MeV) than has previously been reported at \bsix.
Nevertheless, we find it difficult
to rule out alternate analyses, in which the
correlation functions are fit at considerably earlier times.
Such analyses of our data can produce a value for $\fBstat$ as
large as 364 MeV.

At \bspthr, we use an interpolation between the static
and conventional heavy-light results 
in order to estimate the corrections to the asymptotic
scaling law \Ep{scalelaw}. We find large coefficients
({\it e.g.}, $c_1=-1.14(15)\ \GeV$), and thus the correction to the amplitude
$\phi_P$ is significant for the $B$ meson ($\simeq 20\%$) and large for
the $D$ ($\simeq 45\%$). In fact, the $D$ meson is much better simulated
by using the conventional lattice method, as long as suitable adjustments
are made for Wilson fermions at quark masses near the scale of the
inverse lattice spacing ({\it i.e.}, $am\lsim1$). This statement is 
justified, not by a comparison of our results to experiment
(the errors in the experimental measurement of $f_{D_s}$\cite{expt}
are still too large to help us much in this regard), but rather by the 
observation that only when large-$am$ corrections are included in
the calculation do we find that the large-mass extrapolation of the
heavy-light amplitude agrees with the static calculation.
In practical terms, we characterize this agreement (or lack thereof,
if the leading order correction is omitted) by the $\chisqdof$
of the correlated fit to the combined set of results.
 
The systematic error which remains uncorrected in the conventional
calculation, and the extent to which it affects the interpolation 
to the physical amplitudes, is unfortunately difficult to estimate. 
We have argued that, in terms of a non-relativistic expansion,
the leading-order correction is essential but that the next-to-leading
one is difficult---but perhaps not impossible, with good data---to include
in a systematic way, because the kinetic and 
$\sigdotB$ interactions are, for large $am$, incorrectly normalized
with respect to each other in the standard Wilson action.
In Sect. \use{sect.analysis2} we used three methods to estimate the
error incurred in the decay constants, and the results were rather dissimilar.
The first method simply demonstrates that the large-$am$ correction based
on the kinetic term of the non-relativistic expansion ({\it i.e.}, the shift in
$M_P$) has a small effect on the final correlated fit.
The second is an attempt to address directly the unwanted lattice
artifact---the suppression of the $\sigdotB$ interaction. 
Unfortunately, the present data is just not precise enough to support
this type of analysis.  The third approach estimates
this large-$am$ error (and, concurrently, the errors induced by
the perturbative mismatch with the
static computation (Sect. \use{sect.largeam1}) and
the procedure for treating the logarithmic corrections
(Sect. \use{sect.logM})) by dropping all quark masses
with $a\mone>0.17$ and re-analyzing the remaining points under conservative
assumptions.
The results are reported in \Tbl{tab-decay} under ``large-$am$.''
 
Finally, let us return to the conventional results at heavy masses and
the issue of the leading-order correction, the factor of $\ema$ in the
Wilson quark normalization. We note that in the literature, results
exist\cite{elc-6.4} which are in contrast to what we have found
here---namely, that this correction must be included in order to eliminate
the large discrepancy between the static and conventional decay constants.
The difference between our results and those of ref.\cite{elc-6.4}, where
similar methods were used, appears not to be in the static point
(both calculations were at weaker couplings, \bsptwo--\bspfour, and
the static results are in better agreement than at \bsix) but rather in the
conventional amplitude. In short, they see the elimination of the old
discrepancy at these weaker couplings without the
$e^{am}$ corrections, but we do not.
However, as opposed to the analysis of ref.\cite{elc-6.4},
we have consistently used covariant fits throughout so that the effect of
correlations in the data are not missed when attempting to study
numerically the interpolation between the static and heavy-quark regimes.
A conclusive resolution of this discrepancy will clearly
require improved source techniques (and longer lattices) so that
any question of the integrity of the raw lattice measurements is
first addressed.

Questions such as this, however, only pertain to the calculation of
a precise result with the conventional method; the result
thereby obtained is not necessarily an accurate one, as we have
already emphasized.
To remove the ambiguity in the decay constant at $\cO(1/M)$ requires
an improved action (so that the normalization of
$\sigdotB$ relative to the kinetic energy term is corrected) or
better data in the large mass region (so that the difference between
$1/m_2$ and $1/m_3$ can be extracted cleanly), or both. 
Control of these effects, in turn, would facilitate the study
of other key systematics, such as scale violations.
Work along these lines is in progress.

\bigskip\bigskip
\nosechead{Acknowledgements}
We thank A. Kronfeld, P. Lepage, P. Mackenzie, and  S. Sharpe
for very useful discussions, and C.\ M.\ Heard for help with the
computing.  The computations were performed
at the San Diego Supercomputer Center and the National
Energy Research Supercomputer Center; we thank both centers
for their generous help.
The authors C.B., J.L. and A.S. were supported in part by
the U.S. Department of Energy under grant numbers
DE2FG02-91ER40628, DE-FG06-91ER40614 and DE-AC02-76CH0016 respectively.

\vfill\eject
\nosechead{References}
\ListReferences
\vfill\eject


\fullfigure{cheeseburger}
\infiglist{Noise correlator for the static-light channel}
\vskip 6cm
\centerline{\psfig{file=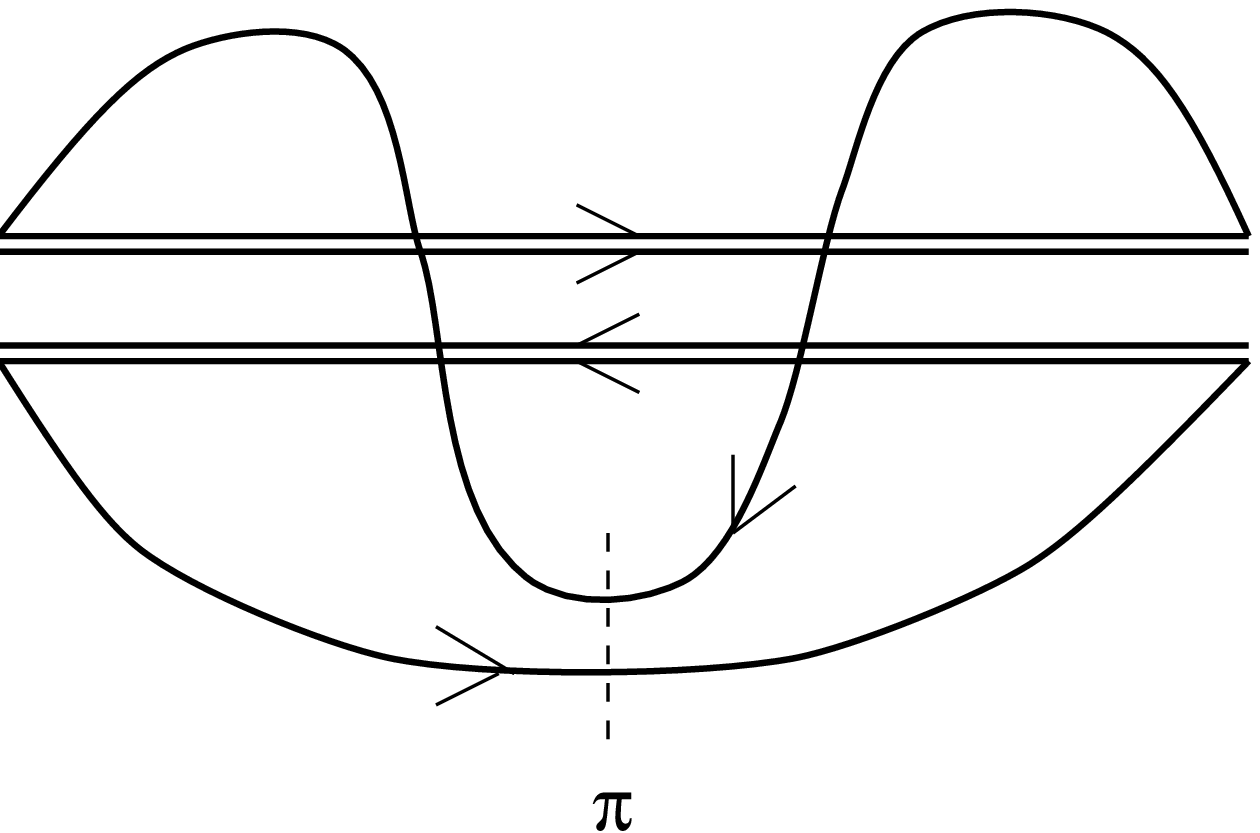,height=8truecm}}
\caption{``Noise correlator'' for the static-light channel}
\endfigure

\fullfigure{fig-emWpion63}
\infiglist{Pion effective mass, \bspthr, wall source}
\vfill
\includegraphics{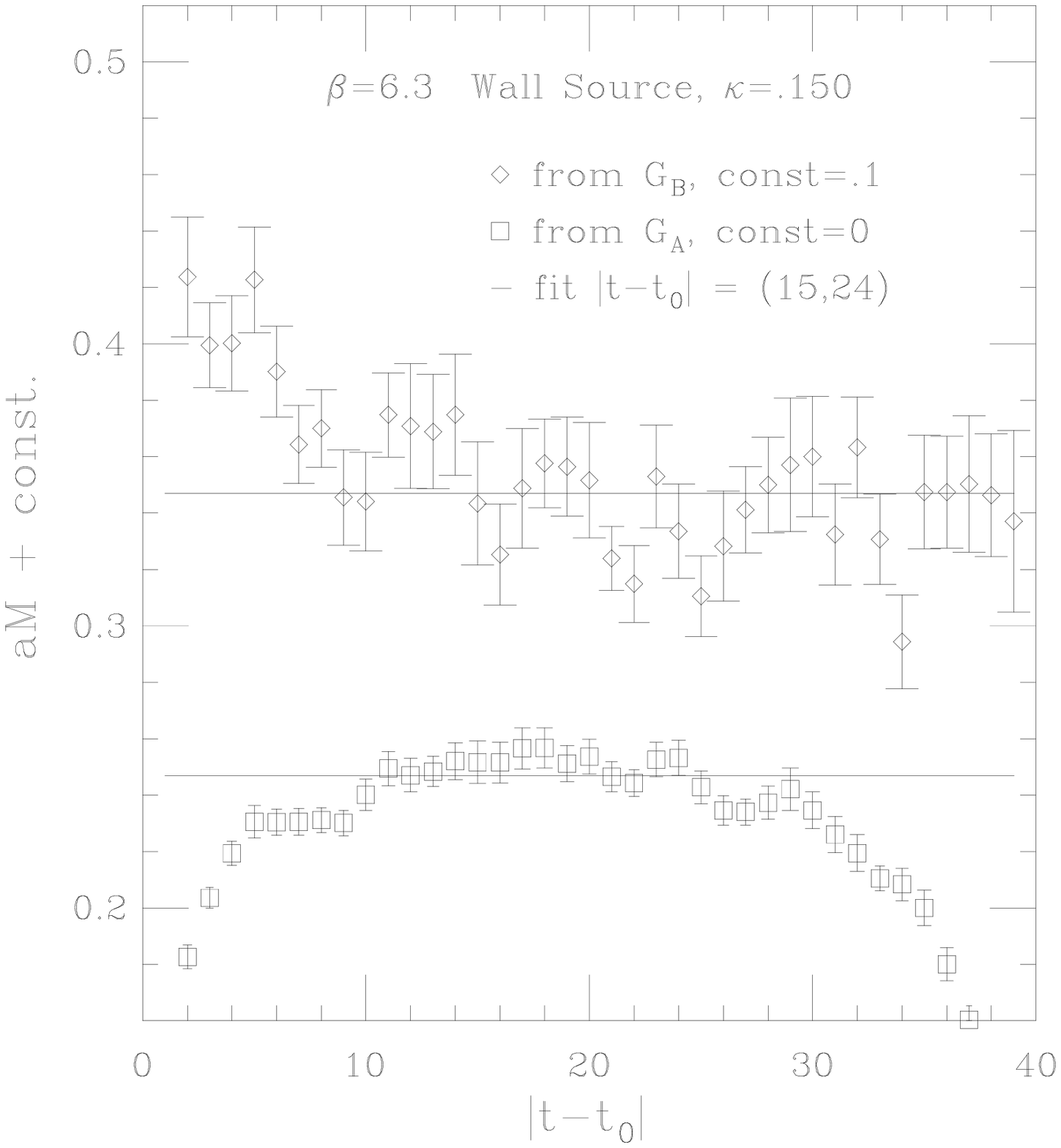}
\Caption
The effective mass of the wall-source pion for $\K=.150$ at \bspthr.
The solid lines indicate the mass from the single-state fit, \Ep{ex3}.
So that both $G_A$ and $G_B$ masses may be shown simultaneously,
the $G_B$ mass is displaced upward by 0.1.
\endCaption
\endfigure

\fullfigure{fig-emPpion63}
\infiglist{Pion effective mass, \bspthr, point source}
\vfill
\includegraphics{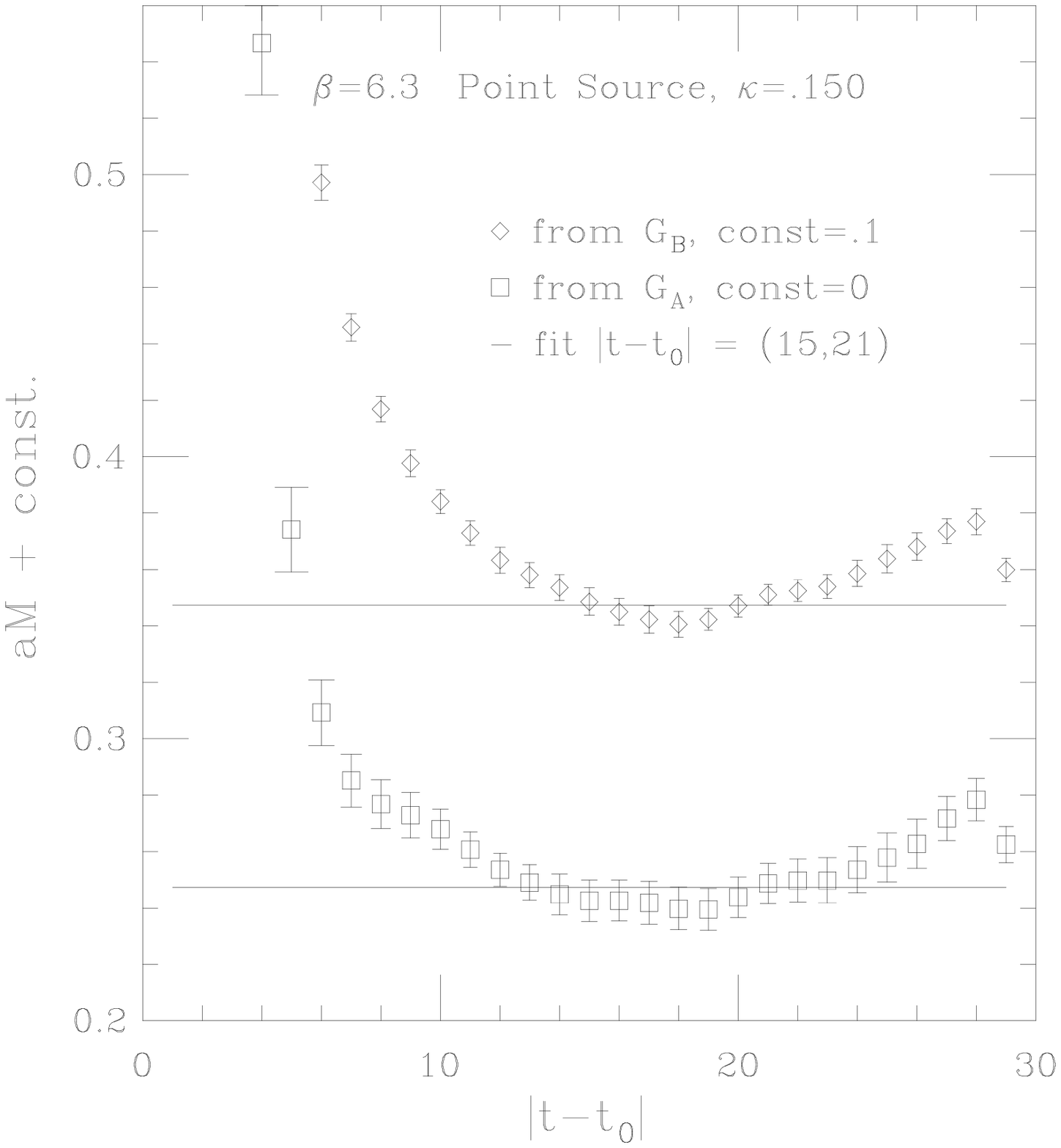}
\Caption
The effective mass of the point-source pion for $\K=.150$ at \bspthr.
The solid lines indicate the mass from the single-state fit, \Ep{ex3}.
\endCaption
\endfigure

\fullfigure{fig-HLmxt63W}
\infiglist{Chiral extrapolations of heavy-light masses, \bspthr}
\vfill
\includegraphics{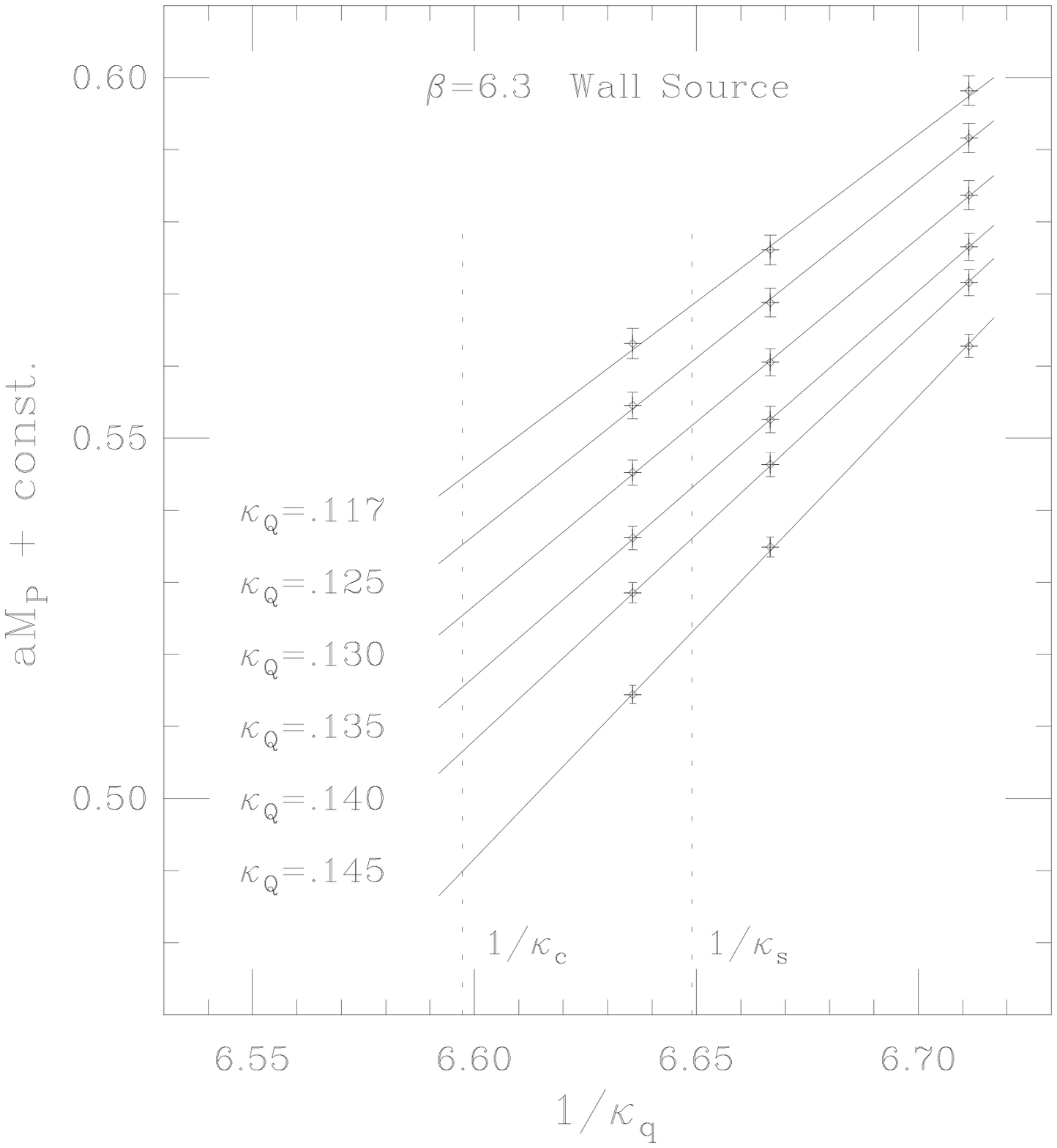}
\Caption
Fits and extrapolations of the heavy-light masses for states
with $M_P>1\GeV$.
The results have been offset by various amounts
to improve the readability of the graph; the constants are
$\K_Q=$.145, +.130; $\K_Q$=.140, 0.; $\K_Q$=135, -.135;
$\K_Q=$.130, -.275; $\K_Q$=.125, -.425; $\K_Q$=117, -.700.
\endCaption
\endfigure

\fullfigure{fig-HLfxt63W}
\infiglist{Chiral extrapolations of heavy-light amplitudes, \bspthr}
\vfill
\includegraphics{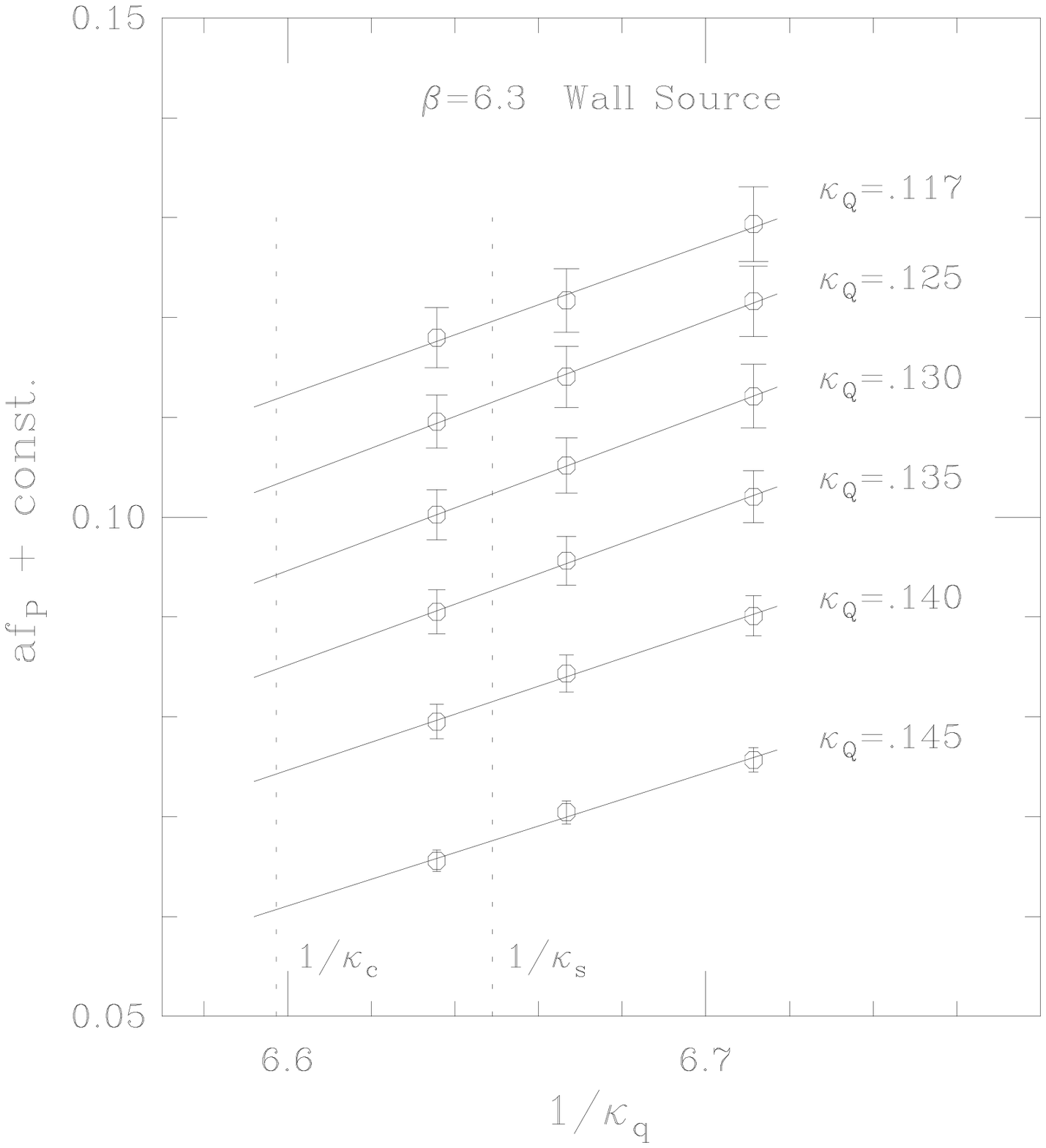}
\Caption
Fits and extrapolations of the heavy-light decay constants for states
with $M_P>1\GeV$. The results have been offset by various amounts
to improve the readability of the graph; the constants are
$\K_Q=$.145, 0.; $\K_Q$=.140, +.01; $\K_Q$=135, +.02;
$\K_Q=$.130, +.03; $\K_Q$=.125, +.04; $\K_Q$=117, +.05.
\endCaption
\endfigure

\fullfigure{fig-fKofpi}
\infiglist{$f_K/f_\pi$ vs. lattice spacing}
\vfill
\includegraphics{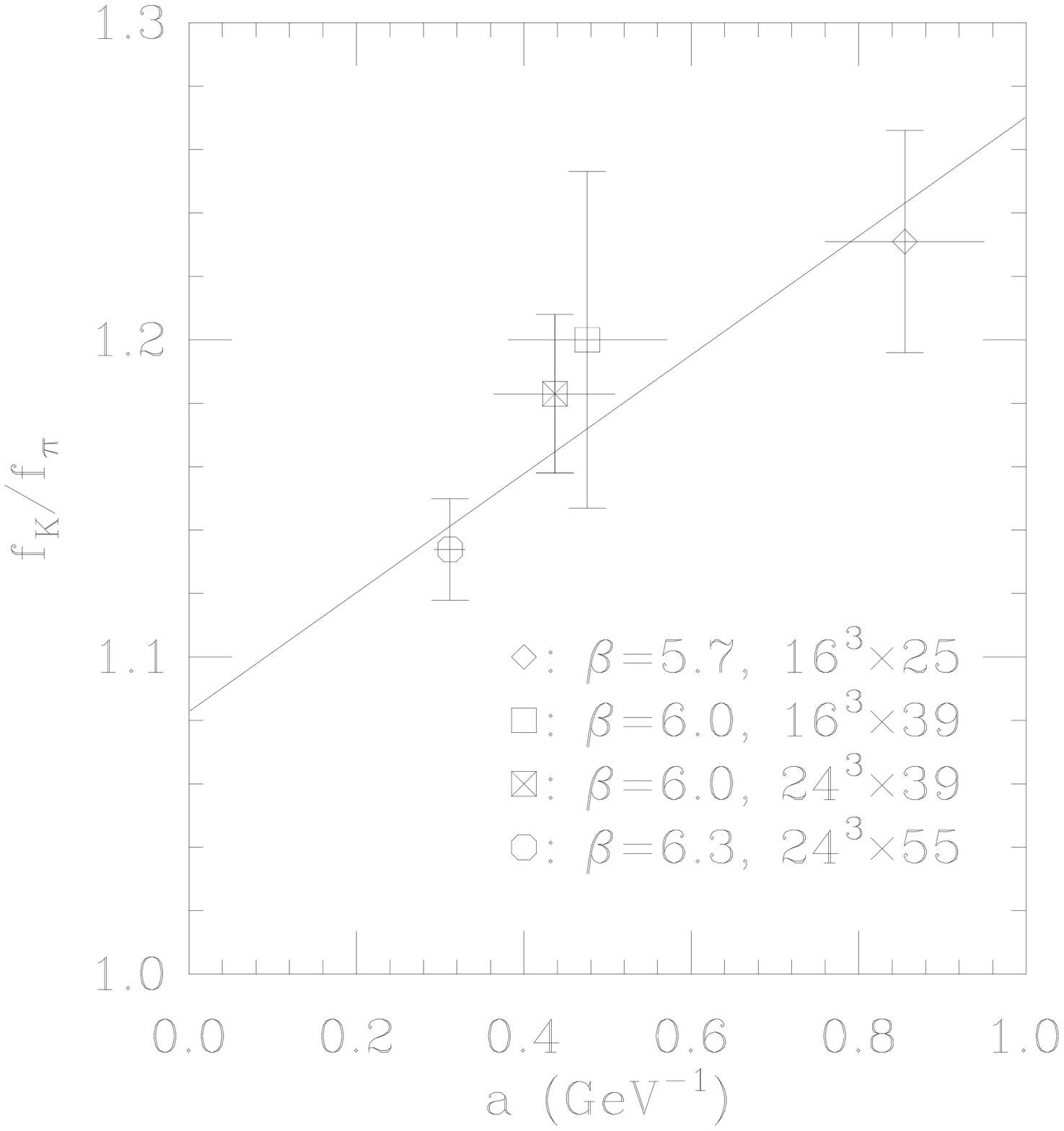}
\Caption
Scaling study of (quenched) $f_K/f_\pi$.
\endCaption
\endfigure

\fullfigure{fig-emWhevy130}
\infiglist{Heavy-Light effective mass, \bspthr, wall source}
\vfill
\includegraphics{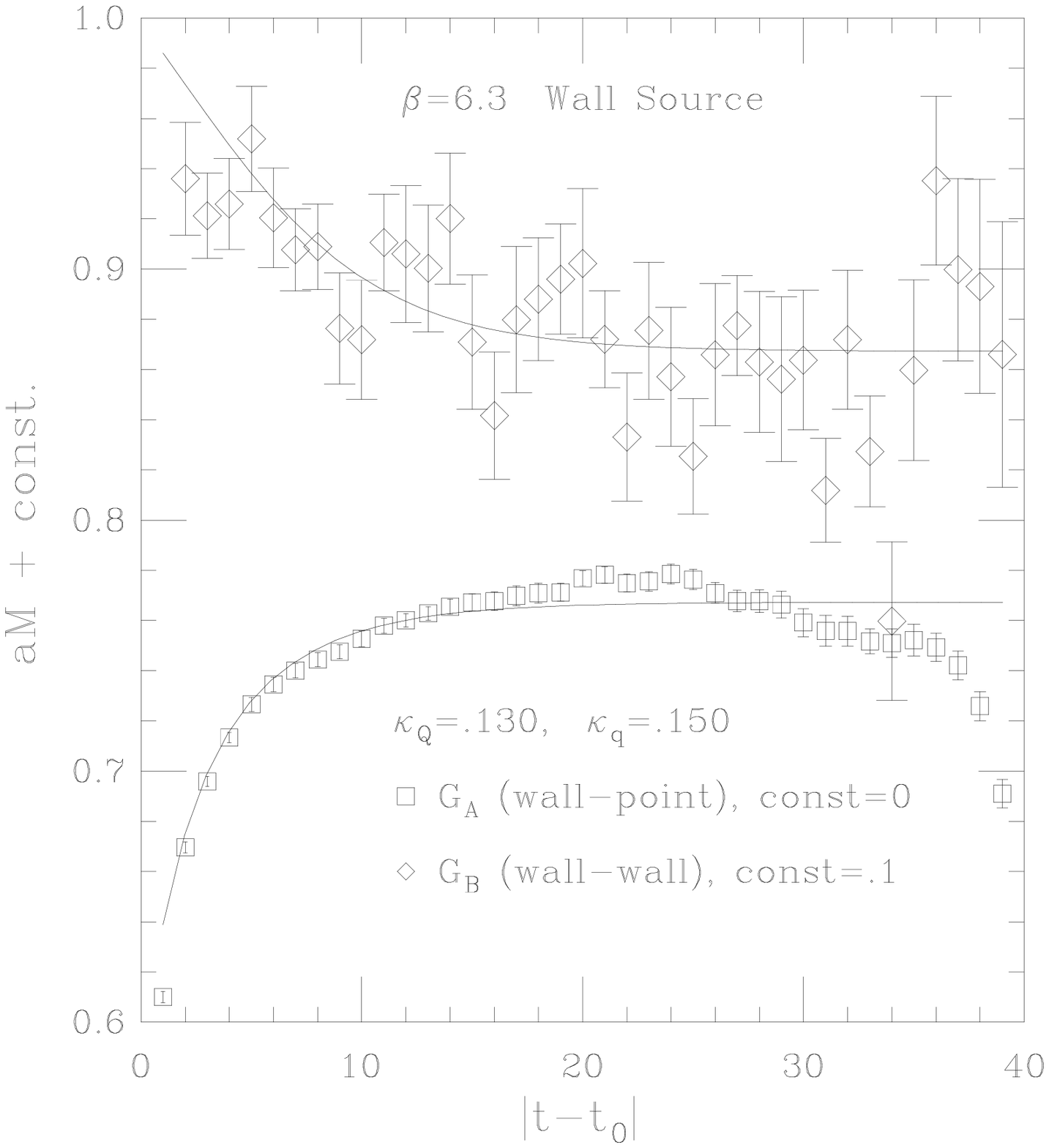}
\Caption
The effective mass of the heavy-light wall-source correlators at \bspthr, 
$\K_Q=.130$ and $\K_q=.150$.
The solid lines are the effective mass computed from the logarithmic
derivative of the fitted function.
The fitted interval was $|t-t_0|$=(3,18).
\endCaption
\endfigure

\fullfigure{fig-emPhevy125}
\infiglist{Heavy-Light effective mass, \bspthr, point source}
\vfill
\includegraphics{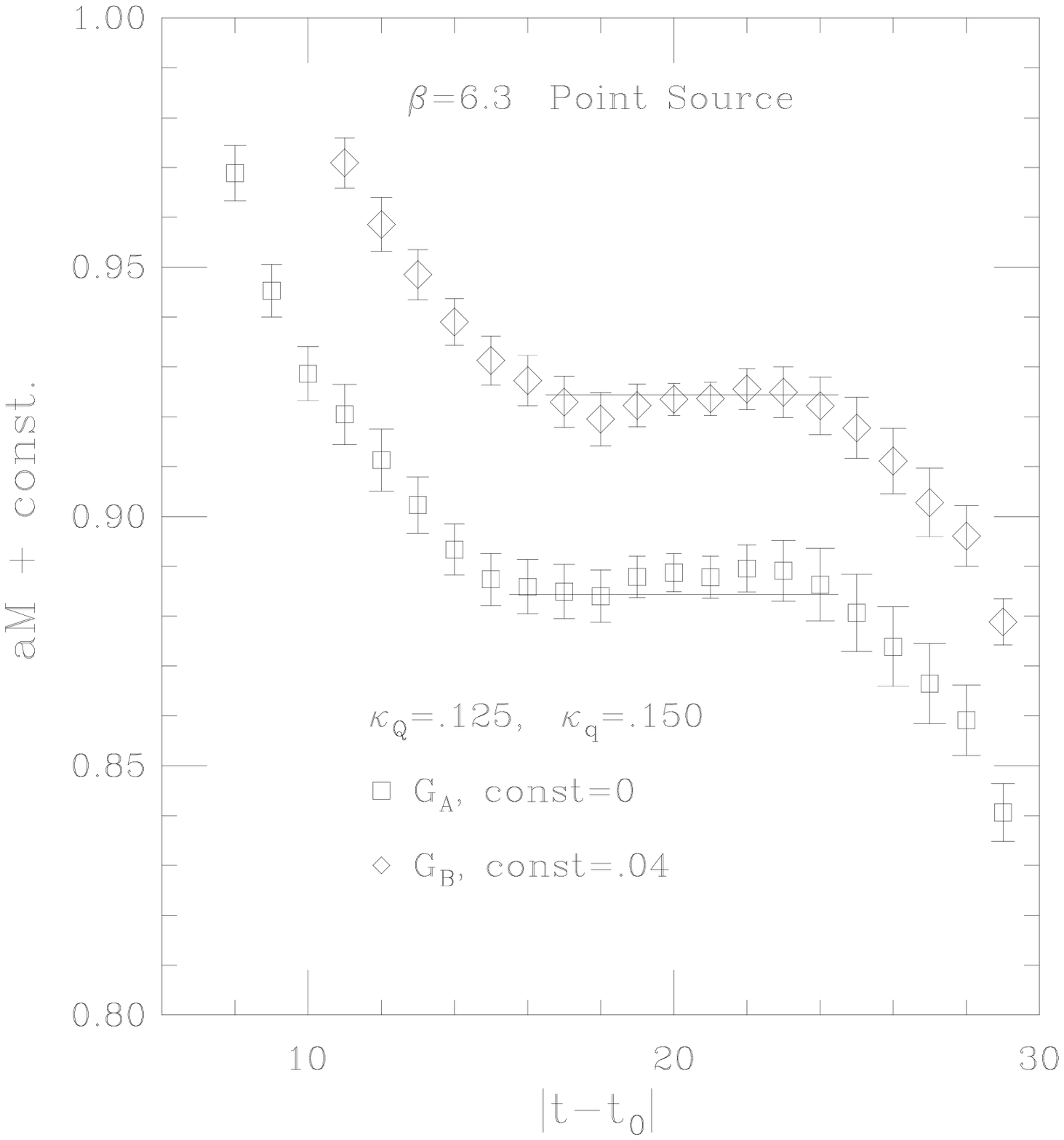}
\Caption
The effective mass of the heavy-light point-source correlators at \bspthr, 
$\K_Q=.125$ and $\K_q=.150$.
The solid lines indicate the fitted mass from the function \Ep{ex3}
and the time intervals that were used.
\endCaption
\endfigure

\fullfigure{fig-emS13P63}
\infiglist{Static-Light effective mass, \bspthr, $V_s=13^3$}
\vfill
\includegraphics{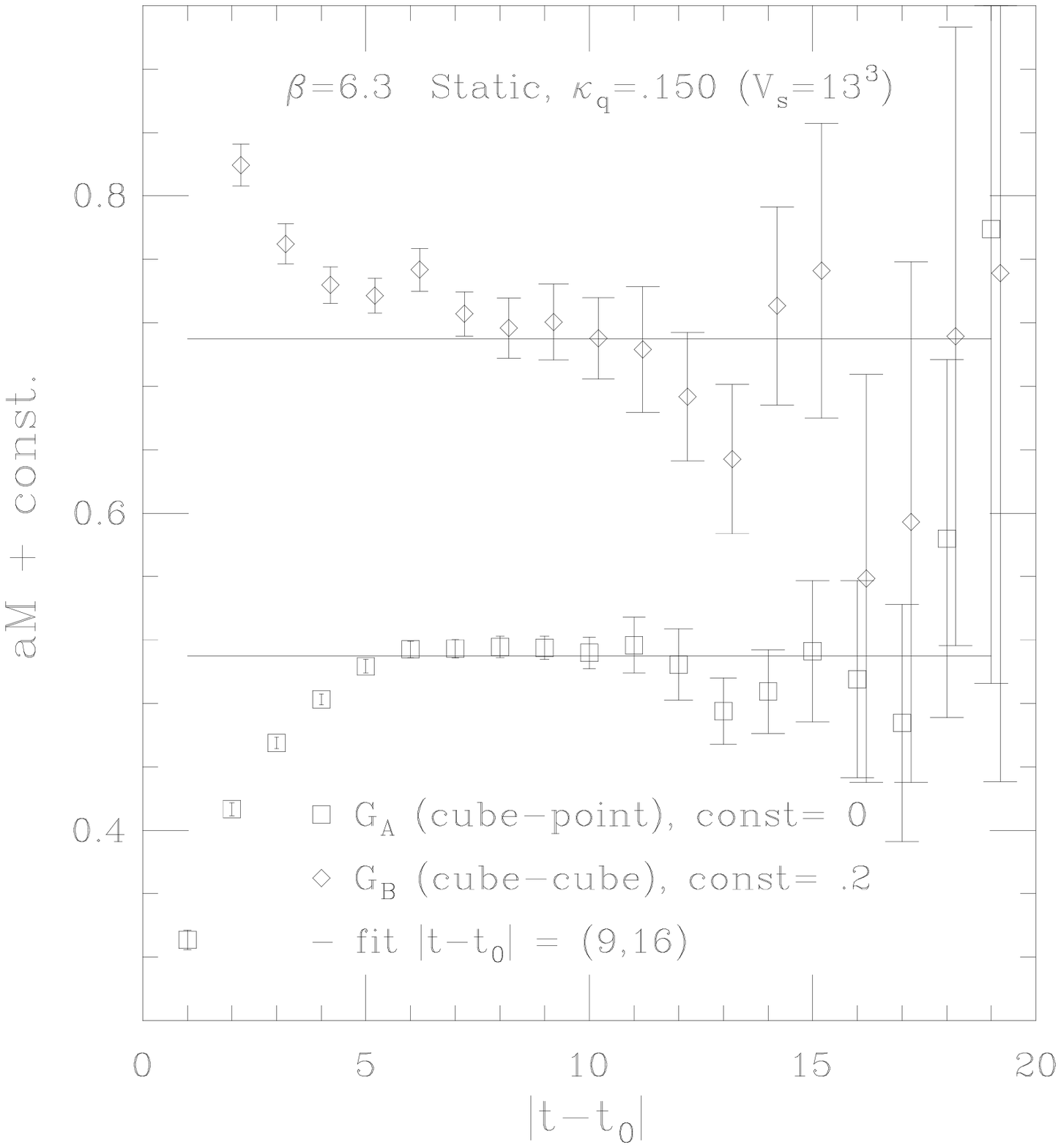}
\Caption
The effective mass of the static correlators at \bspthr, $V_s=13^3$.
\endCaption
\endfigure

\fullfigure{fig-emSW63}
\infiglist{Static-Light effective mass, \bspthr, wall source}
\vfill
\includegraphics{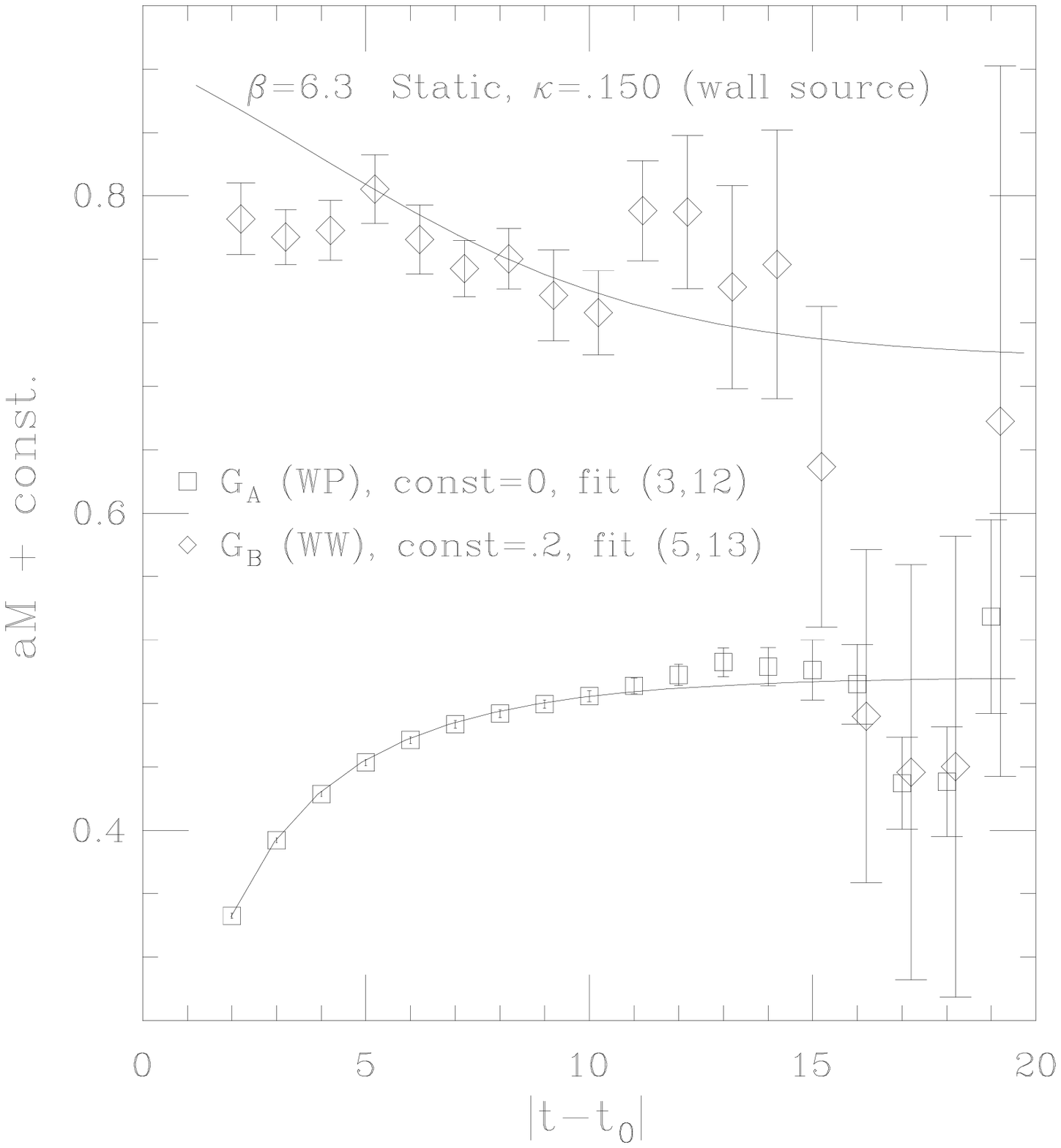}
\Caption
Effective mass of the wall-source static correlators at \bspthr.
\endCaption
\endfigure

\fullfigure{fig-fvsV63}
\infiglist{Smearing study of static amplitude at \bspthr}
\vfill
\includegraphics{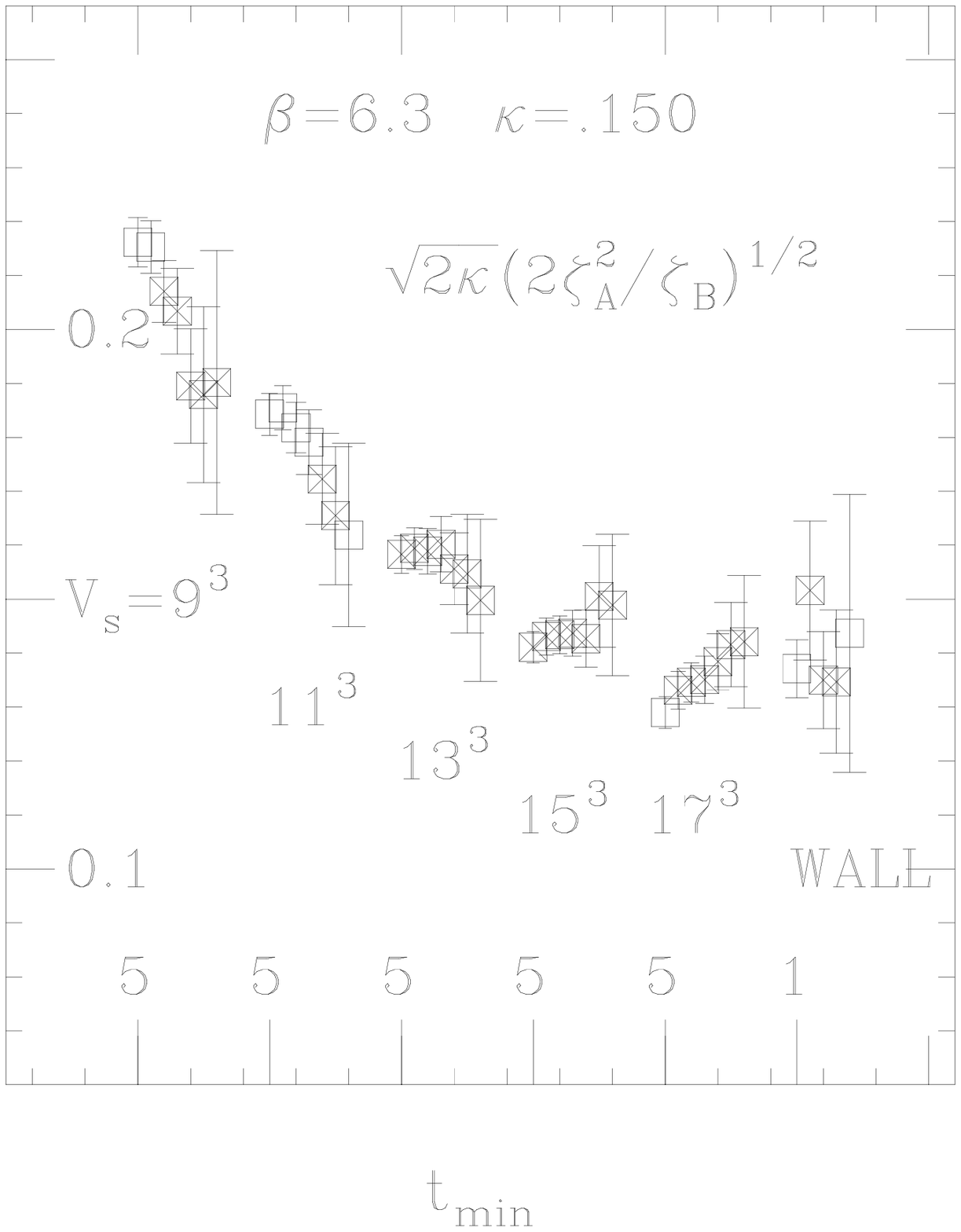}
\Caption
The static-light amplitude at \bspthr, computed using
different smearing sizes and fitting intervals.
For each smearing function, the fit interval for $\Gxx$ is held constant
($t=(10,15)$ for the cube sources, $t=(5,14)$ for the wall) and the interval
for $\Gax$ is indicated by the horizontal axis
($t=(t_{min},t_{min}+T)$, where $T=4$ for the cube sources, and $T=9$
for the wall source).  Fits which passed the criterion $\chisqdof<1$
are marked with $\times$.
\endCaption
\endfigure

\fullfigure{figs-emassV}
\infiglist{Comparison of static-light effective masses, \bsix}
\vfill
\includegraphics{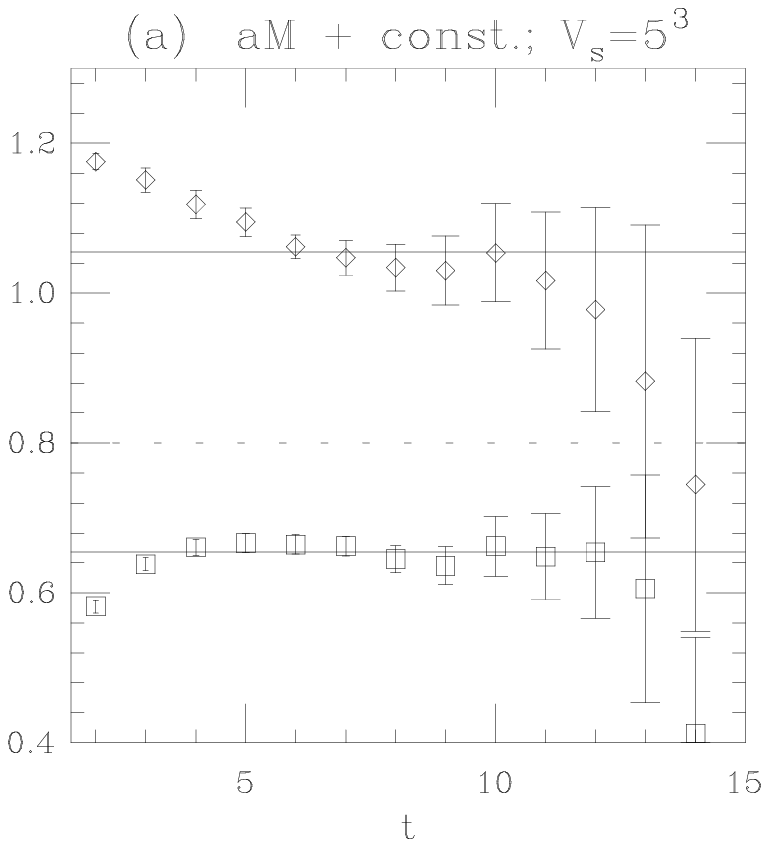}
\includegraphics{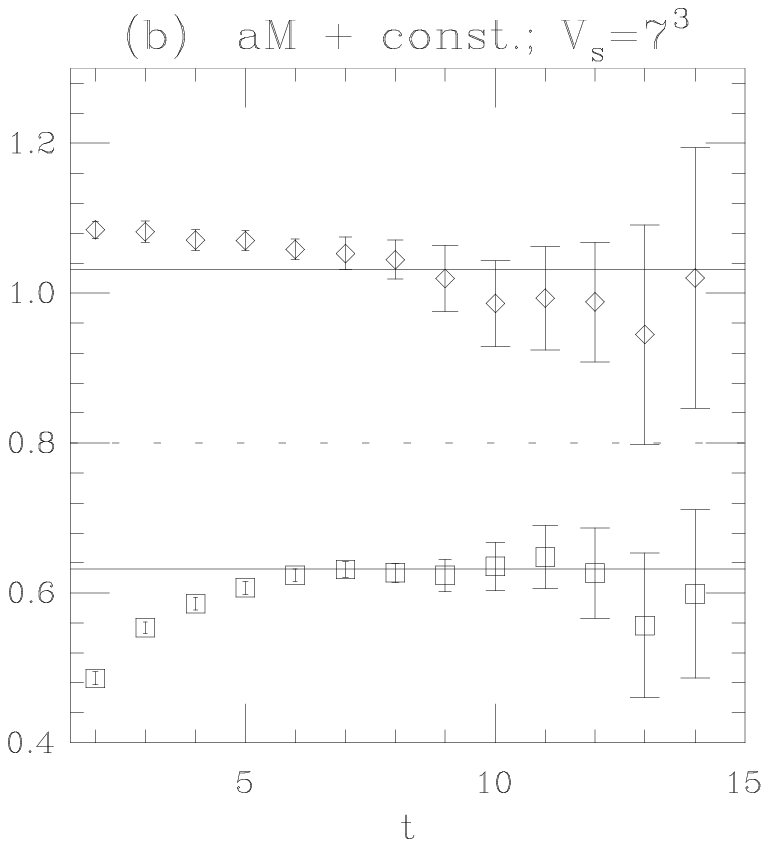}
\includegraphics{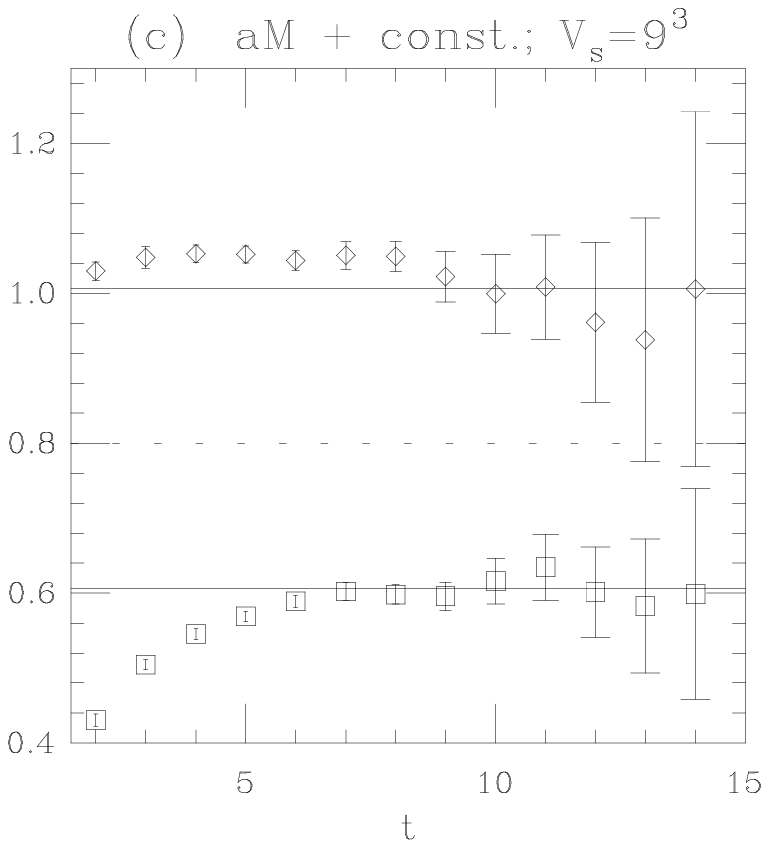}
\includegraphics{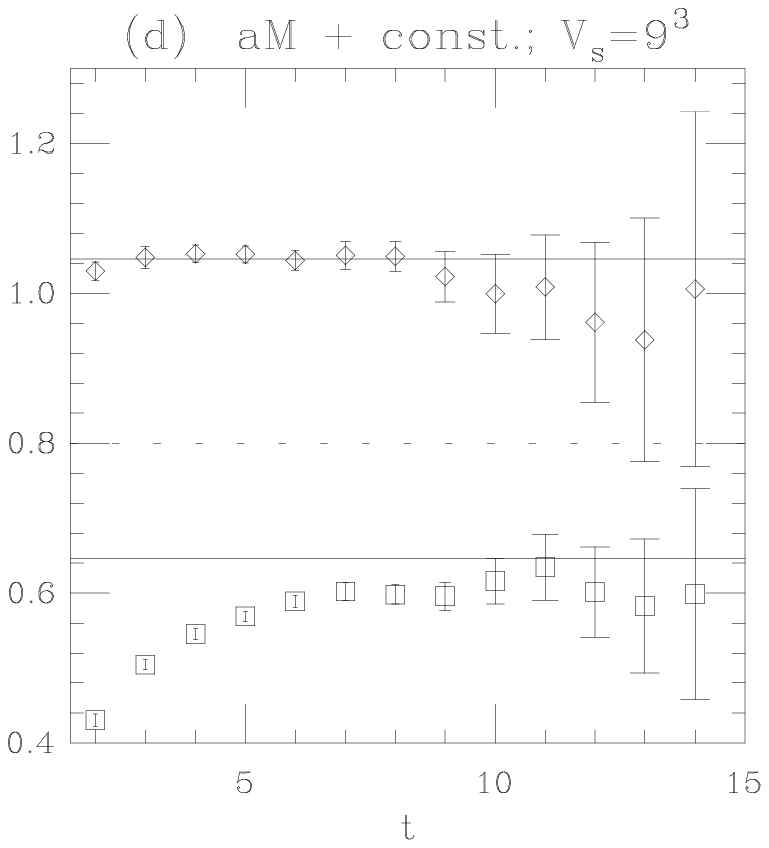}
\Caption
Static-light effective masses 
at \bsix, \latl, $\K_q=.155$, for different smearing volumes.
Diamonds: $\Gxx$ (cube-cube), const. = +.4;
Squares: $\Gax$ (cube-point), const. = 0. 
The solid lines indicate the mass from a coupled single-state
fit. In plots (a)-(c) the fit interval on both correlators is
$t=(5,10)$. In (d), the result for $V_s=9^3$ is repeated,
showing the fitted mass from the ranges $\Gxx(3,7)$ and $\Gax(9,13)$.
\endCaption
\endfigure

\fullfigure{figs-stat60}
\infiglist{Systematics (from smearing) in static amplitude, \bsix}
\vfill
\includegraphics{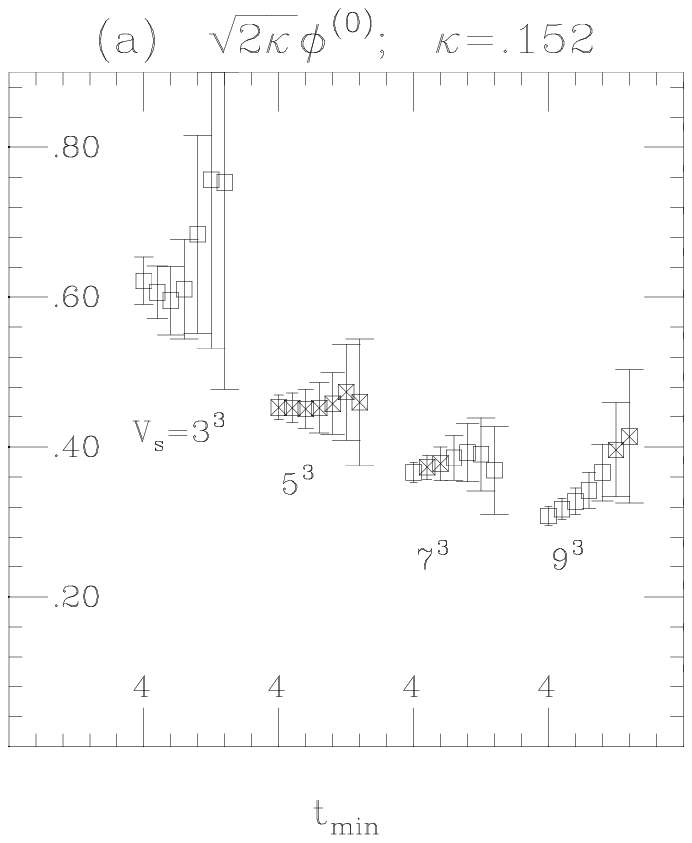}
\includegraphics{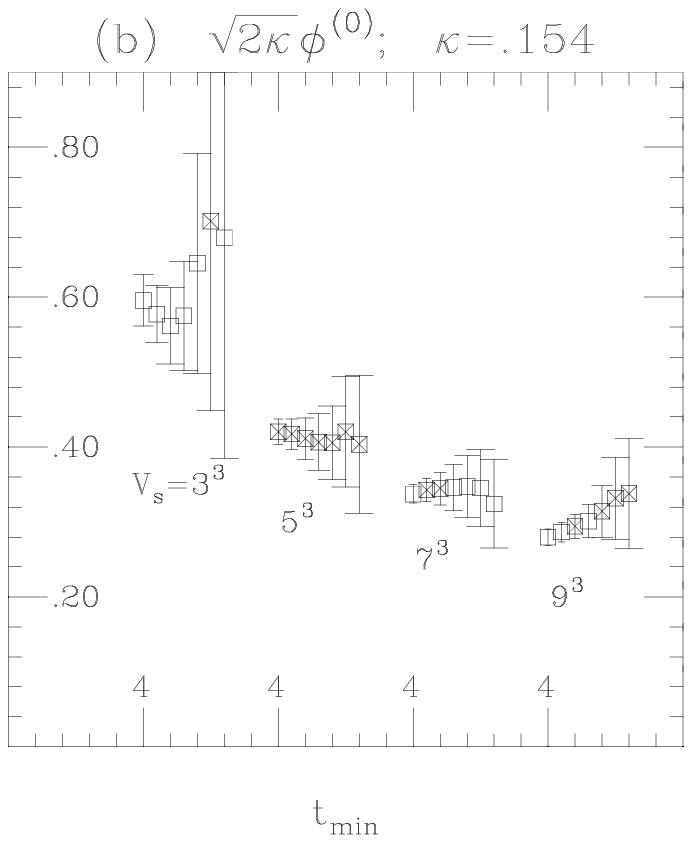}
\includegraphics{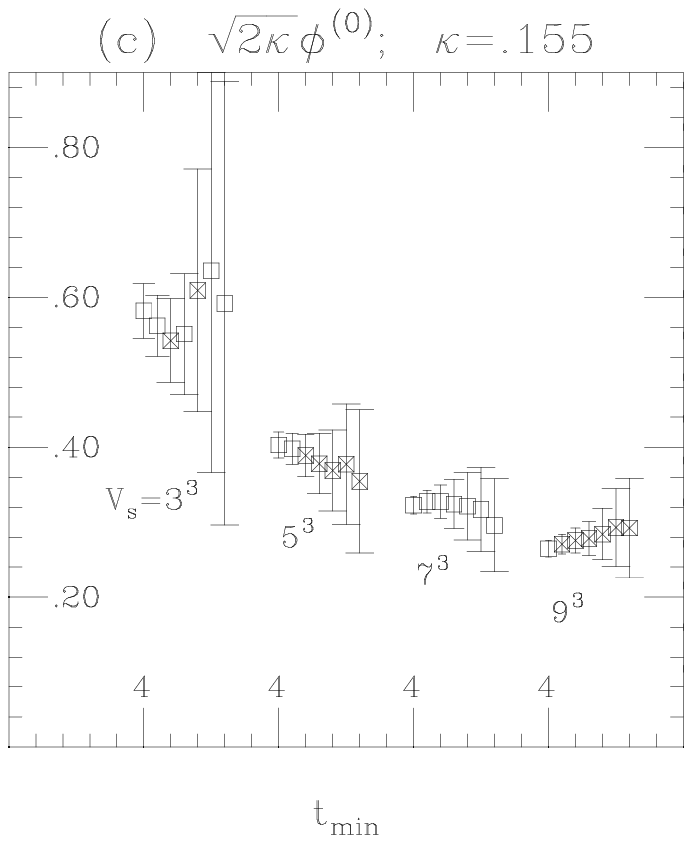}
\includegraphics{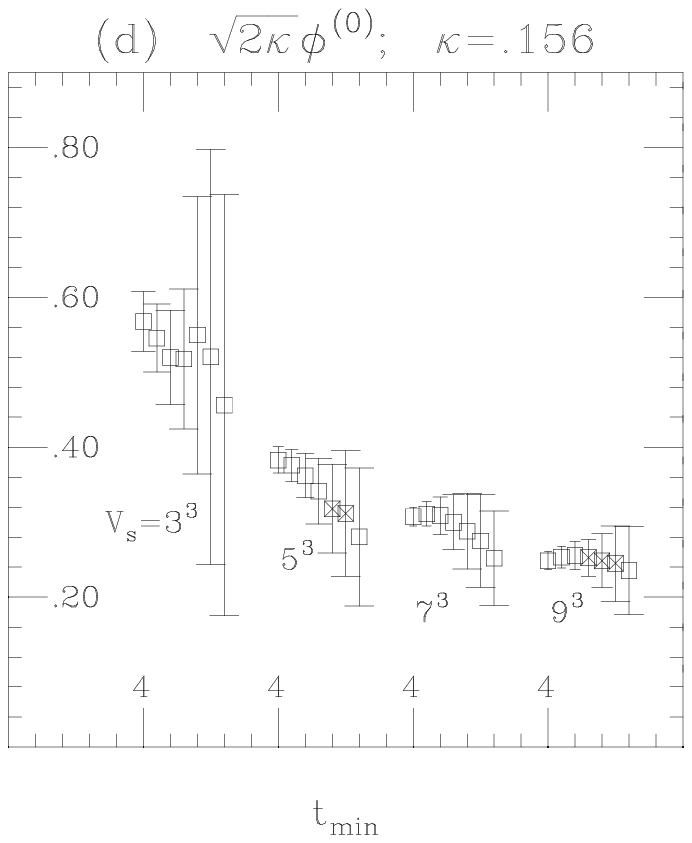}
\Caption
A systematic study of the static-light amplitude at \bsix\ as a
function of the light-quark mass, smearing size and fitting interval.
In the (coupled, single-state) fit that produced each of the results,
the fit intervals were $\Gxx(8,13)$ and $\Gax(t_{\rm min},t_{\rm min}+4)$,
where $t_{\rm min}$ is indicated on the horizontal axis.
The symbol $\times$ indicates $\chisqdof<0.1$ (non-covariant).
\endCaption
\endfigure

\fullfigure{fig-FvsMchE63}
\infiglist{$\phi\ \vs\ 1/M$, \bspthr, chiral extrapolation}
\vfill
\includegraphics{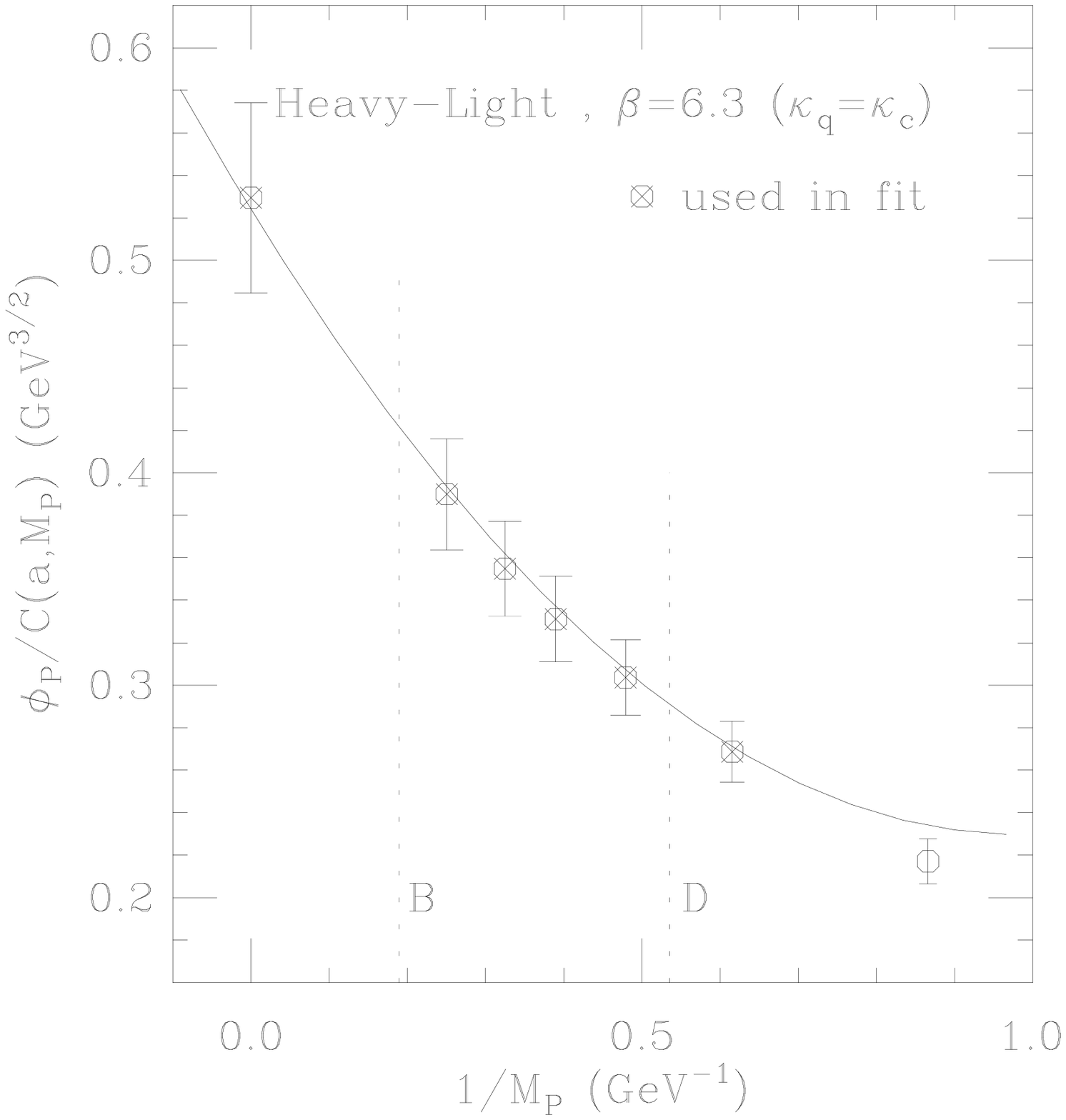}
\Caption
The combined (conventional and static) analysis at \bspthr.
The conventional points are from the wall-source results with the
large-$am$ corrections applied. The static point is from the 
cube source, $V_s=15^3$. The light quark has been extrapolated
to the chiral limit. The (covariant) fit is quadratic in $1/M_P$,
with $\chisqdof=1.9/3$.
\endCaption
\endfigure

\fullfigure{fig-FvsMstrEN63}
\infiglist{Decay amplitude normalization comparison at \bspthr}
\vfill
\includegraphics{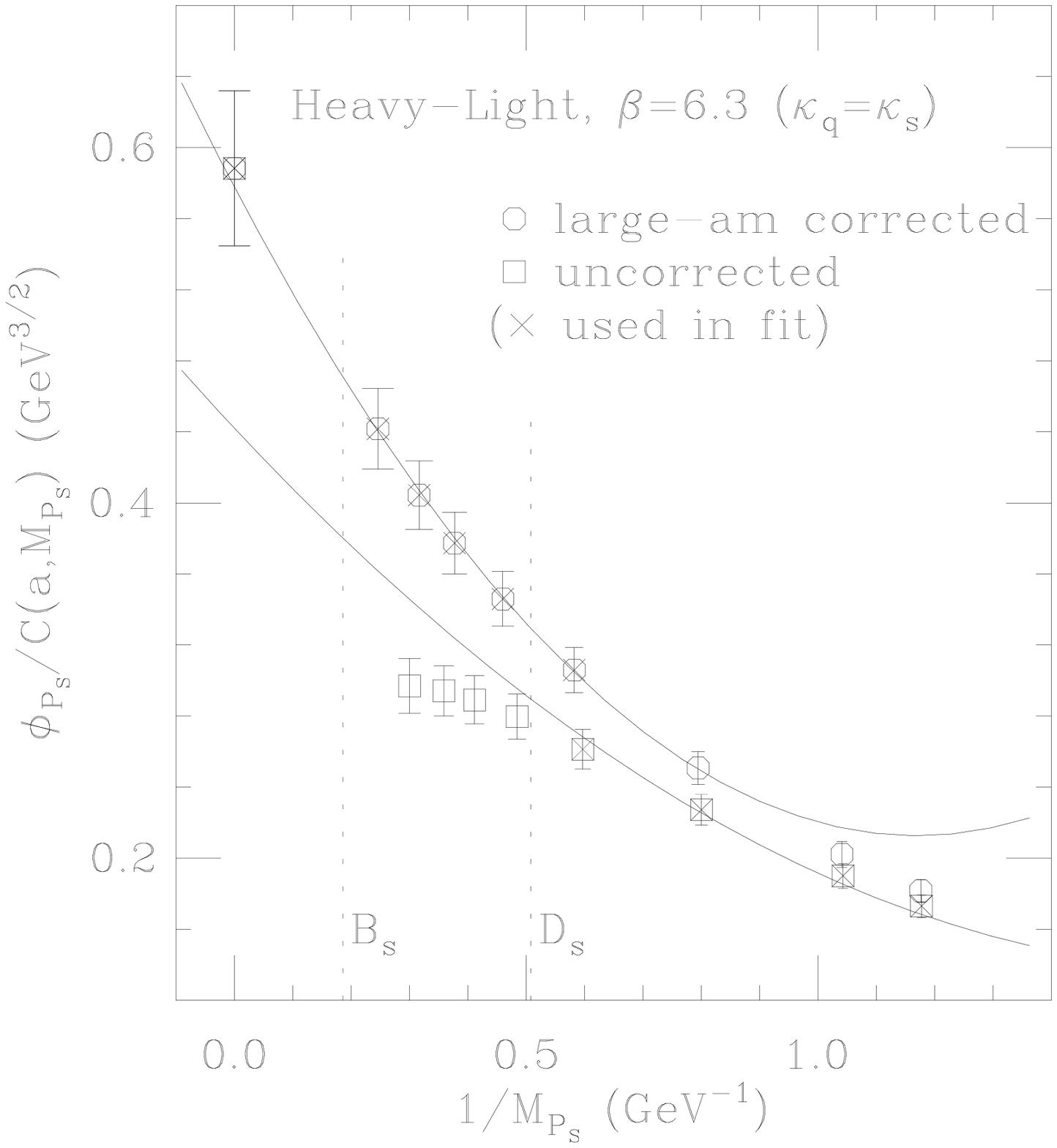}
\Caption
The heavy-light amplitude, with $\K_q=\K_s$. The circles are the results
with large-$am$ corrections included. The squares are the corresponding
results with those corrections removed. The fit to the latter is an
attempt to smoothly interpolate between the static limit and the conventional
results in a region where $\cO(am)$ errors might,
{\it a priori}, have been negligible.
(See text, Sect. \use{sect.analysis2}).
\endCaption
\endfigure

\fullfigure{fig-FvsMuncorr}
\infiglist{Decay amplitude normalization comparison at \bspthr}
\vfill
\includegraphics{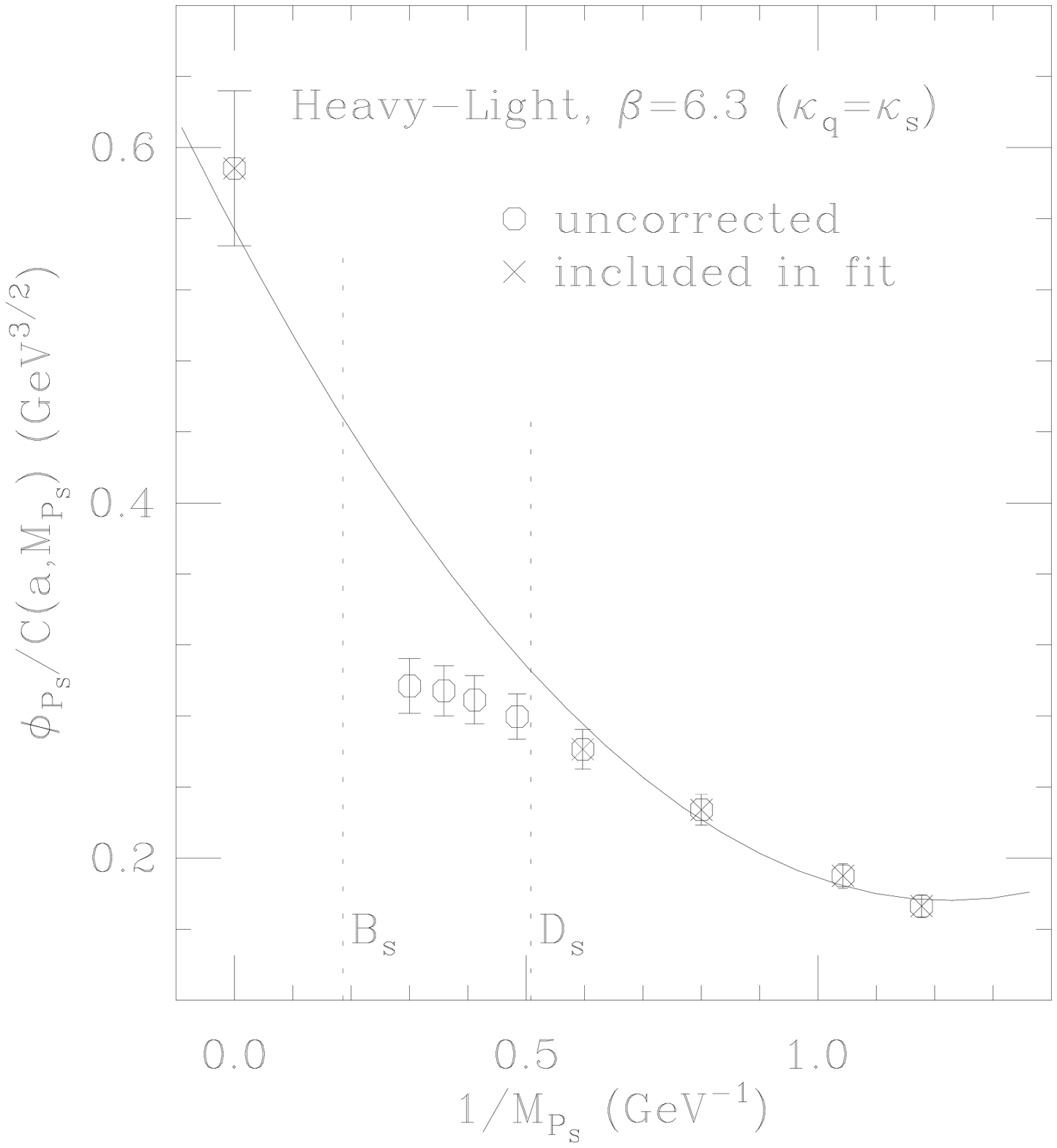}
\Caption
The heavy-light amplitude, again with $\K_q=\K_s$ and without including
the large-$am$ corrections. Here we show the effect of using an 
uncorrelated fit to the data, to be compared with the corresponding
result in \Fig{fig-FvsMstrEN63}.
\endCaption
\endfigure

\fullfigure{fig-new}
\infiglist{Fit to estimate systematic errors at \bspthr}
\vfill
\includegraphics{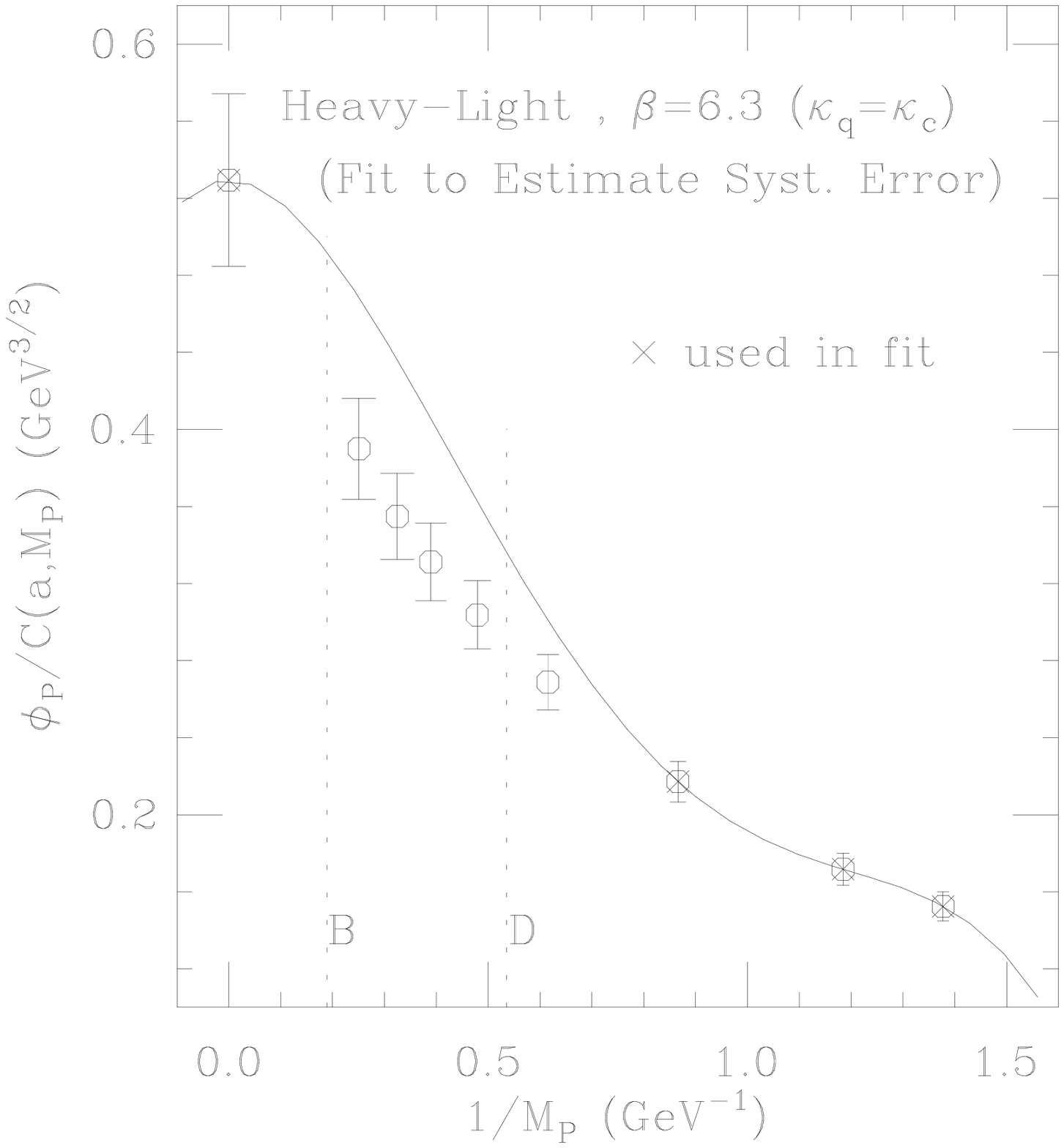}
\Caption
The heavy-light amplitude, with $\K_q=\K_c$, fit as described
in the text in order to estimate a bound on various systematic errors.
\endCaption
\endfigure

\fullfigure{fig-FvsMchE6063}
\infiglist{Scaling of $f\sqrt{M}$, large-$am$ corrected}
\vfill
\includegraphics{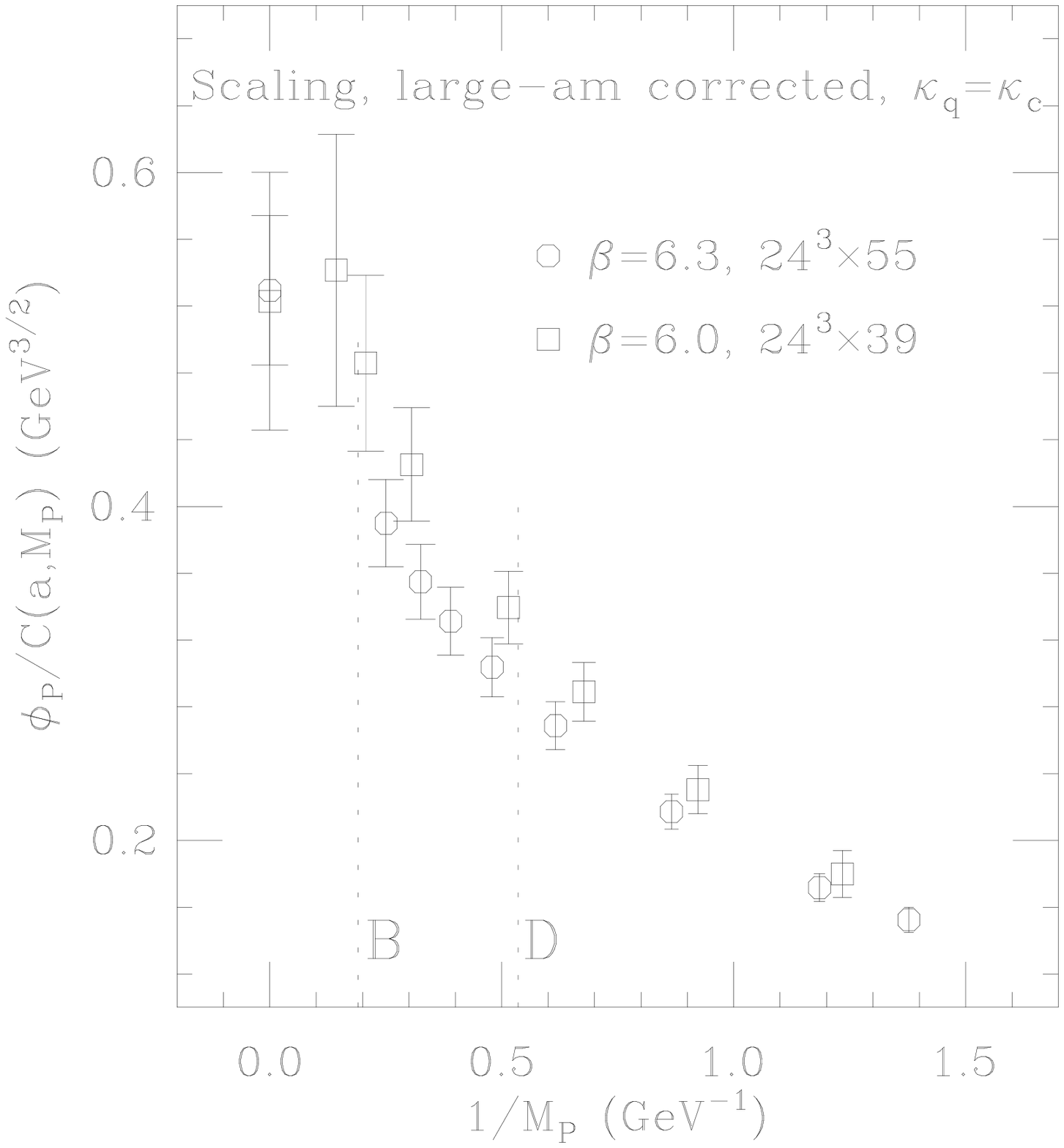}
\Caption
Scaling comparison (\bspthr\ and \bsix) of the heavy-light 
amplitude, computed with the large-$am$ corrections.
\endCaption
\endfigure

\fullfigure{fig-FvsMchN6063}
\infiglist{Scaling of $f\sqrt{M}$, uncorrected}
\vfill
\includegraphics{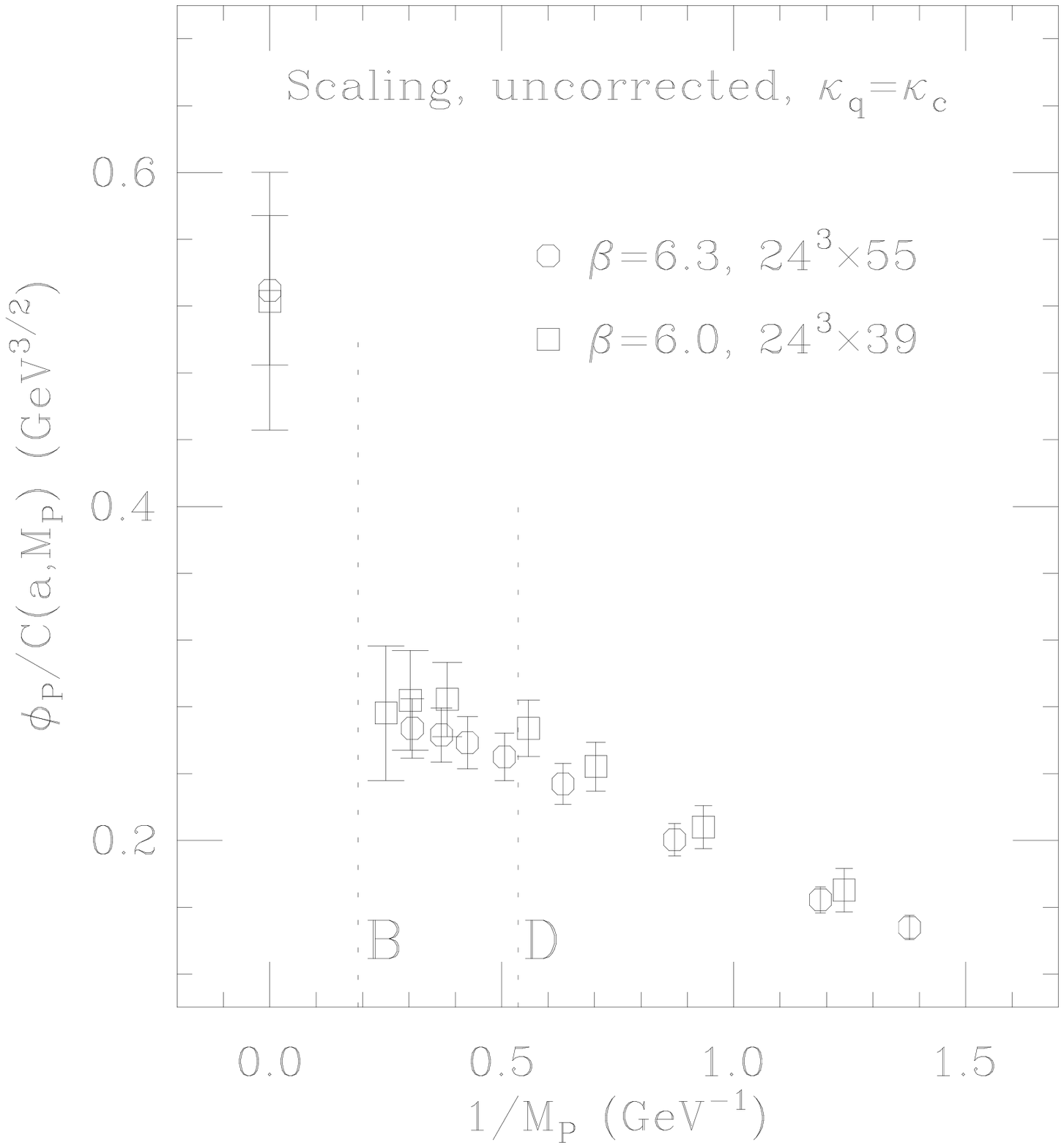}
\Caption
Scaling comparison (\bspthr\ and \bsix) of the heavy-light 
amplitude, computed without large-$am$ corrections.
\endCaption
\endfigure


\fulltable{tab-chir63W}
\intablelist{Light meson results, \bspthr, wall source}
\Caption
Light meson results at \bspthr\ (wall source). Fits were
covariant, single-state, over $t=(15,24)$ for both correlators.
The statistical error is from the jackknife of the fit;
the systematic (fitting) error is estimated as described in the text.
\endCaption
\vfill

\singlespaced
\ruledtable
$\K$ | $\chisqdof$ & $aM$ & $C_A$ & $af^{(0)}$ & 
$af\times10$ \cr
.149 | 2.0 & $.316(2)\pm.003$ & .250 & .263 & $.658(16)\pm.017$~ \crnorule
.150 | 2.8 & $.247(3)\pm.003$ & .245 & .234 & $.573(15)\pm.014$~ \crnorule
.1507 | 2.8 & $.188(4)\pm.004$ & .242 & .203 & $.491(13)\pm.013$~ \crnorule
$\K_c$  | - & $a\times m_\pi [\dag]$ & - & - & $.411(12)\pm.022[\ddag]$ \cr
\multispan{6} $ [\dag] \K_c=0.15158(5)\pm.00007; \quad\quad [\ddag] a^{-1}=3.21(9)\pm0.17 \GeV $ \hfil
\endruledtable
\bigskip\bigskip\bigskip

\singlespaced
\ruledtable
$\K_1$ & $\K_2$ | $\chisqdof$ & $aM$ & $C_A$ & 
$af^{(0)}$ & $af\times 10$ \cr
.149  &  .150   | 2.4 & $.283(2)\pm.003$ & .247 & .249 & $.615(15)\pm.018$ \crnorule
    &  .1507   | 2.9 & $.258(3)\pm.003$ & .246 & .234 & $.574(15)\pm.015$ \crnorule
    &  $\K_c$  |  -  & $.222(2)\pm.005$ & - & - & $.535(14)\pm.024$ \cr
.150  &  .1507   | 3.1 & $.218(3)\pm.004$ & .243 & .218 & $.531(14)\pm.011$ \crnorule
    &  $\K_c$  |  -  & $.176(2)\pm.003$ & - & - & $.491(13)\pm.017$ \cr
.1507  &  $\K_c$  |  -  & $.139(2)\pm.004$ & - & - & $.450(12)\pm.018$ \cr
$\K_s$ & $\K_c$ | - & $a\times m_K [\dag]$ & - & - & $.466(15)\pm.021[\ddag]$ \cr
\multispan{7}  $[\dag] \K_s=0.15043(7)\pm0.00013$; \quad\quad $[\ddag] f_K/f_{\pi}=1.134(6)\pm0.015$ \hfil 
\endruledtable
\vfill
\endtable

\fulltable{tab-chir63P}
\intablelist{Light meson results, \bspthr, point source}
\Caption
Light meson results at \bspthr\ (point source). Fits were
covariant, single-state, over $t=(15,21)$ for both correlators.
\endCaption
\vfill

\singlespaced
\ruledtable
$\K$ | $\chisqdof$ & $aM$ & $C_A$ & $af^{(0)}$ & 
$af\times10$ \cr
.149 | 1.9 & $.316(2)\pm.002$ & .250 & .279 & $.696(25)\pm.022$~ \crnorule
.150 | 1.6 & $.247(2)\pm.002$ & .245 & .243 & $.596(28)\pm.026$~ \crnorule
.1507 | 1.6 & $.192(2)\pm.002$ & .242 & .219 & $.529(31)\pm.019$~ \crnorule
$\K_c$  | - & $a\times m_\pi [\dag]$ & - & - & $.438(38)\pm.025[\ddag]$ \cr
\multispan{6} $ [\dag] \K_c=0.15163(3)\pm.00004; \quad\quad [\ddag] a^{-1}=3.01(26)\pm0.18 \GeV $ \hfil
\endruledtable
\bigskip\bigskip\bigskip

\singlespaced
\ruledtable
$\K_1$ & $\K_2$ | $\chisq$ & $aM$ & $C_A$ & 
$af^{(0)}$ & $af\times 10$ \cr
.149  &  .150   | 1.8 & $.283(2)\pm.002$ & .248 & .263 & $.651(26)\pm.019$ \crnorule
    &  .1507   | 1.6 & $.258(2)\pm.002$ & .246 & .245 & $.604(28)\pm.022$ \crnorule
    &  $\K_c$  |  -  & $.220(2)\pm.003$ & - & - & $.559(31)\pm.024$ \cr
.150  &  .1507   | 1.4 & $.221(2)\pm.002$ & .244 & .228 & $.556(30)\pm.022$ \crnorule
    &  $\K_c$  |  -  & $.178(1)\pm.002$ & - & - & $.505(33)\pm.018$ \cr
.1507 &  $\K_c$  |  -  & $.143(2)\pm.003$ & - & - & $.487(35)\pm.024$ \cr
$\K_s$ & $\K_c$ | - & $a\times m_K [\dag]$ & - & - & $.500(45)\pm.028[\ddag]$ \cr
\multispan{7}  $[\dag] \K_s=0.15029(28)\pm0.00016$; \quad\quad $[\ddag] f_K/f_{\pi}=1.142(13)\pm0.034$ \hfil 
\endruledtable
\vfill
\endtable

\fulltable{tab-chir60l}
\intablelist{Light meson results, \bsix, \latl}
\Caption
Light meson results at \bsix, \latl\ (point source). Fits were
non-covariant, single-state, over $t=(9,14)$ for both correlators.
$\K=.156$ was excluded from the extrapolations to $1/\K_c$ and $1/\K_s$.
\endCaption
\vfill

\singlespaced
\ruledtable
$\K$ | $aM$ & $C_A$ & $af^{(0)}$ & $af\times10$ \cr
.152 | $.489(9)\pm.001$ & .263 & .424 & $1.114(55)\pm.009$~ \crnorule
.154 | $.374(10)\pm.001$ & .254 & .354 & $.898(53)\pm.003$~ \crnorule
.155 | $.311(9)\pm.002$ & .249 & .315 & $.785(64)\pm.009$~ \crnorule
$\K_c$  | $a\times m_\pi [\dag]$ & - & - & $.575(90)\pm.021[\ddag]$ \cr
\multispan{5} $ [\dag] \K_c=0.15701(10)\pm.00018; \quad\quad [\ddag] a^{-1}=2.29(37)\pm.08 \GeV $ \hfil
\endruledtable
\bigskip\bigskip\bigskip

\singlespaced
\ruledtable
$\K_1$ & $\K_2$ | $aM$ & $C_A$ & 
$af^{(0)}$ & $af\times 10$ \cr
.152  &  .154   |  $.434(9)\pm.001$ & .258 & .388 & $1.002(51)\pm.005$ \crnorule
    &  .155   |  $.405(10)\pm.000$ & .256 & .366 & $.935(52)\pm.004$ \crnorule
    &  $\K_c$  | $.339(8)\pm.004$ & - & - & $.824(55)\pm.018$ \cr
.154  &  .155   |  $.343(9)\pm.001$ & .251 & .334 & $.839(57)\pm.008$ \crnorule
    &  $\K_c$  | $.268(7)\pm.003$ & - & - & $.738(67)\pm.013$ \cr
.155  &  $\K_c$  | $.230(5)\pm.002$ & - & - & $.691(78)\pm.010$ \cr
$\K_s$ & $\K_c$ | $a\times m_K [\dag]$ & - & - & $.681(112)\pm.022[\ddag]$ \cr
\multispan{6}  $[\dag] \K_s=0.15528(75)\pm0.00017$; \quad\quad $[\ddag] f_K/f_{\pi}=1.183(21)\pm0.014$ \hfil 
\endruledtable
\vfill
\endtable

\fulltable{tab-chir60A}
\intablelist{Light meson results, \bsix, \latA}
\Caption
Light meson results at \bsix, \latA\ (point source). Fits were
covariant, single-state, over $t=(9,14)$ for both correlators.
$\K=.156$ was excluded from the extrapolations to $1/\K_c$ and $1/\K_s$.
\endCaption
\vfill

\singlespaced
\ruledtable
$\K$ | $\chisqdof$ & $aM$ & $C_A$ & $af^{(0)}$ & $af\times10$ \cr
.152 | 1.0 & $.497(3)\pm.002$ & .263 & .406 & $1.067(44)\pm.023$~ \crnorule
.154 | 1.2 & $.380(3)\pm.005$ & .254 & .384 & $.973(50)\pm.034$~ \crnorule
.155 | 1.2 & $.311(3)\pm.010$ & .249 & .343 & $.855(55)\pm.028$~ \crnorule
$\K_c$  | - & $a\times m_\pi [\dag]$ & - & - & $.628(84)\pm.113[\ddag]$ \cr
\multispan{6} $ [\dag] \K_c=0.15696(8)\pm0.00020; \quad\quad [\ddag] a^{-1}=2.10(27)\pm0.32 \GeV $ \hfil
\endruledtable
\bigskip\bigskip\bigskip

\singlespaced
\ruledtable
$\K_1$ & $\K_2$ | $\chisqdof$ & $aM$ & $C_A$ & 
$af^{(0)}$ & $af\times 10$ \cr
.152  &  .154  | 0.9 & $.443(3)\pm.002$ & .258 & .389 & $1.003(45)\pm.018$ \crnorule
    &  .155   | 1.1 & $.412(3)\pm.004$ & .256 & .382 & $.976(47)\pm.033$ \crnorule
    &  $\K_c$  |  -  & $.346(5)\pm.006$ & - & - & $.924(59)\pm.045$ \cr
.154  &  .155  | 1.2 & $.346(3)\pm.008$ & .251 & .368 & $.924(53)\pm.032$ \crnorule
    &  $\K_c$  |  -  & $.270(3)\pm.007$ & - & - & $.831(65)\pm.052$ \cr
.155 &  $\K_c$  |  -  & $.227(2)\pm.005$ & - & - & $.720(67)\pm.050$ \cr
$\K_s$ & $\K_c$ | - & $a\times m_K [\dag]$ & - & - & $.753(108)\pm.099[\ddag]$ \cr
\multispan{7}  $[\dag] \K_s=0.15479(68)\pm0.00080$; \quad\quad $[\ddag] f_K/f_{\pi}=1.200(25)\pm0.047$ \hfil 
\endruledtable
\vfill
\endtable

\fulltable{tab-chir57J}
\intablelist{Light meson results, \bfps, \latfps}
\Caption
Light meson results at \bfps, \latfps\ (point source). Fits were
covariant, single-state, over $t=(6,9)$ for both correlators.
\endCaption
\vfill

\singlespaced
\ruledtable
$\K$ | $\chisqdof$ & $aM$ & $C_A$ & $af^{(0)}$ & $af\times10$ \cr
.160 | 2.5 & $.690(5)\pm.011$ & .289 & .634 & $1.834(96)\pm.044$~ \crnorule
.164 | 1.3 & $.510(5)\pm.006$ & .272 & .556 & $1.510(92)\pm.052$~ \crnorule
.166 | 0.8 & $.407(7)\pm.006$ & .263 & .515 & $1.354(90)\pm.044$~ \crnorule
$\K_c$  | - & $a\times m_\pi [\dag]$ & - & - & $1.124(95)\pm.070[\ddag]$ \cr
\multispan{6} $ [\dag] \K_c=0.16903(12)\pm0.00006; \quad\quad [\ddag] a^{-1}=1.17(10)\pm0.09 \GeV $ \hfil
\endruledtable
\bigskip\bigskip\bigskip

\singlespaced
\ruledtable
$\K_1$ & $\K_2$ | $\chisqdof$ & $aM$ & $C_A$ & 
$af^{(0)}$ & $af\times 10$ \cr
.160  &  .164   | 1.9 & $.606(5)\pm.007$ & .281 & .595 & $1.673(94)\pm.042$ \crnorule
    &  .166   | 1.6 & $.562(5)\pm.006$ & .277 & .573 & $1.586(92)\pm.042$ \crnorule
    &  $\K_c$  |  -  & $.489(4)\pm.007$ & - & - & $1.467(92)\pm.036$ \cr
.164  &  $\K_c$  |  -  & $.366(4)\pm.006$ & - & - & $1.325(93)\pm.048$ \cr
.166  &  $\K_c$  |  -  & $.306(5)\pm.003$ & - & - & $1.249(91)\pm.064$ \cr
$\K_s$ & $\K_c$ | - & $a\times m_K [\dag]$ & - & - & $1.384(135)\pm.075[\ddag]$ \cr
\multispan{7}  $[\dag] \K_s=0.16228(123)\pm0.00083$; \quad\quad $[\ddag] f_K/f_{\pi}=1.231(28)\pm0.021$ \hfil 
\endruledtable
\vfill
\endtable

\fulltable{tab-hevy63W}
\intablelist{Heavy-light results, \bspthr, wall source}
\Caption
Heavy-light, wall-source results at \bspthr. Fits were
covariant, two-state, over $t=(3,18)$ for both correlators.
\endCaption
\vfill

\singlespaced
\ruledtable
$\K_H$ & $\K_L$ | $\chisqdof$ & $\Delta m_H$ & 
$aM$  & $C_A$ & $a^{3 \over 2}\phi^{(0)}$ & 
$a^{3 \over 2}\phi\times10$ \cr
.148  & .149 | 0.8 & .000492 & $.345(1)\pm.002$ & .252 & .160 & $.403(5)\pm.009$ \crnorule
    & .150 | 0.9 &    & $.314(1)\pm.002$ & .250 & .143 & $.357(5)\pm.008$ \crnorule
    & .1507 | 1.2 &    & $.290(1)\pm.002$ & .248 & .129 & $.320(4)\pm.006$ \crnorule
    & $\K_s$ | -  &    & $.300(2)\pm.003$ &  -  & - & $.334(6)\pm.008$ \crnorule
    & $\K_c$ | -  &    & $.263(2)\pm.003$ &  -  & - & $.283(4)\pm.009$ \cr
.145  & .149 | 0.8 & .00277 & $.433(1)\pm.002$ & .259 & .191 & $.495(8)\pm.011$ \crnorule
    & .150 | 0.9 &    & $.405(1)\pm.002$ & .257 & .174 & $.445(7)\pm.011$ \crnorule
    & .1507 | 1.1 &    & $.384(1)\pm.001$ & .255 & .159 & $.404(6)\pm.009$ \crnorule
    & $\K_s$ | -  &    & $.392(2)\pm.003$ &  -  & - & $.420(8)\pm.011$ \crnorule
    & $\K_c$ | -  &    & $.360(1)\pm.002$ &  -  & - & $.362(7)\pm.012$ \cr
.140  & .149 | 1.0 & .0131 & $.572(1)\pm.002$ & .270 & .223 & $.602(15)\pm.012$ \crnorule
    & .150 | 1.0 &    & $.546(1)\pm.002$ & .268 & .204 & $.546(14)\pm.013$ \crnorule
    & .1507 | 1.2 &    & $.528(1)\pm.002$ & .266 & .189 & $.502(13)\pm.013$ \crnorule
    & $\K_s$ | -  &    & $.535(2)\pm.003$ &  -  & - & $.519(14)\pm.013$ \crnorule
    & $\K_c$ | -  &    & $.507(1)\pm.002$ &  -  & - & $.455(14)\pm.015$ \cr
.135  & .149 | 1.2 & .0343 & $.712(1)\pm.002$ & .281 & .245 & $.688(22)\pm.015$ \crnorule
    & .150 | 1.2 &    & $.688(1)\pm.002$ & .278 & .224 & $.623(20)\pm.015$ \crnorule
    & .1507 | 1.3 &    & $.671(1)\pm.002$ & .276 & .208 & $.574(18)\pm.016$ \crnorule
    & $\K_s$ | -  &    & $.678(2)\pm.003$ &  -  & - & $.593(20)\pm.015$ \crnorule
    & $\K_c$ | -  &    & $.651(2)\pm.002$ &  -  & - & $.519(19)\pm.019$ \cr
.130  & .149 | 1.4 & .0683 & $.859(2)\pm.002$ & .291 & .260 & $.756(31)\pm.020$ \crnorule
    & .150 | 1.3 &    & $.836(1)\pm.002$ & .288 & .237 & $.683(25)\pm.016$ \crnorule
    & .1507 | 1.5 &    & $.820(1)\pm.002$ & .286 & .221 & $.632(22)\pm.018$ \crnorule
    & $\K_s$ | -  &    & $.826(2)\pm.003$ &  -  & - & $.652(25)\pm.017$ \crnorule
    & $\K_c$ | -  &    & $.801(2)\pm.002$ &  -  & - & $.571(23)\pm.021$ \cr
.125  & .149 | 1.6 & .117 & $1.017(2)\pm.003$ & .301 & .272 & $.818(35)\pm.026$ \crnorule
    & .150 | 1.5 &    & $.994(1)\pm.002$ & .298 & .246 & $.734(31)\pm.017$ \crnorule
    & .1507 | 1.6 &    & $.980(1)\pm.003$ & .296 & .231 & $.684(26)\pm.021$ \crnorule
    & $\K_s$ | -  &    & $.985(2)\pm.003$ &  -  & - & $.704(29)\pm.020$ \crnorule
    & $\K_c$ | -  &    & $.960(2)\pm.002$ &  -  & - & $.617(27)\pm.023$ \cr
.117  & .149 | 1.8 & .23 & $1.298(2)\pm.003$ & .316 & .284 & $.898(42)\pm.034$ \crnorule
    & .150 | 1.7 &    & $1.276(2)\pm.002$ & .313 & .257 & $.805(36)\pm.023$ \crnorule
    & .1507 | 1.9 &    & $1.263(2)\pm.003$ & .311 & .244 & $.758(35)\pm.028$ \crnorule
    & $\K_s$ | -  &    & $1.268(2)\pm.003$ &  -  & - & $.777(36)\pm.026$ \crnorule
    & $\K_c$ | -  &    & $1.245(2)\pm.003$ &  -  & - & $.685(36)\pm.027$ 
\endruledtable
\vfill
\endtable

\fulltable{tab-hevy63P}
\intablelist{Heavy-light results, \bspthr, point source}
\Caption
Heavy-light, point-source results at \bspthr. Fits were
covariant, single-state, over $\Gax(17,23)$ and $\Gxx(18,23)$.
\endCaption
\vfill

\singlespaced
\ruledtable
$\K_H$ & $\K_L$ | $\chisqdof$ & $\Delta m_H$ & 
$aM$  & $C_A$ & $a^{3 \over 2}\phi^{(0)}$ & 
$a^{3 \over 2}\phi\times10$ \cr
.148  & .149 | 1.1 & .000514 & $.351(2)\pm.003$ & .252 & .175 & $.442(20)\pm.016$ \crnorule
    & .150 | 1.3 &    & $.320(2)\pm.003$ & .250 & .160 & $.401(20)\pm.015$ \crnorule
    & .1507 | 1.5 &    & $.296(2)\pm.002$ & .248 & .147 & $.365(19)\pm.016$ \crnorule
    & $\K_s$ | -  &    & $.310(9)\pm.007$ &  -  & - & $.385(28)\pm.020$ \crnorule
    & $\K_c$ | -  &    & $.267(2)\pm.004$ &  -  & - & $.327(21)\pm.023$ \cr
.140  & .149 | 0.8 & .0133 & $.578(2)\pm.001$ & .270 & .236 & $.639(20)\pm.015$ \crnorule
    & .150 | 0.8 &    & $.553(2)\pm.001$ & .268 & .216 & $.580(21)\pm.013$ \crnorule
    & .1507 | 0.9 &    & $.534(2)\pm.001$ & .266 & .203 & $.540(21)\pm.014$ \crnorule
    & $\K_s$ | -  &    & $.545(7)\pm.005$ &  -  & - & $.563(29)\pm.021$ \crnorule
    & $\K_c$ | -  &    & $.511(3)\pm.002$ &  -  & - & $.488(23)\pm.016$ \cr
.125  & .149 | 1.4 & .117 & $1.026(2)\pm.002$ & .301 & .287 & $.864(22)\pm.014$ \crnorule
    & .150 | 1.6 &    & $1.004(2)\pm.002$ & .298 & .265 & $.791(23)\pm.014$ \crnorule
    & .1507 | 1.7 &    & $.987(3)\pm.001$ & .296 & .249 & $.737(24)\pm.014$ \crnorule
    & $\K_s$ | -  &    & $.997(7)\pm.003$ &  -  & - & $.768(33)\pm.024$ \crnorule
    & $\K_c$ | -  &    & $.966(4)\pm.003$ &  -  & - & $.672(27)\pm.021$ \cr
.110  & .149 | 2.6 & .371 & $1.588(12)\pm.010$ & .329 & .314 & $1.033(33)\pm.030$ \crnorule
    & .150 | 2.4 &    & $1.572(4)\pm.005$ & .326 & .283 & $.925(31)\pm.017$ \crnorule
    & .1507 | 2.2 &    & $1.557(5)\pm.004$ & .324 & .265 & $.860(35)\pm.017$ \crnorule
    & $\K_s$ | -  &    & $1.566(7)\pm.004$ &  -  & - & $.899(46)\pm.021$ \crnorule
    & $\K_c$ | -  &    & $1.539(8)\pm.006$ &  -  & - & $.765(43)\pm.022$ \cr
.100  & .149 | 2.7 & .652 & $2.077(8)\pm.009$ & .347 & .316 & $1.097(41)\pm.032$ \crnorule
    & .150 | 2.6 &    & $2.058(5)\pm.008$ & .343 & .288 & $.991(45)\pm.037$ \crnorule
    & .1507 | 2.5 &    & $2.045(7)\pm.006$ & .341 & .276 & $.942(55)\pm.027$ \crnorule
    & $\K_s$ | -  &    & $2.053(8)\pm.005$ &  -  & - & $.973(58)\pm.026$ \crnorule
    & $\K_c$ | -  &    & $2.028(9)\pm.006$ &  -  & - & $.849(64)\pm.037$ 
\endruledtable
\vfill
\endtable

\fulltable{tab-hevy60l}
\intablelist{Heavy-light results, \bsix, \latl}
\Caption
Heavy-light results at \bsix, \latl. Fits were non-covariant,
single-state, over $t=(10,15)$ for both correlators.
\endCaption
\vfill

\singlespaced
\ruledtable
$\K_Q$ & $\K_q$ | $\Delta m_Q$ & 
$aM$  & $C_A$ & $a^{3 \over 2}\phi^{(0)}$ & 
$a^{3 \over 2}\phi\times10$ \cr
.152  & .152 |  .00116 & $.488(8)\pm.001$ & .263 & .291 & $.766(39)\pm.006$ \crnorule
    & .154 |     & $.434(9)\pm.001$ & .258 & .254 & $.655(36)\pm.004$ \crnorule
    & .155 |     & $.406(9)\pm.001$ & .256 & .233 & $.595(36)\pm.003$ \crnorule
    & $\K_s$ | & $.399(19)\pm.004$ &  -  & - & $.583(69)\pm.007$ \crnorule
    & $\K_c$ | & $.354(8)\pm.005$ &  -  & - & $.495(37)\pm.010$ \cr
.148  & .152 |  .00602 & $.596(8)\pm.001$ & .272 & .353 & $.961(49)\pm.008$ \crnorule
    & .154 |     & $.546(9)\pm.001$ & .267 & .311 & $.831(43)\pm.007$ \crnorule
    & .155 |     & $.521(10)\pm.001$ & .265 & .288 & $.761(41)\pm.006$ \crnorule
    & $\K_s$ | & $.515(17)\pm.004$ &  -  & - & $.746(70)\pm.010$ \crnorule
    & $\K_c$ | & $.474(10)\pm.004$ &  -  & - & $.643(39)\pm.008$ \cr
.142  & .152 |  .024 & $.757(8)\pm.001$ & .285 & .421 & $1.199(61)\pm.011$ \crnorule
    & .154 |     & $.712(10)\pm.001$ & .280 & .373 & $1.044(55)\pm.010$ \crnorule
    & .155 |     & $.689(11)\pm.001$ & .277 & .346 & $.961(52)\pm.010$ \crnorule
    & $\K_s$ | & $.683(14)\pm.004$ &  -  & - & $.942(73)\pm.014$ \crnorule
    & $\K_c$ | & $.646(12)\pm.003$ &  -  & - & $.818(48)\pm.009$ \cr
.135  & .152 |  .0656 & $.954(8)\pm.001$ & .299 & .475 & $1.422(70)\pm.014$ \crnorule
    & .154 |     & $.912(10)\pm.002$ & .294 & .422 & $1.241(66)\pm.013$ \crnorule
    & .155 |     & $.891(12)\pm.002$ & .291 & .392 & $1.142(64)\pm.014$ \crnorule
    & $\K_s$ | & $.885(13)\pm.004$ &  -  & - & $1.121(77)\pm.018$ \crnorule
    & $\K_c$ | & $.850(13)\pm.003$ &  -  & - & $.973(61)\pm.012$ \cr
.118  & .152 |  .286 & $1.527(8)\pm.003$ & .332 & .558 & $1.850(87)\pm.020$ \crnorule
    & .154 |     & $1.488(11)\pm.003$ & .326 & .494 & $1.609(91)\pm.020$ \crnorule
    & .155 |     & $1.467(13)\pm.004$ & .323 & .457 & $1.476(94)\pm.023$ \crnorule
    & $\K_s$ | & $1.462(12)\pm.004$ &  -  & - & $1.448(89)\pm.028$ \crnorule
    & $\K_c$ | & $1.429(16)\pm.003$ &  -  & - & $1.247(101)\pm.022$ \cr
.103  & .152 |  .669 & $2.209(9)\pm.004$ & .358 & .610 & $2.183(115)\pm.024$ \crnorule
    & .154 |     & $2.171(12)\pm.004$ & .352 & .538 & $1.892(129)\pm.025$ \crnorule
    & .155 |     & $2.150(15)\pm.005$ & .348 & .496 & $1.728(139)\pm.030$ \crnorule
    & $\K_s$ | & $2.145(14)\pm.004$ &  -  & - & $1.695(115)\pm.035$ \crnorule
    & $\K_c$ | & $2.113(20)\pm.004$ &  -  & - & $1.450(160)\pm.031$ \cr
.088  & .152 |  1.31 & $3.154(12)\pm.005$ & .382 & .657 & $2.511(171)\pm.029$ \crnorule
    & .154 |     & $3.116(16)\pm.006$ & .376 & .577 & $2.167(196)\pm.031$ \crnorule
    & .155 |     & $3.095(20)\pm.006$ & .372 & .530 & $1.973(214)\pm.036$ \crnorule
    & $\K_s$ | & $3.090(18)\pm.005$ &  -  & - & $1.933(176)\pm.042$ \crnorule
    & $\K_c$ | & $3.058(26)\pm.005$ &  -  & - & $1.642(253)\pm.038$ 
\endruledtable
\vfill
\endtable

\fulltable{tab-statcomp}
\intablelist{A comparison of static results at \bsix.}
\thicksize=1.2pt
{\singlespaced\tenpoint\tolerance=2000

\ruledtable
Group |\vctr{Smearing} | \vctr{Fitting} |
\multispan{2} $\tilde\phi a^{3/2} \equiv \sqrt{2\K}\hat{\phi}^{(0)}a^{3/2}$ | 
\multispan{3} extrapolation in $1/\K$ \crnorule
\cskip|\cskip|\cskip| \crule | \crule  | \crule | \crule | \crule \crpart
(Lattice)|   |      |$\K\!=.154$ | $\K=\!.155$ | slope   |
$\tilde\phi(\K_c)a^{3/2}$ | $\fBstat$(MeV) \CR

ELC\cite{elc-stat}|Coul. gauge, | $\Gxx(4,9)$;\hfil|
$(\K\!=\!.153)$ | $(\K\!=\!.1545)$ | \cskip | \cskip | \cskip \crpart

$10^2\!\times\!20\!\times\!36$ |cube, $V_s\!=\!7^3$| $\Gax/\Gxx(4,9)$  |
$.356(25)$|$.337(24)$| .31    | .304(24) | 323(26) \cr

Wuppertal\cite{alexandrou}|Coul. gauge, | ``$\Gxx$''$(t\geq2)$;\hfil |
\vctr{$.371(20)$} | \vctr{$.359(20)$} | \vctr{.30} | \vctr{.334(20)} |
\vctr{355(21)} \crnorule

$12^3\!\times\!36$ |exp. WF \hfil | ``$\Gax$''$(t\geq5)$\hfil | | | | |\cr

Lat'90\cite{lat90}|Landau gauge, | $\Gxx(5,10)$,\hfil|
\vctr{$.427(20)$} | \vctr{$.416(20)$} | \vctr{.26} | \vctr{.394(20)} |
\vctr{419(21)} \crnorule

$24^3\!\times\!39$ |cube, $V_s\!=\!5^3$| $\Gax(5,10)$\hfil | | | | | \CR

This work | Landau gauge, | coupled fit,\hfil |
\vctr{$.419(18)$} | \vctr{$.402(18)$} | \vctr{.41} | \vctr{.369(18)} |
\vctr{392(19)} \crnorule

$24^3\!\times\!39$ |cube, $V_s\!=\!5$| $t=(5,10)$\hfil | | | | | \crnorule
\cskip|\crule|\crule| \crule | \crule  | \crule | \crule | \crule \crpart

 |$V_s\!=\!7^3$ | $t=(4,9)$\hfil | $.337(10)$ |$.324(10)$|.31|.299(10) |
\vctr{318(11)} \crnorule
\cskip|\crule|\crule| \crule | \crule  | \crule | \crule | \crule \crpart

 |$V_s\!=\!9^3$ | $t=(5,10)$\hfil | $.282(11)$ |$.268(12)$|.33|.241(12) |
\vctr{256(13)} \crnorule
\cskip|\Crule|\Crule| \Crule | \Crule  | \Crule | \Crule | \Crule \crpart

 |$V_s\!=\!9^3$ | $t=(8,13)$\hfil | $.315(36)$ |$.284(35)$|.74|.223(35) |
\vctr{237(37)} \crnorule
\cskip|\crule|\crule| \crule | \crule  | \crule | \crule | \crule \crpart

 |$V_s\!=\!9^3$ | $\Gxx(2,9)$,\hfil|
\vctr{$.387(23)$} | \vctr{$.372(23)$} | \vctr{.36} | \vctr{.343(23)} |
\vctr{364(24)} \crnorule
 | | $\Gax(10,14)$\hfil | | | | | 

\endruledtable}
\Caption
A comparison of static results at \bsix.
The results at specific $\K$ values are taken directly
from the papers, but
all extrapolations to $\K_c$ are ours, using our
values for $\K_c$, $a^{-1}$, and the perturbative corrections to get
$\fBstat$.
We take the results from\cite{alexandrou}, rather than\cite{newalexandrou},
since both $\kappa=.154$ and $\kappa=.155$ are included
in the former.  The results at $\K = .154$ in\cite{newalexandrou} are
completely consistent with those in\cite{alexandrou}.
\endCaption
\endtable
\thicksize=\thinsize

\midtable{tab-stat63W}
\intablelist{Static results, \bspthr, wall source}
\Caption
Static-light, wall-source results at \bspthr. Fits were
covariant, two-state, over $\Gax(3,12)$ and $\Gxx(5,14)$.
\endCaption
\vfill

\singlespaced
\ruledtable
$\K$ | $\chisqdof$ & $a{\cal E}^{(0)}$ & $\hat{C}_A/C(a,m)$ & 
$a^{3\over2}\hat{\phi}^{(0)}$ & $a^{3\over2}\hat{\phi}\times 10$ \cr
.148 | 0.8 & $.537(3)\pm.004$ & .395 & .290 & $1.15(7)\pm.09$~ \crnorule
.149 | 0.8 & $.518(4)\pm.004$ & .392 & .268 & $1.05(6)\pm.08$~ \crnorule
.150 | 0.7 & $.499(4)\pm.004$ & .388 & .246 & $.96(6)\pm.08$~ \crnorule
.1507 | 0.7 & $.487(6)\pm.006$ & .385 & .231 & $.89(8)\pm.13$~ \crnorule
$\K_s$  | - & $.491(5)\pm.006$ & - & - & $.92(7)\pm.09$ \crnorule 
$\K_c$  | - & $.470(6)\pm.007$ & - & - & $.81(7)\pm.11$
\endruledtable
\vfill
\endtable
\vbox{\vskip 1cm}

\midtable{tab-stat63P}
\intablelist{Static results, \bspthr, smeared $n=15$}
\Caption
Static-light, cube-source results ($V_s=15^3$) at \bspthr. Fits were
covariant, single-state, over $t=(10,16)$ for both correlators.
\endCaption
\vfill
\singlespaced
\ruledtable
$\K$ | $\chisqdof$ & $a{\cal E}^{(0)}$ & $\hat{C}_A/C(a,m)$ & 
$a^{3\over2}\hat{\phi}^{(0)}$ & $a^{3\over2}\hat{\phi}\times 10$ \cr
.149 | 0.7 & $.525(10)\pm.008$ & .392 & .294 & $1.15(6)\pm.06$~ \crnorule
.150 | 0.7 & $.510(11)\pm.009$ & .388 & .275 & $1.07(6)\pm.06$~ \crnorule
.1507 | 0.6 & $.499(11)\pm.009$ & .385 & .259 & $1.00(6)\pm.05$~ \crnorule
$\K_s$  | - & $.503(11)\pm.009$ & - & - & $1.02(6)\pm.06$ \crnorule 
$\K_c$  | - & $.486(12)\pm.010$ & - & - & $0.92(6)\pm.06$ \cr
\multispan{6} $ \hat{\phi}(\K_c)/\hat{\phi}(\K_s)=0.900(16)\pm0.018 $ \hfil
\endruledtable
\vfill
\endtable
\vbox{\vskip 1cm}

\midtable{tab-stat60l}
\intablelist{Static results, \bsix, non-covariant fits}
\Caption
Static-light, cube-source ($V_s=9^3$) results at \bsix. Fits were
non-covariant, single-state, over $t=(8,13)$ for both correlators.
\endCaption
\vfill

\singlespaced
\ruledtable
$\K$ |  $a{\cal E}^{(0)}$ & $\hat{C}_A/C(a,m)$ & 
$a^{3\over2}\hat{\phi}^{(0)}$ & $a^{3\over2}\hat{\phi}\times 10$ \cr
.152 |  $.676(19)\pm.015$ & .390 & .666 & $2.60(25)\pm.21$~ \crnorule
.154 |  $.633(21)\pm.010$ & .384 & .566 & $2.17(22)\pm.11$~ \crnorule
.155 |  $.608(24)\pm.005$ & .380 & .509 & $1.93(21)\pm.05$~ \crnorule
$\K_s$  |  $.603(26)\pm.006$ & - & - & $1.88(23)\pm.06$ \crnorule 
$\K_c$  |  $.566(28)\pm.007$ & - & - & $1.51(21)\pm.09$
\endruledtable
\vfill
\endtable


\midtable{tab-scalesys}
\intablelist{Numerical study of scale systematics at \bspthr}
\Caption
Interpolations from $\phi\ \vs\ 1/M$ with shifted scales (\bspthr)
\bigskip\bigskip\bigskip
\endCaption

\singlespaced
\ruledtable
Analysis & $\ainv$ (GeV) & $\K_q$ & $\chisqdof$ & $f_{cq}$ (MeV) & $f_{bq}$ (MeV) \cr
Large-$am$      & 3.44   &  $\K_c$  & 1.9/3 & 220(7) & 202(8) \crnorule
Corrected       &        &  $\K_s$  & 3.5/3 & 248(6) & 229(8) \cr
Uncorrected     & 3.44   &  $\K_c$  & 18/2  & 204(5) & 168(6) \crnorule
                &        &  $\K_s$ & 17/2 & 224(4) & 189(6) 
\endruledtable
\endtable

\table{tab-decay}
\intablelist{Decay constants and systematic error estimates}
\caption{Decay constants and systematic error estimates}

\singlespaced
\ruledtable
\vctr{meson} | \vctr{$f$ (MeV)} & fitting & scale & large-$am$ \crnorule
                | & \& extrap. & ($\ainvMrho$) & \cr
$B$             |  187(10)      & $\pm12$ & $\pm15$ & $\pm32$ \crnorule
$B_s$           |  207(9)       & $\pm10$ & $\pm22$ & $\pm32$ \crnorule
$D$             |  208(9)       & $\pm11$ & $\pm12$ & $\pm33$ \crnorule
$D_s$           |  230(8)       & $\pm10$ & $\pm18$ & $\pm28$
\endruledtable
\endtable

\table{tab-ratios}
\intablelist{Jackknifed Ratios}
\caption{Jackknifed Ratios}

\singlespaced
\ruledtable
\vctr{Ratios}| \vctr{central} & fitting  & scale & large-$am$ \crnorule
             |     & \& extrap. & ($\ainvMrho$) & \cr
$f_B/f_D$         |.90(3) & $\pm.02$ & $\pm.02$ & $\pm.01$ \crnorule
$f_{B_s}/f_{D_s}$ |.90(2) & $\pm.02$ & $\pm.02$ & $\pm.02$ \crnorule
$f_{B}/f_{B_s}$   |.90(2) & $\pm.03$ & $\pm.02$ & $\pm.02$ \crnorule
$f_D/f_{D_s}$     |.90(2) & $\pm.02$ & $\pm.02$ & $\pm.03$
\endruledtable
\endtable
\bye